\documentclass[12pt]{article}
\usepackage{amssymb,amsmath}
\usepackage{comment}
\usepackage{graphicx}
\usepackage{bm}
\usepackage{color}
\usepackage{here}
\usepackage{enumerate}
\usepackage{here}
\usepackage{subfigure} 
\DeclareMathOperator*{\Tr}{{\rm Tr}}
\DeclareMathOperator*{\tr}{{\rm tr}}

\newcommand{\Str}{\mathrm{Str}}
\newcommand{\Sdet}{\mathrm{Sdet}}
\newcommand{\Det}{\mathrm{Det}}
\newcommand{\ket}[1]{\left\vert #1 \right\rangle}

\numberwithin{equation}{section}

\begin{document}

\thispagestyle{empty}
\begin{flushright}
DCPT-17/39 \\
NCTS-TH/1716
\\

\end{flushright}
\vskip2cm
\begin{center}
{\Large \bf Matrix supergroup Chern-Simons models \\ 
\vskip0.25cm
for vortex-antivortex systems
}

\vskip1.5cm
Tadashi Okazaki\footnote{tadashiokazaki@phys.ntu.edu.tw} 

\bigskip
{\it 
Department of Physics and Center for Theoretical Sciences,\\
National Taiwan University, Taipei 10617, Taiwan}
\\
\bigskip
and
\\
\bigskip
Douglas J. Smith\footnote{Douglas.Smith@durham.ac.uk}

\bigskip
{\it Department of Mathematical Sciences, Durham University,\\
Lower Mountjoy, Stockton Road, Durham DH1 3LE, UK}

\end{center}

\vskip1cm
\begin{abstract}
We study a $U(N|M)$ supermatrix Chern-Simons model with an $SU(p|q)$ internal symmetry. 
We propose that the model describes a system consisting of $N$ vortices  and $M$ antivortices 
involving $SU(p|q)$ internal spin degrees of freedom. 
We present both classical and quantum ground state solutions, and demonstrate
the relation to Calogero models.
We present evidence that a large $N$ limit describes $SU(p|q)$ WZW models.
In particular, we derive $\widehat{\mathfrak{su}}(p|q)$ Kac-Moody algebras.
We also present some results on the calculation of the partition function 
involving a supersymmetric generalization of the Hall-Littlewood polynomials,
indicating the mock modular properties.
\end{abstract}

\newpage
\setcounter{tocdepth}{2}
\tableofcontents


\section{Introduction}
\label{sec_intro}

Matrix Chern-Simons models \cite{Polychronakos:2001mi, Dorey:2016mxm} are
gauged quantum mechanics models
whose Lagrangian is first order in
time derivatives. 
They have remarkable connections to a wide range of topics in physics and mathematics.

Susskind \cite{Susskind:2001fb} proposed that 
the infinite-dimensional matrix Chern-Simons quantum mechanical model,
as the non-commutative Chern-Simons theory on a plane,
could describe the Laughlin theory \cite{Laughlin:1983fy} 
in such a way that the positions of an infinite number of electrons moving in a two-dimensional plane influenced under strong magnetic field 
correspond to infinite matrices. 
Polychronakos \cite{Polychronakos:2001mi} subsequently proposed that 
the finite-dimensional $U(N)$ matrix Chern-Simons model, as the regularized version of the Susskind model,
could describe the fractional quantum Hall effect for $N$ electrons. 
In fact, the matrix Chern-Simons theories share many features with the Laughlin theory in the lowest Landau level 
\cite{Polychronakos:2001mi, 
Hellerman:2001rj, 
Karabali:2001xq, 
Karabali:2001pw, 
Cappelli:2004xk, 
Rodriguez:2008nz, 
Dorey:2016hoj, 
Tong:2015xaa, 
Dorey:2016mxm}. 
The classical ground state of the theory describes  
an incompressible homogeneous state at the Laughlin filling fractions. 
Level quantization leads to the specific values of the filling fraction of the Laughlin states \cite{Polychronakos:2001mi}. 
There exists a formal mapping between the quantum physical states of the matrix model and the Laughlin states \cite{Hellerman:2001rj}. 
Further extension has been studied in \cite{Dorey:2016mxm, Dorey:2016hoj} 
by introducing an $SU(p)$ global symmetry in the Polychronakos model. 
It has been argued that this extended model describes the non-Abelian quantum Hall effect with internal spin degrees of freedom. 

Another remarkable application of the $U(N)$ matrix Chern-Simons model has been proposed by Tong \cite{Tong:2003vy}. 
From the brane construction in type IIB string theory \cite{Hanany:2003hp} 
and Manton's analysis \cite{Manton:1997tg} of vortices in non-relativistic Chern-Simons theory, 
he conjectured that the $U(N)$ matrix Chern-Simons model can be viewed as a description 
of the low-energy dynamics of $N$ vortices in non-relativistic Abelian Chern-Simons matter theories. 
In addition, a further generalization has been argued in \cite{Dorey:2016mxm} that 
the $U(N)$ matrix Chern-Simons model with an $SU(p)$ global symmetry is 
the effective description of $N$ vortices in non-relativistic $U(p)$ Chern-Simons matter theories.

As shown in \cite{Polychronakos:2001mi}, 
the $U(N)$ matrix Chern-Simons model is equivalent to the Calogero model, 
which is an integrable system of $N$ non-relativistic particles with pairwise inverse-square interaction.  
This relation can be achieved by identifying the eigenvalues of the matrix 
with the coordinates of the particles on a line. 
Using the relation to the Calogero model, 
the spectrum of the $U(N)$ matrix Chern-Simons model has been examined \cite{Karabali:2001pw}.

An exciting link between the $U(N)$ matrix Chern-Simons model with $SU(p)$ global symmetry 
and the $SU(p)$ Wess-Zumino-Witten (WZW) model has been established in \cite{Dorey:2016hoj}. 
In the large $N$ limit, the current operators constructed from the matrix degrees of freedom 
realize the affine Lie algebra $\widehat{\mathfrak{su}}(p)$,
and the partition function of the matrix model turns out to be proportional to the character of the affine Lie algebra. 
This reflects rather rich mathematical structures of the matrix Chern-Simons model.   

We will study a new type of generalization of the matrix Chern-Simons theory, 
that is a $U(N|M)$ supermatrix Chern-Simons quantum mechanics with an $SU(p|q)$ global symmetry.
Mostly we will consider the case where $N\ge M$ and $p\ge q$.

In section \ref{secsgmqm} we introduce the model and argue that 
the model describes the dynamics of $N$ vortices and $M$ antivortices 
in the non-relativistic Chern-Simons matter theory, motivated by Tong's proposal \cite{Tong:2003vy}. 
The supermatrix field $\widehat{Z}_{AB}$ consists of an
$N\times N$ bosonic matrix field $Z_{ab}$, an $M\times M$ bosonic matrix field $\widetilde{Z}_{\alpha\beta}$, 
an $N\times M$ fermionic matrix field $A_{a\alpha}$ and an $M\times N$ fermionic matrix field $B_{\alpha a}$. 
While the bosonic fields $Z_{ab}$ and $\widetilde{Z}_{\alpha \beta}$ correspond to the positions on a plane 
of $N$ vortices and $M$ antivortices respectively, 
the fermionic $A_{a\alpha}$ and $B_{\alpha a}$ describe interactions between 
a vortex and an antivortex. 
In addition, the supervector field $\widehat{\Phi}_{AI}$ describes their internal spin degrees 
of freedom of the two-dimensional system.  
In a purely theoretical setup we argue that it can be realized by two types of multilayered structure 
characterized by strong magnetic fields in opposite directions. 

In section \ref{sec2classical} we study the classical ground states as 
lowest energy solutions to classical equations of motion. 
The model turns out to be related to the generalized Calogero model with $SU(p|q)$ spin degrees of freedom. 
We find two types of classical ground states. 
Both configurations admit non-trivial configuration for $Z_{ab}$ which forms 
a circular droplet of $(N-M)$ vortices as in \cite{Polychronakos:2001mi}, however, 
they are distinguished by the positive or negative contributions of vortex-antivortex pairs to the energy. 
In fact, these are similar to the two types of energy contributions of vortex-antivortex pairs due to different polarizations of the pairs of 
vortices and antivortices. 

In section \ref{sec2quantize} we study the quantization of the theory. 
We represent the quantum ground state in terms of a superdeterminant operator. 

In section \ref{seckc} we examine a connection to the WZW model. 
Following the idea of \cite{Dorey:2016hoj}, 
we construct the current operators from matrix degrees of freedom and 
demonstrate that this provides the left-moving $\widehat{\mathfrak{su}}(p|q)$ affine Lie superalgebra.  

In section \ref{secpfn} we study the spectrum by studying the partition function. 
We present a general integral expression of the partition function. 
We argue that it admits an expression in terms of a supersymmetric
generalization of the Hall-Littlewood polynomials, indicating potential
mock modularity. 
In particular, for ordinary gauge group we obtain an explicit expression of the partition function 
in terms of Kostka polynomials and supersymmetric Schur polynomials. 
From the resulting partition function we show that the ground state energy in section \ref{sec2quantize} can be correctly reproduced.

In section \ref{sec_dis} we conclude and discuss future directions.

\section{Supermatrix Chern-Simons model}
\label{secsgmqm}

\subsection{Model}
\label{subsecsgmqm}

We consider a $U(N|M)$ supermatrix Chern-Simons model whose action is given by
\begin{align}
\label{sgqm1a}
S&=
\int dt 
\left[
i\ \mathrm{Str}\left(
\widehat{Z}^{\dag}D_{t}\widehat{Z}
\right)
+i \sum_{I,A} (-1)^I \widehat{\Phi}_{IA}^{\dag}D_{t}\widehat{\Phi}_{AI}
-\kappa\ \mathrm{Str}\widehat{\alpha}
-\omega\ \mathrm{Str}\widehat{Z}^{\dag}\widehat{Z}
\right]
\end{align}
where indices $A \in \{1,\cdots, N, N+1, \cdots, N+M \}$ denote $U(N|M)$ gauge symmetry  
and $I \in \{1,\cdots, p, p+1, \cdots p+q \}$ are $SU(p|q)$ flavor symmetry. 
Also, $(-1)^I = 1$ for $I \in \{1,\cdots, p \}$ and
$(-1)^I = -1$ for $I \in \{p+1, \cdots p+q \}$.
Here 
\begin{align}
\label{z1a}
\widehat{Z}_{AB}&=
\left(
\begin{array}{cc}
Z_{ab}&A_{a\beta}\\
B_{\alpha b} &\widetilde{Z}_{\alpha\beta}\\
\end{array}
\right),&  
\widehat{Z}^{\dag}_{AB}&=
\left(
\begin{array}{cc}
Z_{ba}^{*}&B^{*}_{b \alpha}\\
A^{*}_{\beta a} &\widetilde{Z}^{*}_{\beta\alpha}\\
\end{array}
\right),& 
\widehat{\alpha}_{AB}&=
\left(
\begin{array}{cc}
\alpha_{ab}&\lambda_{a\beta}\\
\widetilde{\lambda}_{\alpha b}&\widetilde{\alpha}_{\alpha\beta}\\
\end{array}
\right)
\end{align}
are the $(N+M)\times (N+M)$ supermatrices 
where the indices $a,b=1,\cdots, N$ and $\alpha,\beta=1,\cdots, M$ 
label the bosonic subgroups $U(N)$ and $U(M)$ of the supergauge group $U(N|M)$. 
The supermatrix $\widehat{Z}_{AB}$ involves a
bosonic $U(N)$ adjoint complex scalar field $Z_{ab}$, 
a bosonic $U(M)$ adjoint complex scalar $\widetilde{Z}_{\alpha\beta}$, and
fermionic bi-fundamental fields $A_{a\beta}$ and $B_{\alpha b}$.
The supermatrix $\widehat{\alpha}$ is the $U(N|M)$ supergroup gauge field. 
It contains a $U(N)$ bosonic gauge field $\alpha_{ab}$,
a $U(M)$ bosonic gauge field $\widetilde{\alpha}_{\alpha\beta}$, 
and fermionic bi-fundamental parts $\lambda_{a\beta}$, $\widetilde{\lambda}_{\alpha b}$ of the supergroup gauge field.
Since the gauge field $\widehat{\alpha}$ is Hermitian, so are $\alpha$ and
$\widetilde{\alpha}$, while $\widetilde{\lambda} = \lambda^{\dagger}$.
The fields 
\begin{align}
\label{phi1a}
\widehat{\Phi}_{AI}&=
\left(
\begin{array}{cc}
\phi_{ai}&\psi_{a \lambda}\\
\widetilde{\psi}_{\alpha i}&\widetilde{\phi}_{\alpha \lambda}\\
\end{array}
\right),& 
\widehat{\Phi}^{\dag}_{IA}&=
\left(
\begin{array}{cc}
\phi^{\dagger}_{ia}&
\widetilde{\psi}^{\dagger}_{i \alpha}
\\
\psi^{\dagger}_{\lambda a}
&\widetilde{\phi}^{\dagger}_{\lambda\alpha}\\
\end{array}
\right)
 = \left(
\begin{array}{cc}
\phi^{*}_{ai}&
\widetilde{\psi}^{*}_{\alpha i}
\\
\psi^{*}_{a \lambda}
&\widetilde{\phi}^{*}_{\alpha \lambda}\\
\end{array}
\right)
\end{align}
are 
arrays of complex $(N|M)$ supervectors and $(p|q)$ supervectors 
where the indices $i=1,\cdots,p$ and 
$\lambda=1,\cdots, q$ label the $SU(p)$ and $SU(q)$ global symmetry subgroups.
The full global $SU(p|q)$ transformations are 
\begin{align}
\label{sgqmsupq}
\widehat{\Phi}_{AI} & \rightarrow 
\widehat{\Phi}_{AJ} \widehat{M}_{JI}
\end{align}
where $\widehat{M} \in SU(p|q)$.
The covariant derivatives are defined by
\begin{align}
\label{sgqm1b}
D_{t}\widehat{Z}&=\dot{\widehat{Z}}-i[\widehat{\alpha},\widehat{Z}],& 
D_{t}\widehat{\Phi}&=\dot{\widehat{\Phi}}-i\widehat{\alpha}\widehat{\Phi}. 
\end{align}
The gauge transformations are  
\begin{align}
\label{sgauge1a}
\widehat{Z}_{AB}&\rightarrow 
\widehat{U}_{AC}\widehat{Z}_{CD}\widehat{U}_{DB}^{\dag},\\
\label{sgauge1b}
\widehat{\Phi}_{AI}&\rightarrow \widehat{U}_{AB}\widehat{\Phi}_{BI},\\
\label{sgauge1c}
\widehat{\alpha}_{AB}&\rightarrow 
\widehat{U}_{AC}\widehat{\alpha}_{CD}\widehat{U}_{DB}^{\dag}
+i\widehat{U}_{AC}\dot{\widehat{U}}^{\dag}_{CB}
\end{align}
where
\begin{align}
\label{sgauge1d}
\widehat{U}_{AB}
&=\left(
\begin{array}{cc}
U_{ab}&V_{a \beta}\\
W_{\alpha b}&\widetilde{U}_{\alpha\beta}\\
\end{array}
\right)\in U(N|M)
\end{align}
and since $\widehat{U}$ is unitary
\begin{align}
\label{sgauge1e}
\widehat{U}^{\dagger} = \widehat{U}^{-1}
&=\left(
\begin{array}{cc}
U^{-1}(I - V \widetilde{U}^{-1} W U^{-1})^{-1} & -U^{-1} V \widetilde{U}^{-1} (I - W U^{-1} V \widetilde{U}^{-1})^{-1} \\
-\widetilde{U}^{-1} W U^{-1}(I - V \widetilde{U}^{-1} W U^{-1})^{-1} & \widetilde{U}^{-1} (I - W U^{-1} V \widetilde{U}^{-1})^{-1} \\
\end{array}
\right) .
\end{align}

Note that we have chosen the $SU(p|q)$ transformations in (\ref{sgqmsupq})
to act from the right on
$\widehat{\Phi}$. This is so that these transformations commute with the
$U(N|M)$ gauge transformations (\ref{sgauge1b}) acting from the left. This is
only necessary when $M \ne0$ and $q\ne 0$ since in this general case not all
components of $\widehat{M}$ commute with all components of $\widehat{U}$, as
some pairs anti-commute.

In terms of the elements of the matrices 
(\ref{z1a}) and (\ref{phi1a}), 
the action (\ref{sgqm1a}) is expressed as 
\begin{align}
\label{sgqm1c}
S=
\int dt 
\Biggl[
i \Tr 
&\Biggl(
Z^{\dag}DZ
+B^{\dag}DB
-iZ^{\dag}\left(\lambda B
-A\widetilde{\lambda}\right)
-iB^{\dag}\left(\widetilde{\lambda}Z-\widetilde{Z}\widetilde{\lambda}\right)\nonumber\\
-&\widetilde{Z}^{\dag}D\widetilde{Z}
-A^{\dag}DA
+i\widetilde{Z}^{\dag}\left(\widetilde{\lambda}A-B\lambda\right)
+iA^{\dag}\left(\lambda \widetilde{Z}-Z\lambda\right)
\Biggr)\nonumber\\
+i\sum\Biggl(
&\phi^{\dag}D\phi-i\phi^{\dag}\lambda\widetilde{\psi}
-\psi^{\dag}D\psi+i\psi^{\dag}\lambda \widetilde{\phi}
-\widetilde{\phi}^{\dag}D\widetilde{\phi}
+i\widetilde{\phi}^{\dag}\widetilde{\lambda}\psi
+\widetilde{\psi}^{\dag}D\widetilde{\psi}
-i\widetilde{\psi}^{\dag}\widetilde{\lambda}\phi
\Biggr)
\nonumber\\
-\kappa \Tr&\left( 
\alpha-\widetilde{\alpha}
\right)
-\omega 
\Tr 
\left( 
Z^{\dag}Z+B^{\dag}B-\widetilde{Z}^{\dag}\widetilde{Z} - A^{\dag}A
\right)
\Biggr]
\end{align}
where the covariant derivatives are defined by 
\begin{align}
DZ&=\dot{Z}-i[\alpha,Z],& 
D\widetilde{Z}&=\dot{\widetilde{Z}}-i[\widetilde{\alpha},\widetilde{Z}], \nonumber\\
DA&=\dot{A}-i\alpha A+iA\widetilde{\alpha},& 
DB&=\dot{B}-i\widetilde{\alpha}B+iB\alpha, \nonumber\\
D\phi&=\dot{\phi}-i\alpha\phi,& 
D\widetilde{\phi}&=\dot{\widetilde{\phi}}-i\widetilde{\alpha}\widetilde{\phi}\nonumber\\  
D\psi&=\dot{\psi}-i\alpha\psi, & 
D\widetilde{\psi}&=\dot{\widetilde{\psi}}-i\widetilde{\alpha}\widetilde{\psi}.
\end{align}
The gauge transformations (\ref{sgauge1a}) of the fields $\widehat{Z}_{AB}$ are expressed by
\begin{align}
\label{sgauge2a}
Z&\rightarrow UZU^{\dag}+VBU^{\dag}+UAV^{\dag}+V\widetilde{Z}V^{\dag}, \\
\label{sgauge2b}
\widetilde{Z}&\rightarrow \widetilde{U}\widetilde{Z}\widetilde{U}^{\dag}
+WA\widetilde{U}^{\dag}+\widetilde{U}BW^{\dag}+WZW^{\dag}, \\
\label{sgauge2c}
A&\rightarrow UA\widetilde{U}^{\dag}+UZW^{\dag}+V\widetilde{Z}\widetilde{U}^{\dag}+VBW^{\dag}, \\
\label{sgauge2d}
B&\rightarrow \widetilde{U}BU^{\dag}
+WZU^{\dag}+\widetilde{U}\widetilde{Z}V^{\dag}+WAV^{\dag}, 
\end{align}
those of the fields $\widehat{\Phi}_{AI}$ are
\begin{align}
\label{sgauge2e}
\phi&\rightarrow U\phi+V\widetilde{\psi},& 
\widetilde{\phi}&\rightarrow 
\widetilde{U}\widetilde{\phi}+W\psi,\\
\label{sgauge2f}
\psi&\rightarrow U\psi+V\widetilde{\phi},& 
\widetilde{\psi}&\rightarrow \widetilde{U}\widetilde{\psi}+W\phi,
\end{align}
and those of the gauge fields $\widehat{\alpha}$ are
\begin{align}
\label{sgauge2g}
\alpha&\rightarrow 
U\alpha U^{\dag}
+V\widetilde{\lambda}U^{\dag}
+U\lambda V^{\dag}+V\widetilde{\alpha}V^{\dag}
+iU\dot{U}^{\dag}+iV\dot{V}^{\dag}, \\
\label{sgauge2h}
\widetilde{\alpha}&\rightarrow 
\widetilde{U}\widetilde{\alpha}\widetilde{U}^{\dag}
+W\lambda \widetilde{U}^{\dag}
+\widetilde{U}\widetilde{\lambda} W^{\dag}
+W\alpha W^{\dag}+i\widetilde{U}\dot{\widetilde{U}}^{\dag}+iW\dot{W}^{\dag}, \\
\label{sgauge2i}
\lambda&\rightarrow 
U\lambda \widetilde{U}^{\dag}
+U\alpha W^{\dag}+V\widetilde{\alpha}\widetilde{U}^{\dag}
+V\widetilde{\lambda}W^{\dag}+iU\dot{W}^{\dag}+iV\dot{\widetilde{U}}^{\dag}, \\
\label{sgauge2j}
\widetilde{\lambda}&\rightarrow 
\widetilde{U}\widetilde{\lambda}U^{\dag}
+W\alpha U^{\dag}+\widetilde{U}\widetilde{\alpha}V^{\dag}
+W\lambda V^{\dag}
+i W\dot{U}^{\dag}+i\widetilde{U}\dot{V}^{\dag}. 
\end{align}

When $M=0$ and $q=0$, our supermatrix Chern-Simons model (\ref{sgqm1a}) becomes 
the ordinary matrix Chern-Simons model in \cite{Polychronakos:2001mi, Dorey:2016mxm, Dorey:2016hoj}
\begin{align}
\label{qmx1}
S&=\int dt 
\left[
i\Tr \left(
Z^{\dag}D_{t}Z
\right)
+i\sum_{i=1}^{p}\phi_{i}^{\dag}D_{t}\phi_{i}
-\kappa \Tr \alpha
-\omega \Tr Z^{\dag}Z
\right]
\end{align}
where $Z$ is a complex adjoint scalar 
and $\phi_{i}$, $i=1,\cdots,p$ are $p$ fundamental complex scalars.  
Here the covariant derivatives are 
\begin{align}
\label{qmx2}
D_{t}Z&=\dot{Z}-i[\alpha,Z],& 
D_{t}\phi_{i}&=\dot{\phi}_{i}-i\alpha\phi_{i}
\end{align}
and the trace is taken over the $U(N)$ gauge indices. 
The gauge symmetry transformations (\ref{sgauge2a})-(\ref{sgauge2f}) reduce to 
\begin{align}
\label{qmxgauge1}
Z&\rightarrow UZU^{\dag},& 
\phi_{i}&\rightarrow U\phi_{i}
\end{align}
for $U\in U(N)$. 
This ordinary matrix Chern-Simons model (\ref{qmx1}) is considered as an effective theory 
of the fractional quantum Hall states composed of $N$ electrons in the lowest Landau level 
\cite{Polychronakos:2001mi, Hellerman:2001rj, Dorey:2016mxm}. 
Although the matrix $Z$ is not diagonalized, it describes positions of $N$ electrons on the plane. 
The vectors $\phi_{i}$ describe the internal spin, which is called pseudospin, 
degrees of freedom of $N$ electrons \cite{Dorey:2016mxm}.

\subsection{Vortex-antivortex system in multilayers}
\label{secvaml}

\subsubsection{Chern-Simons vortex quantum mechanics}
\label{subsec_csvortex}
The matrix Chern-Simons theory with a $U(N)$ gauge symmetry 
and an $SU(p)$ flavor symmetry has been proposed as 
an effective theory of $N$ vortices in non-relativistic $U(p)$ Chern-Simons matter theory. 
We will review the discussion in 
\cite{Manton:1997tg, Romao:2000ru, Romao:2004df, Tong:2003vy, Tong:2015xaa, Dorey:2016mxm}.

Let us consider a Chern-Simons matter theory with gauge group 
\begin{align}
\label{cs_v_gauge}
U(p)_{k',k}&=\frac{U(1)_{k'}\times SU(p)_{k}}{\mathbb{Z}_{p}}
\end{align}
with the relation
\begin{align}
\label{cs_v_level}
k'-kp\in p^{2}\mathbb{Z}
\end{align}
and the following Lagrangian \cite{Dorey:2016mxm}
\begin{align}
\label{cs_v1}
S&=S_{\mathrm{CS}}+S_{\mathrm{matter}}, \nonumber\\
S_{\mathrm{CS}}
&=-\int d^{3}x \left[
\frac{k'}{4\pi}\epsilon^{\mu\nu\rho}a_{\mu}\partial_{\nu}a_{\rho}
+\frac{k}{4\pi}\Tr \epsilon^{\mu\nu\rho}
\left(
A_{\mu}\partial_{\nu}A_{\rho}
-\frac{2i}{3}A_{\mu}A_{\nu}A_{\rho}
\right)-\mu a_{0}
\right], \nonumber\\
S_{\mathrm{matter}}&=
\int d^{3}x 
\left[
i\phi_{i}^{\dag}D_{0}\phi_{i}
-\frac{1}{2m}D_{\alpha}\phi_{i}^{\dag}D_{\alpha}\phi_{i}
-\frac{\pi}{m}
\left\{
\frac{1}{k'} \left(\phi_{i}^{\dag}\phi_{i}\right)^{2}
 + \frac{1}{k} \left(\phi_{i}^{\dag}t^{\alpha}\phi_{i}\right)^2
\right\}
\right]
\end{align}
where $\mu,\cdots=0,1,2$ are space-time indices, 
$\alpha,\cdots=1,2$ are spatial indices. 
Here $a_{\mu}$ is the $U(1)$ gauge field, $A_{\mu}$ is the $SU(p)$ gauge field 
and $\phi_{i}$, $i=1,\cdots,p$ are the $p$ fundamental complex bosonic fields. 
Note that the matter is non-relativistic, having first order time derivatives
and obeying Schr\"odinger-like equations of motion. 
The action has BPS equations which give the vortex equations
\begin{align}
\label{cs_v_eom1}
f_{12}&=\frac{2\pi}{k'}\left(
|\phi_{i}|^{2}-\mu
\right),& 
F_{12}^{\alpha}&=\frac{2\pi}{k}\phi_{i}^{\dag}t^{\alpha}\phi_{i},\\
\label{cs_v_eom2}
D_{z}\phi_{i}&=0
\end{align}
where $f_{12}=\partial_{1}a_{2}-\partial_{2}a_{1}$ and 
$F_{12}=\partial_{1}A_{2}-\partial_{2}A_{1}-i[A_{1},A_{2}]$. 
The solutions to the vortex equations (\ref{cs_v_eom1}) and (\ref{cs_v_eom2}) are not unique 
and the most general solutions with the vortex number $N$ have $2pN$ parameters \cite{Hanany:2003hp}. 
The space of solutions is the vortex moduli space, $\mathcal{M}_{p, N}$, 
in which the solutions are parametrized by $2pN$ collective coordinates $X^{a}, a=1,\cdots, 2pN$ as 
$\phi_{i}(x; X)$ and $A_{\alpha}(x; X)$. 

In order to describe the vortex dynamics, it is important to note that 
the non-relativistic action (\ref{cs_v1}) is first order in time derivatives. 
This implies that the vortex moduli space $\mathcal{M}_{p,N}$ is 
not the configuration space but rather the phase space. 
In the relativistic theory with second order time derivatives 
the moduli space is the configuration space 
and the soliton dynamics is addressed by 
geodesic motion of a slowly moving particle on the moduli space 
with respect to the metric $g_{ab}(X)$ \cite{Manton:1981mp}
\begin{align}
\label{Manton_1}
S&=\int dt 
\left[ g_{ab}\dot{X}^{a}\dot{X}^{b}-V(X)
\right]
\end{align}
where $V(X)$ is some potential term. 
Meanwhile, in the non-relativistic theory with first order time derivatives, 
the moduli space is the phase space. 
In general, the low-energy effective description of 
such soliton dynamics is given by \cite{Manton:1997tg}
\begin{align}
\label{Manton_2}
S&=\int dt \left[
\mathcal{A}_{a}(X)\dot{X}^{a}-V(X)
\right]. 
\end{align}
Here $\mathcal{A}_{a}$ is the connection one-form on the moduli space which obeys \cite{Manton:1997tg, Romao:2000ru}
\begin{align}
\label{Manton_3}
d\mathcal{A}&=\Omega
\end{align}
where $\Omega$ is the K\"{a}hler form with respect to the metric $g$ on the moduli space. 
This fact relates the vortices in the Chern-Simons theories and those in the Yang-Mills theories 
so that the corresponding effective descriptions for both obey a similar relationship.

\begin{figure}
\begin{center}
\includegraphics[width=8cm]{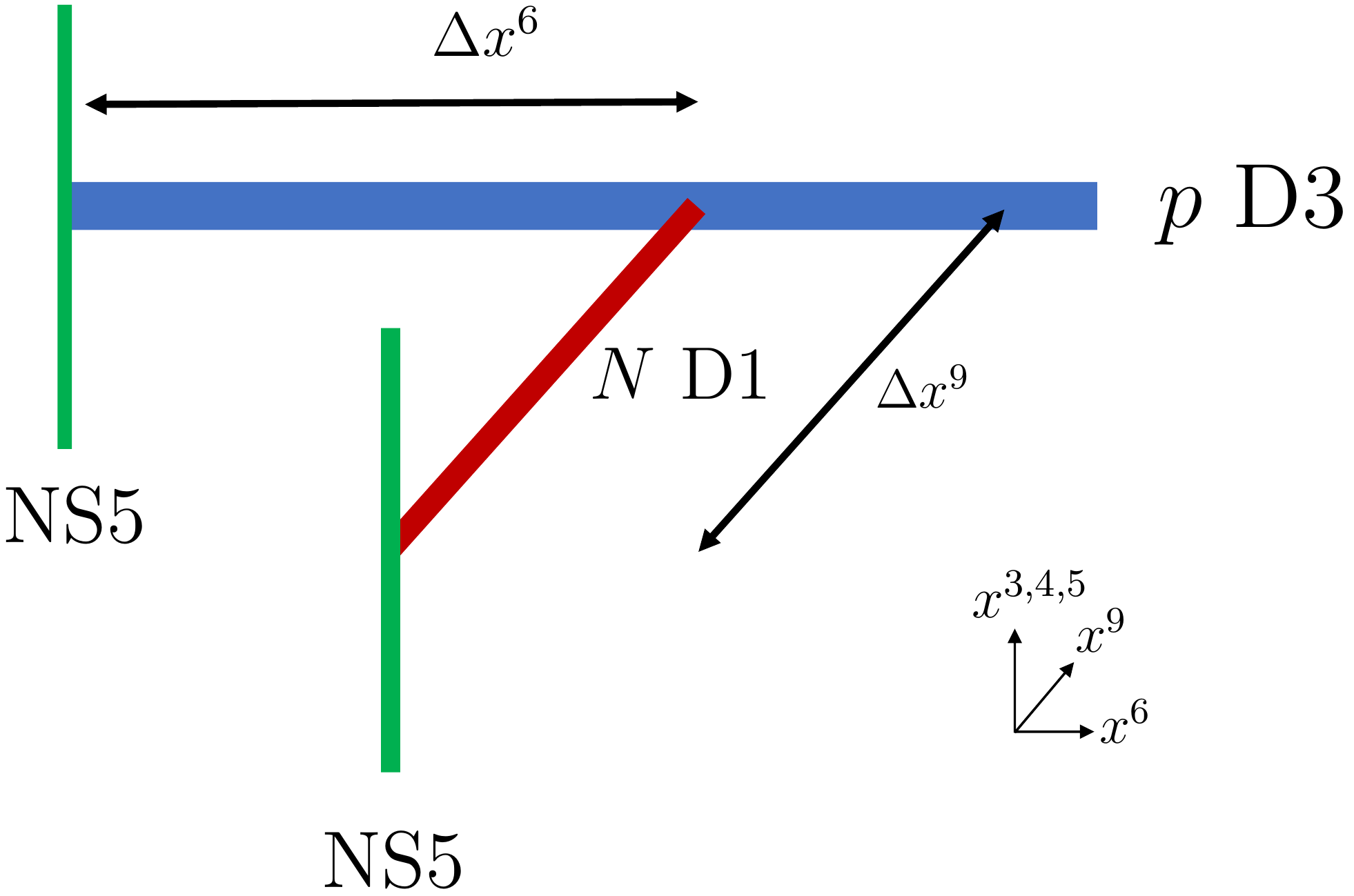}
\caption{$N$ Vortices appear as $N$ D1-branes in the 3d $\mathcal{N}=4$ $U(p)$ Yang-Mills-Higgs theory on $p$ D3-branes attached on NS5-branes.}
\label{fig_hananytong}
\end{center}
\end{figure}

To extract such a relationship, 
Tong \cite{Tong:2003vy} uses the construction of vortices in Yang-Mills-Higgs theories 
via the brane configuration in type IIB string theory \cite{Hanany:2003hp}
\begin{align}
\label{braneconf1a2}
\begin{array}{ccccccccccc}
&0&1&2&3&4&5&6&7&8&9\\
\textrm{$p$ D3}&\circ&\circ&\circ&&&&\circ&&&\\
\textrm{NS5}&\circ&\circ&\circ&\circ&\circ&\circ&&&&\\
\textrm{$N$ D1}&\circ&&&&&&&&&\circ \\
\end{array}
\end{align}
which is depicted in Figure \ref{fig_hananytong}. 
The $p$ D3-branes and the NS5-branes provide the 3d $\mathcal{N}=4$ $U(p)$ Yang-Mills-Higgs theory 
and $N$ vortices are realized as the $N$ D1-branes. 
The dynamics of the $N$ D1-branes is given by the $U(N)$ gauged quantum mechanics, which includes
a gauge field $\alpha$, 
the real adjoint scalar fields $\sigma^{I}$, $I=3,4,5$ 
describing the positions of D1-branes in the $x^{3,4,5}$ directions, 
the complex adjoint scalar fields $Z$ as $N\times N$ complex matrices, 
describing the positions of D1-branes in the two-dimensional $x^1$-$x^{2}$ plane, 
and the fundamental complex scalars $\phi$ as $p\times N$ matrices arising from the D1-D3 strings. 
The bosonic part of the Lagrangian is given by 
\begin{align}
\label{HT_v1}
\mathcal{L}&=
\tr \Biggl[
\frac{1}{2g^2}
D_{t}\sigma^{I}D_{t}\sigma^I+D_{t}Z^{\dag}D_{t}Z+D_{t}\phi D_{t}\phi^{\dag}\nonumber\\
&-\frac{1}{2g^2}\left[\sigma^I, \sigma^J\right]^2
-\left|[Z,\sigma^{I}] \right|^{2}
-\phi\phi^{\dag}\sigma^I \sigma^I
-\frac{g^2}{2}\left(
\phi\phi^{\dag}+[Z,Z^{\dag}]-r\mathbb{I}
\right)
\Biggr]
\end{align}
where 
\begin{align}
D_{t}Z&=\dot{Z}-i[\alpha, Z],& 
D_{t}\sigma^{I}&=\dot{\sigma}^{I}-i[\alpha,\sigma^{I}],& 
D_{t}\phi=\dot{\phi}-i\alpha \phi
\end{align}
and the gauge coupling $g$ and the FI parameter $r$ are encoded by the positions of the D3-branes and NS5-branes 
\begin{align}
\label{HT_v2}
\frac{1}{g^{2}}&=\frac{2\pi l_{s}^2 \Delta x^{9}}{g_{s}},& 
r&=\frac{\Delta x^6}{g_s}. 
\end{align}
The decoupling limit of the D3-brane theory can be achieved in the strong coupling limit $g^{2}\rightarrow \infty$. 
This leads to the D-term constraints from the leading terms in the Lagrangian (\ref{HT_v1})
\begin{align}
\label{HT_v3}
[Z,Z^{\dag}]+\phi\phi^{\dag}-r\mathbb{I}&=0.
\end{align}
According to the $N^2$ constraints (\ref{HT_v3}), 
the $(N^2+pN)$ original matrix degrees of freedom from 
the matrices $Z$ and $\phi$ reduce to $pN$ complex ($2pN$ real) degrees of freedom 
as required from the dimensions of the vortex moduli space $\mathcal{M}_{p,N}$. 

From the above analysis via string theory, 
we see that 
the dynamics of $N$ vortices in $U(p)$ Yang-Mills theory 
is captured by the matrix model (\ref{HT_v1}) with the constraints (\ref{HT_v3}). 
To find the matrix model of Chern-Simons vortices, 
we observe the following facts: 
\begin{enumerate}
\item The K\"{a}hler form $\Omega$ on $\mathcal{M}_{p,N}$ can be constructed 
from the canonical K\"{a}hler form on the space of unconstrained $Z$ and $\phi$ 
by imposing the non-trivial constraint (\ref{HT_v3}) via 
the symplectic quotient construction. 
\item The general action (\ref{Manton_2}) is first order in time derivatives. 
\end{enumerate}
It then turns out that 
the above necessary properties follow from 
the matrix $U(N)$ Chern-Simons models (\ref{qmx1}) with $SU(p)$ flavor symmetry 
in such a way that the auxiliary gauge field $\alpha$ plays a role of a Lagrange multiplier
which yields the Gauss law constraints as (\ref{HT_v3}).

This fact further instructs us to consider our supermatrix $U(N|M)$ Chern-Simons model (\ref{sgqm1a}) 
with an $SU(p|q)$ flavor symmetry as the microscopic description of the system 
which involves $N$ vortices and $M$ anti-vortices with internal $SU(p|q)$ spin degrees of freedom. 
We will provide supporting evidence for this interpretation.


\subsubsection{Vortices and antivortices in multilayers}
\label{subsec_VAmlayers}

\paragraph{Vortex-antivortex pairs}
A vortex and an antivortex are distinguished by the winding number or vortex number 
in such a way that a vortex carries the winding number $+1$ and an antivortex does $-1$. 
When a vortex and an antivortex meet, they can form a vortex-antivortex pair. 
Below a certain temperature, the thermal energy is not enough to generate vortices, 
however, the lower energy vortex-antivortex pairs can occur. 
Vortex-antivortex pairs can be localized configurations.
The two-dimensional superfluid phase that is characterized by 
the existence of vortex-antivortex pairs is called the Berezinskii-Kosterlitz-Thouless (BKT) phase \cite{Kosterlitz:1973xp} 
\footnote
{
In the quantum Hall states, a vortex-antivortex can be viewed as a quasiparticle-hole \cite{oshikawa2007topological}. 
}. 
In addition to winding number, 
vortices and antivortices are also characterized by polarity $p$ \cite{papanicolaou1999semitopological}. 
The polarity is an out-of plane magnetization at the vortex-core which can either point up $(p>0)$ or down $(p<0)$. 
The winding number $N$ and polarity $p$ specify the circulation or vorticity $q$ by \cite{papanicolaou1999semitopological}
\begin{align}
\label{vortex_circ}
q&=-2\pi Np. 
\end{align}
In general the topology and the dynamics of pairs of vortices depend on the circulation. 
Let $q_1$ and $q_2$ be circulations of vortices. 
The kinetic energy of the pairs per unit mass in the plane is given by \cite{barenghi2016primer}
\begin{align}
\label{va_pair_kin}
E&=\pi
\left[
q_{1}^{2}\ln \frac{R_{0}}{a_{0}}
+q_{2}^{2}\ln \frac{R_{0}}{a_{0}}
+2q_{1}q_{2}\ln \frac{R_{0}}{d}
\right]
\end{align}
where $R_{0}$ is the size of container, $a_{0}$ is the vortex core radius and $d$ is the separation of pairs. 
For vortex-antivortex pairs the winding numbers are taken to be opposite 
and therefore the circulations are determined by their polarities. 
Dynamics of pairs of vortices have been studied in \cite{barenghi2016primer} 
and vortex-antivortex pairs have been studied numerically in 
\cite{komineas2007rotating, komineas2007dynamics, papanicolaou1999semitopological}. 
Let us briefly review the properties of vortex-antivortex pairs.

\begin{enumerate}

\item \textbf{Parallel polarized vortex-antivortex pairs}

When the vortex and antivortex cores are polarized parallel to each other, 
they have the opposite circulation according to (\ref{vortex_circ}). 
Then the interaction energy in (\ref{va_pair_kin}) is negative binding energy. 
Qualitatively this is because the flow fields of vortices and antivortices tend to cancel in the bulk 
and the total kinetic energy is reduced.
Consequently the cores of vortices approach each other on spiraling orbits and meet in the center. 

\item \textbf{Antiparallel polarized vortex-antivortex pairs}

Because of (\ref{vortex_circ}), 
when the vortex and antivortex cores are polarized antiparallel to each other, 
they have the same circulation. 
From (\ref{va_pair_kin}) the interaction energy is positive in this case 
as the flow fields tend to enlarge in the bulk and the total kinetic energy increases. 
This indicates that after the creation of a vortex-antivortex pair, 
the antivortex quickly moves towards the original vortex in a rapid process and they then annihilate each other 
\footnote{See \cite{hertel2006exchange} for the detail of the magnetization dynamics of such annihilation process.}. 
It has been shown that such vortex-antivortex annihilation is connected 
with the emission of sound waves 
\cite{lee2005radiation, van2006magnetic, hertel2007ultrafast, tretiakov2007vortices, barenghi2016primer}. 

\end{enumerate}
We will see in section \ref{subsec2cl_ground} that 
the classical ground states in our supermatrix Chern-Simons model (\ref{sgqm1a})  
support these two different types of vortex-antivortex pairs.

\begin{figure}
\begin{center}
\includegraphics[width=10cm]{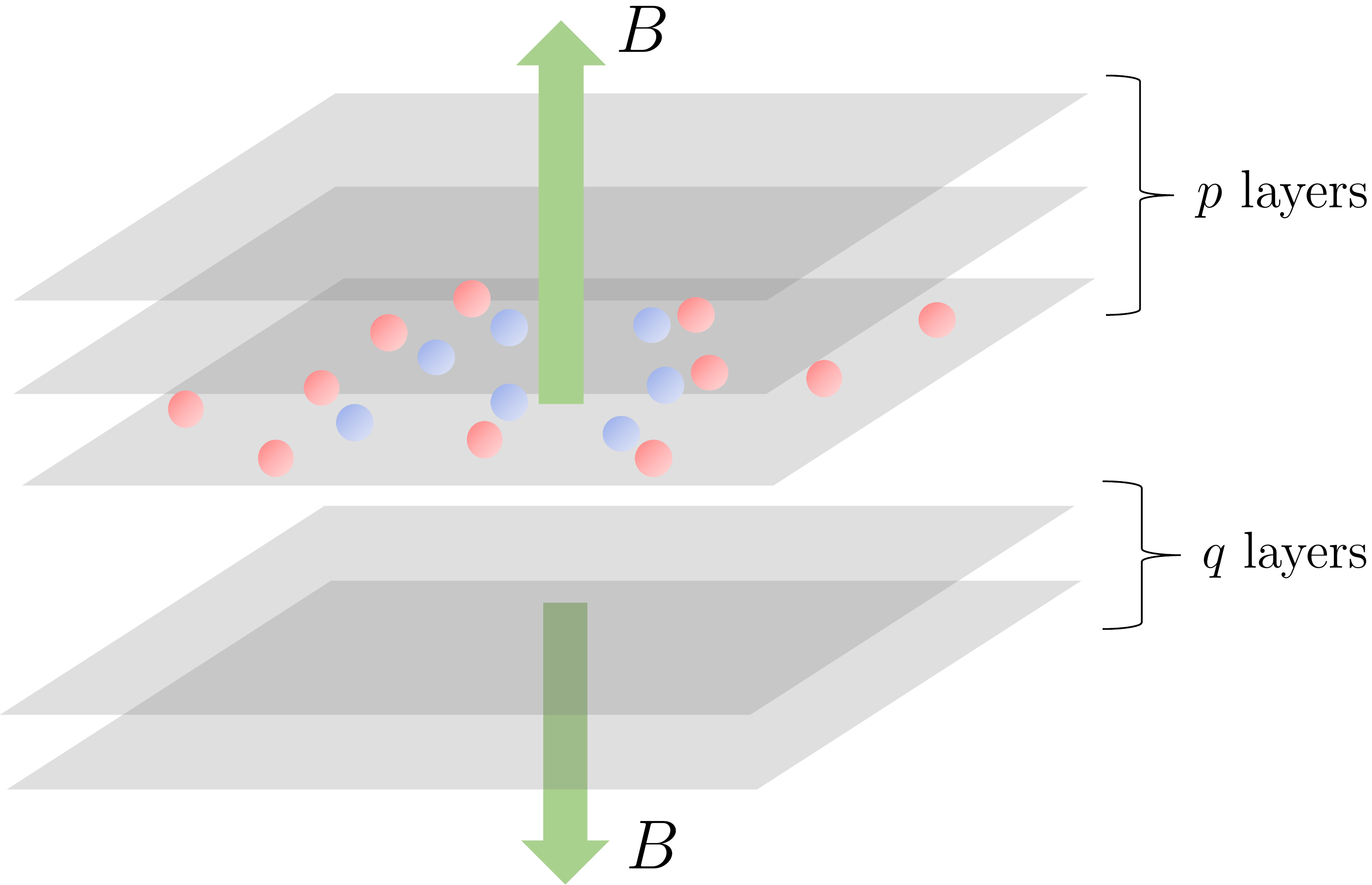}
\caption{$N$ vortices (red) and $M$ antivortices (blue) in multiple $p$ layers under the influence of vertically upward magnetic field $B$ 
and $q$ layers under that of the downward magnetic field. }
\label{figbilayer}
\end{center}
\end{figure}

\paragraph{Vortices in multilayers}
Multilayered quantum Hall systems and vortices have been constructed and studied in theoretical and experimental setup  
\cite{eisenstein1992new, suen1994origin, murphy1994many, chae2012direct, hierro2017deterministic}. 
In this case electrons or vortices may occupy several layers and carry different spins 
in such a way that additional layer indices label them with internal spins. 
Recently it has been proposed in \cite{Dorey:2016mxm} that 
the fractional quantum Hall states or vortices in $p$-component systems can be described by 
the matrix Chern-Simons model (\ref{qmx1}) with an internal $SU(p)$ symmetry. 
Here we will consider a generalization of this model with an internal $SU(p|q)$ symmetry. 
It is expected that our generalized model may be theoretically realized in 
a system which consists of $N$ vortices and $M$ antivortices in two sets of multilayers as shown in Figure \ref{figbilayer} 
where one set of $p$ multilayers is put in a perpendicular magnetic field pointing upward 
whereas the other $q$ are in the downward magnetic field.

\section{Classical solutions}
\label{sec2classical}

We first derive the classical equations of motion and Gauss law constraints for
the general case. We note that the model is related to generalized Calogero
models. We then investigate classical ground states and their physical
interpretation, focussing on various special cases.

The classical equations of motion for dynamical scalar fields and fermions from $\widehat{Z}$ read
\begin{align}
\label{seom1a}
i(DZ)_{ab}&=A_{a\alpha}\widetilde{\lambda}_{\alpha a}-\lambda_{a\alpha}B_{\alpha b}+\omega Z_{ab}, \\
\label{seom1b}
i(D\widetilde{Z})_{\alpha\beta}
&=B_{\alpha a}\lambda_{a\alpha}-\widetilde{\lambda}_{\alpha a}A_{a\beta}+\omega \widetilde{Z}_{\alpha\beta}, \\
\label{seom1c}
i(DA)_{a\alpha}&=Z_{ab}\lambda_{b\alpha}
-\lambda_{a\beta}\widetilde{Z}_{\beta\alpha}
+\omega A_{a\alpha}, \\
\label{seom1d}
i(DB)_{\alpha a}&=\widetilde{Z}_{\alpha\beta}\widetilde{\lambda}_{\beta a}-
\widetilde{\lambda}_{\alpha b}Z_{ba}
+\omega B_{\alpha a}
\end{align}
and those from $\widehat{\Phi}$ are 
\begin{align}
\label{seom1e}
i(D\phi)_{ai}&= -\lambda_{a\alpha}\widetilde{\psi}_{\alpha i}, & 
i(D\widetilde{\phi})_{\alpha \lambda}&= -\widetilde{\lambda}_{\alpha a}\psi_{a \lambda}, \\
\label{seom1f}
i(D\psi)_{a \lambda}&= -\lambda_{a\alpha}\widetilde{\phi}_{\alpha \lambda}, & 
i(D\widetilde{\psi})_{\alpha i}&= -\widetilde{\lambda}_{\alpha a}\phi_{ai}.
\end{align}
For $M=0$ and $q=0$
the classical equations of motion for the scalar fields are
\begin{align}
\label{eom1a}
iD_{t}Z&=\omega Z,& 
D_{t}\phi_{i}&=0.
\end{align}

The equations of motion for gauge fields lead to the Gauss law constraints.
Using notation $(-1)^A = 1$ if $A = a$ and $(-1)^A = -1$ if $A = \alpha$
etc.\ the relevant part of the action is
\begin{align}
S_{\mathrm{gauge}} & = \int dt \left( (-1)^A \alpha_{AB} \left[ \widehat{Z}, \widehat{Z}^{\dag} \right]_{BA} + (-1)^I \widehat{\Phi}^{\dag}_{IA}\alpha_{AB}\widehat{\Phi}_{BI} -\kappa (-1)^A \alpha_{AA} \right) \nonumber \\
 & = \int dt \; \alpha_{AB} (-1)^A \left( \left[ \widehat{Z}, \widehat{Z}^{\dag} \right]_{BA} + (-1)^{A + I + (A+B)(A+I)} \widehat{\Phi}^{\dag}_{IA}\widehat{\Phi}_{BI} -\kappa \delta_{AB} \right) \nonumber \\
 & = \int dt \; \alpha_{AB} (-1)^A \left( \left[ \widehat{Z}, \widehat{Z}^{\dag} \right]_{BA} + \widehat{\Phi}_{BI} \widehat{\Phi}^{\dag}_{IA} -\kappa \delta_{AB} \right). 
\label{gaugeaction}
\end{align}
The Gauss law constraints are therefore
\begin{align}
\label{gauss_sg}
\left[ \widehat{Z}, \widehat{Z}^{\dag} \right]_{AB} + \widehat{\Phi}_{AI} \widehat{\Phi}^{\dag}_{IB} -\kappa \delta_{AB} & = 0 ,
\end{align}
or in component form
\begin{align}
\label{gaussbb}
[Z,Z^{\dag}] + A A^{\dag}-B^{\dag} B+\sum_{i}\phi_{i}\phi_{i}^{\dag}+\sum_{\lambda}\psi_{\lambda}\psi_{\lambda}^{\dag}-\kappa\mathbb{I}&=0,\\ 
\label{gaussff}
[\widetilde{Z},\widetilde{Z}^{\dag}]+ B B^{\dag}- A^{\dag} A+\sum_{\lambda}\widetilde{\phi}_{\lambda}\widetilde{\phi}_{\lambda}^{\dag}
+\sum_{i}\widetilde{\psi}_{i}\widetilde{\psi}_{i}^{\dag}-\kappa\mathbb{I}&=0,\\
\label{gaussfb}
BZ^{\dag}-\widetilde{Z}^{\dag}B+\widetilde{Z}A^{\dag}-A^{\dag}Z
+\sum_{i}\widetilde{\psi}_{i}\phi^{\dag}_{i}+\sum_{\lambda}\widetilde{\phi}_{\lambda}\psi^{\dag}_{\lambda}&=0,\\
\label{gaussbf}
ZB^{\dag}-B^{\dag}\widetilde{Z}
-Z^{\dag}A+A\widetilde{Z}^{\dag}
+\sum_{i}\phi_{i}\widetilde{\psi}^{\dag}_{i}+\sum_{\lambda}\psi_{\lambda}\widetilde{\phi}^{\dag}_{\lambda}&=0.
\end{align}
Note that (\ref{gaussbf}) follows from the Hermitian conjugation of (\ref{gaussfb}).

Tracing the $U(N)$ and $U(M)$ parts of the Gauss law constraints (\ref{gaussbb}) and (\ref{gaussff}) give respectively
\begin{align}
\label{Gtrace_bb}
\sum_{a}\sum_{\alpha}
(A_{a\alpha} A_{\alpha a}^{\dag}+B_{\alpha a}B^{\dag}_{a \alpha})
+\sum_{i}\sum_{\alpha}
\phi_{i a}^{\dag}\phi_{ai}
-\sum_{\lambda}\sum_{a}\psi_{\lambda a}^{\dag}\psi_{a \lambda}&=\kappa N,
\\
\label{Gtrace_ff}
\sum_{a}\sum_{\alpha}
(A_{a\alpha} A_{\alpha a}^{\dag}+B_{\alpha a}B^{\dag}_{a \alpha})
+\sum_{\lambda}\sum_{\alpha}\widetilde{\phi}^{\dag}_{\lambda \alpha}\widetilde{\phi}_{\alpha \lambda}
-\sum_{i}\sum_{\alpha}\widetilde{\psi}^{\dag}_{i \alpha}\widetilde{\psi}_{\alpha i}&=\kappa M
\end{align}
and the difference of these equations is the supertrace of the Gauss law constraints
\begin{align}
\label{G_strace}
\sum_{i}\sum_{\alpha}
\phi_{i a}^{\dag}\phi_{ai}
-\sum_{\lambda}\sum_{a}\psi_{\lambda a}^{\dag}\psi_{a \lambda}
-\sum_{\lambda}\sum_{\alpha}\widetilde{\phi}^{\dag}_{\lambda \alpha}\widetilde{\phi}_{\alpha \lambda}
+\sum_{i}\sum_{\alpha}\widetilde{\psi}^{\dag}_{i\alpha}\widetilde{\psi}_{\alpha i}&=\kappa (N-M).
\end{align}

Now, we can
find explicit solutions after first gauge fixing. The simplest is to choose the
temporal gauge $\widehat{\alpha}=0$. Then
the equations of motion for the supervector field $\widehat{\Phi}$, 
(\ref{seom1e}) and (\ref{seom1f}) become $\dot{\widehat{\Phi}}=0$. 
Thus $\widehat{\Phi}$ should be constant. 
Meanwhile the equations of motion for the supermatrix field $\widehat{Z}$, 
(\ref{seom1a})-(\ref{seom1d}) become 
\begin{align}
\label{seom1a1}
\dot{Z}_{ab}&=-i\omega Z_{ab}, \\
\label{seom1b1}
\dot{\widetilde{Z}}_{\alpha\beta}&=-i\omega \widetilde{Z}_{\alpha\beta}, \\
\label{seom1a2}
\dot{A}_{a\alpha}&=-i\omega A_{a\alpha}, \\
\label{seom1a3}
\dot{B}_{\alpha a}&=-i\omega B_{\alpha a}
\end{align}
and therefore each block matrix has a simple time dependence given by a factor
$e^{-i\omega t}$. In the Gauss law constraints (\ref{gaussbb})-(\ref{gaussbf})
these time-dependent phases cancel, so in the temporal gauge the classical
solutions correspond to time-independent solutions of the Gauss law constraints.
Note that we still have the residual gauge symmetry of arbitrary
time-independent gauge transformations.

\subsection{Generalized Calogero models}
\label{subsec2Calogero}

The Chern-Simons supermatrix quantum mechanics models 
(\ref{sgqm1a}) can also be related to generalized Calogero
models. To do this, first split $\widehat{Z}$ into its Hermitian and
anti-Hermitian parts as
\begin{align}
\widehat{Z} = \frac{1}{\sqrt{2}} \left( \widehat{X} + i \widehat{Y} \right)
\end{align}
where $\widehat{X}$ and $\widehat{Y}$ are both Hermitian.
Then the $U(N|M)$ symmetry can be used to diagonalize $\widehat{X}$ which we
write as $\widehat{X}_{AB} = x_A \delta_{AB}$. In this gauge we find
\begin{align}
\label{CalogZZCom}
\left[ \widehat{Z}, \widehat{Z}^{\dagger} \right]_{AB} & = -i(x_A - x_B) Y_{AB}
\end{align}
so the Gauss law constraints (\ref{gauss_sg}) become
\begin{align}
\label{CalogGauss}
(x_A - x_B) \widehat{Y}_{AB} & = i\kappa\delta_{AB} -i \widehat{\Phi}_{AI} \widehat{\Phi}_{IB}^{\dagger} .
\end{align}
Clearly the LHS vanishes for $A=B$ so we see that for all $A$ (but summing
over $I$)
\begin{align}
\label{CalogPhiNorm}
(-1)^{I+A} \widehat{\Phi}_{IA}^{\dagger} \widehat{\Phi}_{AI} & =
  \widehat{\Phi}_{AI} \widehat{\Phi}_{IA}^{\dagger} = \kappa
\end{align}
and there is no constraint on the diagonal elements of $\widehat{Y}$ so we can
label $\widehat{Y}_{AA} = y_A$.

Note that up to total derivative terms the kinetic term in the Lagrangian is
\begin{align}
\label{CalogKinetic}
i\Str \left( \widehat{Z}^{\dagger} \partial_t \widehat{Z} \right) & \simeq
  \Str \left( \widehat{Y} \partial_t \widehat{X} \right) =
  \sum_A (-1)^A y_A \dot{x}_A
\end{align}
so $(-1)^A y_A$ is the conjugate momentum to $x_A$. Since the coordinates
$x_A$ are unconstrained, they are generically distinct so we can just divide
by $(x_A - x_B)$ in (\ref{CalogGauss}) to find the off-diagonal components
of $\widehat{Y}_{AB}$. We can then write
\begin{align}
\label{CalogYAB}
\widehat{Y}_{AB} & = y_A \delta_{AB} + \frac{i}{x_A - x_B} \left( \kappa \delta_{AB} - \widehat{\Phi}_{AI} \widehat{\Phi}_{IB}^{\dagger} \right)
\end{align}
with the understanding that the second term vanishes for $A=B$ due to the
constraint (\ref{CalogPhiNorm}).

We can now write the Hamiltonian in terms of the coordinates $x_A$ and their
conjugate momenta as
\begin{align}
\label{CalogHam}
H & = \omega \Str \left( \widehat{Z}^{\dagger} \widehat{Z} \right) =
 \frac{\omega}{2} \Str \left( \widehat{X}^2 + \widehat{Y}^2 \right) =
 \frac{\omega}{2} \sum_a (x_a^2 + y_a^2) - \frac{\omega}{2} \sum_{\alpha} (x_{\alpha}^2 + y_{\alpha}^2) + V , \\
\label{CalogPot}
V & = \frac{\omega}{2} \sum_{A \ne B} \frac{(-1)^A}{(x_A - x_B)^2}
  \widehat{\Phi}_{AI} \widehat{\Phi}_{IB}^{\dagger}
  \widehat{\Phi}_{BJ} \widehat{\Phi}_{JA}^{\dagger}. 
\end{align}

It is possible to interpret this as a model of $N+M$ particles. However, while
$N$ have a standard kinetic term, $M$ have the wrong sign for the kinetic term.
At the level of equations of motion this is not a problem but it is likely
problematic to treat the quantum system. Note, however, that the original
system had only first order derivative terms which are well-defined for either
sign of kinetic term. In fact, we will see in Section~\ref{subsecquantize}
that it can be quantized to produce a Hamiltonian bounded from below. Clearly,
this indicates some subtleties in relating the matrix quantum mechanics to the
Calogero model. Nevertheless, it may be possible to interpret the model as a
coupling of an $N$-particle Calogero model to an $M$-particle model with a
specific interaction potential. In particular we see that
\begin{align}
\label{CalogNMHam}
H & = H_N + H_M + V_{\textrm{int}} , \\
\label{CalogNMHamN}
H_N & = 
 \frac{\omega}{2} \sum_a (x_a^2 + y_a^2) +
 \frac{\omega}{2} \sum_{a, b} \frac{1}{(x_a - x_b)^2}
  \widehat{\Phi}_{aI} \widehat{\Phi}_{Ib}^{\dagger}
  \widehat{\Phi}_{bJ} \widehat{\Phi}_{Ja}^{\dagger} , \\
\label{CalogNMHamM}
H_M & = 
 - \frac{\omega}{2} \sum_{\alpha} (x_{\alpha}^2 + y_{\alpha}^2) -
 \frac{\omega}{2} \sum_{\alpha, \beta} \frac{1}{(x_{\alpha} - x_{\beta})^2}
  \widehat{\Phi}_{\alpha I} \widehat{\Phi}_{I\beta}^{\dagger}
  \widehat{\Phi}_{\beta J} \widehat{\Phi}_{J\alpha}^{\dagger} , \\
V_{\textrm{int}} & = \frac{\omega}{2} \sum_{a, \beta} \frac{1}{(x_a - x_{\beta})^2}
  \widehat{\Phi}_{aI} \widehat{\Phi}_{I\beta}^{\dagger}
  \widehat{\Phi}_{\beta J} \widehat{\Phi}_{Ja}^{\dagger}
 - \frac{\omega}{2} \sum_{\alpha, b} \frac{1}{(x_{\alpha} - x_b)^2}
  \widehat{\Phi}_{\alpha I} \widehat{\Phi}_{Ib}^{\dagger}
  \widehat{\Phi}_{bJ} \widehat{\Phi}_{J\alpha}^{\dagger} \nonumber \\
\label{CalogNMIntPot}
 & = \omega \sum_{a, \beta} \frac{1}{(x_a - x_{\beta})^2}
  \widehat{\Phi}_{aI} \widehat{\Phi}_{I\beta}^{\dagger}
  \widehat{\Phi}_{\beta J} \widehat{\Phi}_{Ja}^{\dagger} .
\end{align}

This reduces to the usual Calogero model in the case $M = q = 0$ and $p=1$. 
The constraints (\ref{CalogPhiNorm}) for the single fundamental scalar $\phi$
give (up to unimportant phases which cancel in the potential)
\begin{align}
\label{Calog1phiNorm}
\phi_a & = \sqrt{k} \;\; \forall a .
\end{align}
We then recognize the usual Calogero model
\begin{align}
\label{Calog1phiHam}
H & = \frac{\omega}{2} \sum_a (x_a^2 + y_a^2) + \frac{\omega}{2} \sum_{a \ne b} \frac{\kappa^2}{(x_a - x_b)^2}. 
\end{align}

If we allow arbitrary $p$, but still $M = q = 0$, the potential is generalized
to
\begin{align}
\label{CalogpphiPot}
V & = \frac{\omega}{2} \sum_{a \ne b} \frac{\kappa^2}{(x_a - x_b)^2} (\phi_{ai} \phi_{ib}^{\dagger}) (\phi_{bj} \phi_{ja}^{\dagger})
\end{align}
with the constraints
\begin{align}
\label{CalogpphiNorm}
\phi_{ai} \phi^{\dagger}_{ia} & = \kappa \;\; \forall a .
\end{align}

Further generalizing to $q \ne 0$ we get a potential containing terms quadratic
and quartic in fermions $\psi_{\lambda a}$:
\begin{align}
\label{CalogphipsiPot}
V & = \frac{\omega}{2} \sum_{a \ne b} \frac{\kappa^2}{(x_a - x_b)^2} (\phi_{ai} \phi_{ib}^{\dagger} + \psi_{a \lambda} \psi_{\lambda b}^{\dagger}) (\phi_{bj} \phi_{ja}^{\dagger} + \psi_{b \rho} \psi_{\rho a}^{\dagger})
\end{align}
with the constraints
\begin{align}
\label{CalogphipsiNorm}
\phi_{ai} \phi^{\dagger}_{ia} + \psi_{a \lambda} \psi_{\lambda a}^{\dagger}& = \kappa \;\; \forall a .
\end{align}
Such generalized $N$-particle Calogero models with $SU(p|q)$ internal degrees of
freedom, and the related spin chain models -- so-called supersymmetric
Polychronakos models -- which arise as their freezing limit, have been studied
in \cite{Brink:1997zi, BasuMallick:1999vm, Hikami:1999twa, Basu-Mallick:2008bsa}. 
We also note that this model can be embedded in specific angular momentum
sectors of the ordinary bosonic $U(N)$ matrix model without any
vector-like fields, i.e.\ with $p=q=0$ or Chern-Simons term. In particular,
upon quantization the conserved angular momenta $[Z, Z^{\dagger}]$ become
integer representation $SU(N)$ generators with vanishing $\mathbb{Z}_N$ charge. However,
any such representation of $SU(N)$ can be obtained from a set of bosonic and
fermionic oscillators by the Schwinger construction. These oscillators
correspond to the fields $\phi_{ai}$ and $\psi_{a\lambda}$ in
(\ref{CalogphipsiPot}) while (\ref{CalogphipsiNorm}) imposes the integrality
condition \footnote{We thank A.\ Polychronakos for pointing this out.}.

\subsection{Classical ground state}
\label{subsec2cl_ground}
The classical ground state is the classical solution of least energy, so we
need to find the time-independent solution to the Gauss law constraints which minimizes the
Hamiltonian
\begin{align}
H & = \omega \Str \left( \widehat{Z}^{\dagger} \widehat{Z} \right). 
\end{align}
The Gauss law constraint can be imposed using a Lagrange multiplier. As such a
time-independent Lagrange multiplier appears in exactly the same way as the
original gauge field, we use the same notation $\widehat{\alpha}$. So, we must
minimize
\begin{align}
H & = \Str \left( \omega \widehat{Z}^{\dagger} \widehat{Z} - \widehat{\alpha} \left( [\widehat{Z}, \widehat{Z}^{\dagger}] 
+ \widehat{\Phi}_I\widehat{\Phi}_I^{\dagger} - \kappa I \right) \right). 
\end{align}
Varying with respect to $\widehat{Z}^{\dagger}$ we see that the result is that
the ground state is given by a solution to the Gauss law constraint where also
for some $\widehat{\alpha}$
\begin{align}
\label{ZgsConst}
\omega \widehat{Z} & = \left[ \widehat{\alpha}, \widehat{Z} \right]. 
\end{align}

As finding the general solution is not simple,
let us now consider explicitly the case where $p=1$ and $q=1$. 
Up to gauge transformations, we take the following generic configuration

\begin{align}
\label{csol1a}
\phi_{a}&=
\left(
\begin{array}{c}
0 \\
\vdots \\
0 \\
x \\
\end{array}
\right), &
\widetilde{\psi}_{\alpha}&=
\left(
\begin{array}{c}
0 \\
\vdots \\
0 \\
0 \\
\end{array}
\right),& 
\widetilde{\phi}_{\alpha}&=
\left(
\begin{array}{c}
0 \\
\vdots \\
0 \\
z \\
\end{array}
\right), &
\psi_{a}&=
\left(
\begin{array}{c}
0 \\
\vdots \\
0 \\
y \\
\end{array}
\right). 
\end{align}
Then the Gauss law constraints become 
\begin{align}
\label{Gauss_AB}
[\widehat{Z}, \widehat{Z}^{\dagger}]_{AB} & = \left( \kappa - (x^{\dagger}x - y^{\dagger}y)\delta_{AN} - z^{\dagger}z\delta_{A(N+M)} \right) \delta_{AB} -yz^{\dagger}\delta_{AN}\delta_{B(N+M)} - zy^{\dagger}\delta_{A(N+M)}\delta_{BM} \nonumber \\
 & = \left( \kappa - z^{\dagger}z (\delta_{AN} + \delta_{A(N+M)}) - \kappa(N-M)\delta_{AN} \right) \delta_{AB} -yz^{\dagger}\delta_{AN}\delta_{B(N+M)} - zy^{\dagger}\delta_{A(N+M)}\delta_{BN}
\end{align}
where the second line takes into account the constraint
\begin{align}
x^{\dagger}x - y^{\dagger}y - z^{\dagger}z = \kappa(N-M)
\end{align}
on $x$, $y$ and $z$
imposed by (\ref{G_strace}), the supertrace of the Gauss law constraints (\ref{gauss_sg}).

We can also use the residual $U(N-1|M-1)$ symmetry to partially
diagonalize $\widehat{\alpha}$:
\begin{align}
\label{csol1b}
\alpha_{ab}&=\beta_{a}\delta_{ab},& a, b \ne N, \\
\widetilde{\alpha}_{\alpha\beta}&=\widetilde{\beta}_{\alpha}\delta_{\alpha\beta},& \alpha, \beta \ne M. 
\end{align}
For generic $\beta_a$ and $\widetilde{\beta}_{\alpha}$ this reduces the
symmetry $U(N-1|M-1)$ to its maximal Abelian subgroup, i.e.\ the Cartan subgroup
so that the diagonal components correspond to the generators of the Cartan
subalgebra. Of course, the residual symmetry will be enhanced to a
non-Abelian subgroup of $U(N|M)$ if there is any degeneracy in the values of
$\beta_a$ and $\widetilde{\beta}_{\alpha}$.

Finally, varying with respect to $\widehat{\Phi}_I^{\dagger}$ gives
\begin{align}
\label{csol1b2}
\widehat{\alpha}\widehat{\Phi}_I = 0
\end{align}
which for the above configuration for $\widehat{\Phi}$ imposes additional
constraints on $\widehat{\alpha}$ and $\widehat{\Phi}$ which can be described in
terms of three cases:
\begin{enumerate}
\item $\widehat{\alpha}_{AN} = \widehat{\alpha}_{A(M+N)} = 0$ so that
$\widehat{\alpha}$ is completely diagonal
\footnote{
However, note that this solution with $\alpha_{NN}=\widetilde{\alpha}_{MM}=0$
involving the same values for $\widehat{\alpha}$ is not desirable if we do not
want any enhanced symmetry. 
}
but there are no further restrictions on $\widehat{\Phi}$.
\item $x = y = \widehat{\alpha}_{A(M+N)} = 0$. 
\item $z = \widehat{\alpha}_{AN} = 0$. 
\end{enumerate}

We comment on the first case in appendix~\ref{app_classicsol}.
In case 2 both $\phi$ and $\psi$ 
vanish so this is equivalent to
considering solutions in the case $p=0$ and $q=1$.
We now consider the third case, but as we will see, we can find solutions
taking the more restrictive ansatz
\begin{align}
\label{csol1b3}
\alpha_{NN} &= 0,& 
z&=\widetilde{\phi}_{M} =0. 
\end{align}

Then the equations for $\widehat{Z}_{AB}$, 
(\ref{ZgsConst}) reduce to  
\begin{align}
\label{csol1c}
\left(\beta_{a}-\beta_{b}-\omega\right)Z_{ab}&=0,& 
\left(\widetilde{\beta}_{\alpha}-\widetilde{\beta}_{\beta}-\omega\right)\widetilde{Z}_{\alpha\beta}&=0,\\
\label{csol1c2}
\left(\beta_{a}-\widetilde{\beta}_{\alpha}-\omega\right)A_{a\alpha}&=0,& 
\left(\widetilde{\beta}_{\alpha}-\beta_{a}-\omega\right)B_{\alpha a}&=0. 
\end{align}
It follows that the diagonal parts of $Z$ and $\widetilde{Z}$ are zero. 
Let $\bm{\alpha}^{a}$ be the diagonal matrices associated to the simple roots 
which form a complete set of diagonal matrices
\begin{align}
\label{simpler_1a}
\bm{\alpha}^{a}&=\bm{\alpha}^{a}_{i}H_{i}
=\frac12 \mathrm{diag}
\left(
0,0,\cdots,0,\underbrace{1}_{a}, \underbrace{-1}_{a+1},0\cdots,0
\right)
\end{align}
where 
\begin{align}
\label{simpler_1b}
H_{m}&=
\mathrm{diag}\left(
\underbrace{1,1,\cdots, 1}_{m}, -m, 0,\cdots, 0
\right)
\end{align}
is the generator of the Cartan subalgebra 
and the diagonal parts $\alpha_{D}$ of the gauge fields $\alpha$ 
can be written as 
\begin{align}
\label{simpler_1c}
\alpha_{D}&=\sum_{a=1}^{N-1}c_{a}\bm{\alpha}^{a}. 
\end{align}
It follows that 
\begin{align}
\label{simpler_1d}
\Tr(\alpha_{D}\cdot \bm{\alpha}^{a})
&=\frac12 c_{a}=\beta_{a}-\beta_{a+1}
\end{align}
where 
\begin{align}
\label{simpler_1e}
\alpha_{D}&=\mathrm{diag}\left(
\beta_1, \beta_2, \cdots, \beta_N
\right). 
\end{align}
Thus $\left\{\beta_{a}-\beta_{b}=\omega\right\}$ corresponds to simple roots 
$\left\{\epsilon_{a}-\epsilon_{b}\right\}$ of $A_{N-1}$ 
so that the associated canonical variables $Z_{ab}$ take values in $\mathbb{C}E_{ab}$ 
where $\left\{\epsilon_{1},\cdots, \epsilon_{N}\right\}$ is the basis of the root space 
and $E_{ab}$ is the matrix with $ab$-entry being one and all others being zeros. 
Similarly, $\left\{\widetilde{\beta}_{\alpha}-\widetilde{\beta}_{\beta}=\omega\right\}$ corresponds to 
simple roots $\left\{\delta_{\alpha}-\delta_{\beta}\right\}$ of $A_{M-1}$ 
so that $\widetilde{Z}_{\alpha\beta}$ take values in $\mathbb{C}E_{\alpha\beta}$. 

We consider the case in which 
there is no enhanced gauge symmetry with $(N-1)$ different values of $\beta_{a}$ 
and $(M-1)$ different values of $\widetilde{\beta}_{\alpha}$. 

In contrast to the Lie algebra, 
there are many inequivalent simple root systems in the basic Lie superalgebra 
consisting of the even roots
\begin{align}
\label{even_root1}
\Delta_{\overline{0}}&=
\left\{
\epsilon_{a}-\epsilon_{b}|1\le a\neq b\le N
\right\}\cup 
\left\{
\delta_{\alpha}-\delta_{\beta}|
1\le \alpha\neq \beta\le M
\right\}
\end{align}
and the odd roots 
\begin{align}
\label{odd_root1}
\Delta_{\overline{1}}&=
\left\{
\pm \left(
\epsilon_{a}-\delta_{\alpha}
\right)|
1\le a\le N, 1\le \alpha\le M
\right\}
\end{align}
where $\left\{\epsilon_{1},\cdots, \epsilon_{N}; \delta_{1},\cdots, \delta_{M}\right\}$ is the basis 
of the root space with the bilinear forms $(\epsilon_{a}, \epsilon_{b})=\delta_{ab}$, 
$(\delta_{\alpha},\delta_{\beta})=-\delta_{\alpha\beta}$ and 
$(\epsilon_{a}, \delta_{\alpha})=0$. 
All the simple root systems are given by \cite{Kac:1977em}
\begin{align}
\label{ssim_root}
\Pi&=
\pm 
\left\{
\epsilon_1-\epsilon_2, 
\epsilon_2-\epsilon_3, 
\cdots, 
\epsilon_{s_1}-\delta_1, 
\delta_1-\delta_2, 
\cdots, 
\delta_{t_1}-\epsilon_{s_{1}+1}, 
\cdots
\right\}
\end{align}
up to the Weyl equivalence 
where $S=\left\{s_1<s_2<\cdots\right\}$ and $T=\left\{t_1<t_2< \cdots\right\}$ 
are two increasing sequences. 
Correspondingly $\left\{\beta_{a}-\beta_{b}\right\}$, 
$\left\{\widetilde{\beta}_{\alpha}-\widetilde{\beta}_{\beta}\right\}$, 
$\left\{\beta_{a}-\widetilde{\beta}_{\alpha}\right\}$ 
and 
$\left\{\widetilde{\beta}_{\alpha}-\beta_{a}\right\}$ 
admit different configurations for non-trivial valued $Z_{ab}$, $\widetilde{Z}_{\alpha\beta}$, $A_{a\alpha}$ and $B_{\alpha a}$.

According to the configuration (\ref{csol1b3}), 
the Gauss law conditions (\ref{gaussbb})-(\ref{gaussbf}) reduce to 
\begin{align}
\label{c_gauss_bb}
[Z,Z^{\dag}]_{ab}+A_{a\alpha}A_{\alpha a}^{\dag}
-B_{a \alpha}^{\dag}B_{\alpha a}+\kappa (N-M)\delta_{aN}\delta_{bN}&=\kappa \delta_{ab}, \\
\label{c_gauss_ff}
[\widetilde{Z},\widetilde{Z}]_{\alpha\beta}
-A^{\dag}_{\alpha a}A_{a\beta}+B_{\alpha a}B_{a\beta}^{\dag}&=\kappa \delta_{\alpha \beta}, \\
\label{c_gauss_fb}
Z_{ab}B_{b\beta}^{\dag}-B_{a\alpha}^{\dag}\widetilde{Z}_{\alpha\beta}
-Z^{\dag}_{ab}A_{b\beta}+A_{a\alpha}\widetilde{Z}_{\alpha \beta}^{\dag}&=0,\\
\label{c_gauss_bf}
B_{\alpha a}Z_{ab}^{\dag}-\widetilde{Z}_{\alpha\beta}^{\dag}B_{\beta b}
+\widetilde{Z}_{\alpha \beta}A_{\beta b}^{\dag}-A_{\alpha a}^{\dag}Z_{ab}&=0
\end{align}
and the trace conditions (\ref{Gtrace_bb}), (\ref{Gtrace_ff}) and the supertrace condition (\ref{G_strace}) become
\begin{align}
\label{c_gtrace_bb}
\sum_{a,\alpha}\left(
A_{a\alpha}A_{\alpha a}^{\dag}+B_{\alpha a}B_{a\alpha}^{\dag}
\right) +|x|^{2}-y^{\dag}y&=\kappa N, \\
\label{c_gtrace_ff}
\sum_{a,\alpha}\left(
A_{a\alpha}A_{\alpha a}^{\dag}+B_{\alpha a}B_{a\alpha}^{\dag}
\right)&=\kappa M, \\
\label{c_gstrace}
|x|^{2}-y^{\dag}y&=\kappa (N-M). 
\end{align}

Let us take, for some integer $r$ with $1 \le r \le M \le N$,
\begin{align}
\label{Beta_conf1}
\beta_a&=
\begin{cases}
(N+r+1-2a)\omega& a=1,\cdots, r\cr
(N-a)\omega&  a=r+1, \cdots, N \cr
\end{cases} \nonumber\\
\widetilde{\beta}_{\alpha}&=
\begin{cases}
(N+r-2\alpha)\omega& \alpha=1,\cdots, r\cr
(N+M+r-\alpha)\omega& \alpha=r+1,\cdots, M\cr
\end{cases}
\end{align}
for which we have distinct values for $\beta_{a}$, $a=1,\cdots, N$ 
and $\widetilde{\beta}_{\alpha}$, $\alpha=1,\cdots, M$ 
as we are considering the case with no enhanced symmetry. 
These diagonal parts correspond to the set of roots 
\begin{align}
\label{G_root1}
\Delta_{\overline{0}}&=
\left\{
\epsilon_{r+1}-\epsilon_{r+2}, 
\epsilon_{r+2}-\epsilon_{r+3}, 
\cdots, 
\epsilon_{N-1}-\epsilon_{N}
\right\}\cup 
\left\{
\delta_{r+1}-\delta_{r+2}, 
\delta_{r+2}-\delta_{r+3}, 
\cdots, 
\delta_{M-1}-\delta_{M}
\right\} ,
\nonumber\\
\Delta_{\overline{1}}&=
\left\{
\epsilon_{1}-\delta_{1}, 
\epsilon_{2}-\delta_{2}, 
\cdots, 
\epsilon_{r}-\delta_{r}
\right\}
\cup 
\left\{
\delta_{M+1}-\epsilon_{1}, 
\delta_{1}-\epsilon_{2}, 
\delta_{2}-\epsilon_{3}, 
\cdots, 
\delta_{r}-\epsilon_{r+1}
\right\}. 
\end{align}
The number of available independent components is 
$(N-r-1)+(M-r-1)+r+(r+1)=N+M-1$. 
According to (\ref{Beta_conf1}), 
the component fields of supermatrix $\widehat{Z}$ take general forms as 
\begin{align}
\label{Gr_form1}
Z_{ab}&=Z_{a (a+1)},& a&=r+1,\cdots, N-1, \nonumber\\
\widetilde{Z}_{\alpha \beta}&=\widetilde{Z}_{\alpha (\alpha+1)},& \alpha&=r+1,\cdots, M-1, \nonumber\\
A_{a\alpha}&=A_{\alpha \alpha}\delta_{a \alpha},& a&=1,\cdots, r, \nonumber\\
B_{\alpha a}&=B_{\alpha (\alpha+1)}\delta_{a (\alpha+1)}+B_{M1}\delta_{\alpha M}\delta_{a 1}, & \alpha&=1,\cdots, r. 
\end{align}
Then the Gauss law constraints (\ref{c_gauss_bb}) become
\begin{align}
\label{G_gauss_bb1}
A_{11}A_{11}^{\dag}&=B_{1M}^{\dag}B_{M1}+\kappa, \\
\label{G_gauss_bb2}
A_{aa}A_{aa}^{\dag}&=B_{a (a-1)}^{\dag}B_{(a-1) a}+\kappa,& a&=2,\cdots, r, \\
\label{G_gauss_bb3}
|Z_{(r+1)(r+2)}|^{2}&=
B_{(r+1) r}^{\dag}B_{r (r+1)}+\kappa, \\
\label{G_gauss_bb4}
|Z_{a (a+1)}|^{2}&=|Z_{(a-1) a}|^{2}+\kappa,& a&=r+2, \cdots , N-1,\\
\label{G_gauss_bb5}
|Z_{(N-1) N}|^{2}&=\kappa(N-M-1). 
\end{align}
and we find that 
\begin{align}
\label{G_sol0a}
|Z_{a (a+1)}|^{2}&=\kappa(a-M)
\end{align}
for $a=r+1,\cdots, N-1$. 
Due to the positivity of the equation (\ref{G_sol0a}), it follows that 
$r \ge M-1$ and so $r$ must be either $M-1$ or $M$.

\begin{enumerate}

\item \textbf{Parallel polarized vortex-antivortex state}

Let us consider the case for $r=M-1$, in which the configuration (\ref{Beta_conf1}) reduces to 
\begin{align}
\label{Beta_conf1a}
\beta_a&=
\begin{cases}
(N+M-2a)\omega& a=1,\cdots, M-1\cr
(N-a)\omega&  a=M, \cdots, N \cr
\end{cases} \nonumber\\
\widetilde{\beta}_{\alpha}&=
\begin{cases}
(N+M-2\alpha)\omega& \alpha=1,\cdots, M-1\cr
(N+M-1)\omega& \alpha=M\cr
\end{cases}
\end{align}
and the corresponding simple root system is 
\begin{align}
\Delta_{\overline{0}}&=
\left\{
\epsilon_{M}-\epsilon_{M+1}, 
\epsilon_{M+2}-\epsilon_{M+2}, 
\cdots, 
\epsilon_{N-1}-\epsilon_{N}
\right\}
\nonumber\\
\Delta_{\overline{1}}&=
\left\{
\epsilon_{1}-\delta_{1}, 
\epsilon_{2}-\delta_{2}, 
\cdots, 
\epsilon_{M-1}-\delta_{M-1}
\right\}
\cup 
\left\{
\delta_{M+1}-\epsilon_{1}, 
\delta_{1}-\epsilon_{2}, 
\delta_{2}-\epsilon_{3}, 
\cdots, 
\delta_{M-1}-\epsilon_{M}
\right\}. 
\end{align}
Thus the classical configuration does not admit non-trivial values for $\widetilde{Z}$. 
From the configurations (\ref{G_sol0a}) and  
the Gauss law conditions (\ref{G_gauss_bb3}) one finds that 
\begin{align}
\label{R_sol_1}
B^{\dag}_{M (M-1)}B_{(M-1) M}&=-\kappa. 
\end{align}
Then the second set (\ref{c_gauss_ff}) of the Gauss law constraints become 
\begin{align}
\label{G_gauss_ff1}
-A_{\alpha\alpha}^{\dag}A_{\alpha\alpha}+B_{\alpha (\alpha+1)}B_{(\alpha+1) \alpha}^{\dag}&=\kappa,& 
\alpha&=1,\cdots, M-2, \\
\label{G_gauss_ff2}
A_{(M-1) (M-1)}^{\dag}A_{(M-1) (M-1)}&=0, \\
\label{G_gauss_ff3}
B_{M1}B_{1M}^{\dag}&=\kappa. 
\end{align}
Combining these with (\ref{G_gauss_bb1})-(\ref{G_gauss_bb5}) and (\ref{G_sol0a}), 
we obtain 
\begin{align}
\label{Rad_sola}
Z&=\sqrt{\kappa}
\left(
\begin{array}{ccccc|cccccc}
&&&&&&&&& \\
&&\textrm{\Large 0}&&&&&&\textrm{\Large 0}& \\
&&&&&&&& \\ \hline 
&&&&&0&\sqrt{1}&&& \\
&&&&&&0&\sqrt{2}&& \\
&&\textrm{\Large 0}&&&&&\ddots&\ddots& \\
&&&&&&&&&\sqrt{N-M-1} \\
&&&&&&&&&0 \\
\end{array}
\right), \nonumber\\
B&=
\left(
\begin{array}{ccccc|ccccc}
0&B_{12}&&&\\
\vdots&&B_{23}&&\\
&&&\ddots&&&&\textrm{\Large 0}&&\\
0&&&&B_{(M-1) M} \\ 
B_{M 1}&0&\cdots&&0 \\
\end{array}
\right), \nonumber\\
\widetilde{Z}&=0,& A&=0,\nonumber\\
\phi&=\left(
\begin{array}{c}
0\\
0\\
\vdots\\
\vdots\\
0\\
x\\
\end{array}
\right),& 
\psi&=\left(
\begin{array}{c}
0\\
0\\
\vdots\\
\vdots\\
0\\
y\\
\end{array}
\right)
\end{align}
so that 
\begin{align}
\label{Rad_solb}
B_{(\alpha+1) \alpha}^{\dag}B_{\alpha (\alpha + 1)} &=-\kappa,& 
B_{1M}^{\dag}B_{M1}&=-\kappa,\nonumber\\
|x|^{2}-y^{\dag}y&=\kappa (N-M). 
\end{align}
The fermionic Gauss law conditions (\ref{c_gauss_fb}) and (\ref{c_gauss_bf}) hold 
for the above static configurations.

\begin{figure}
\begin{center}
\includegraphics[width=9.5cm]{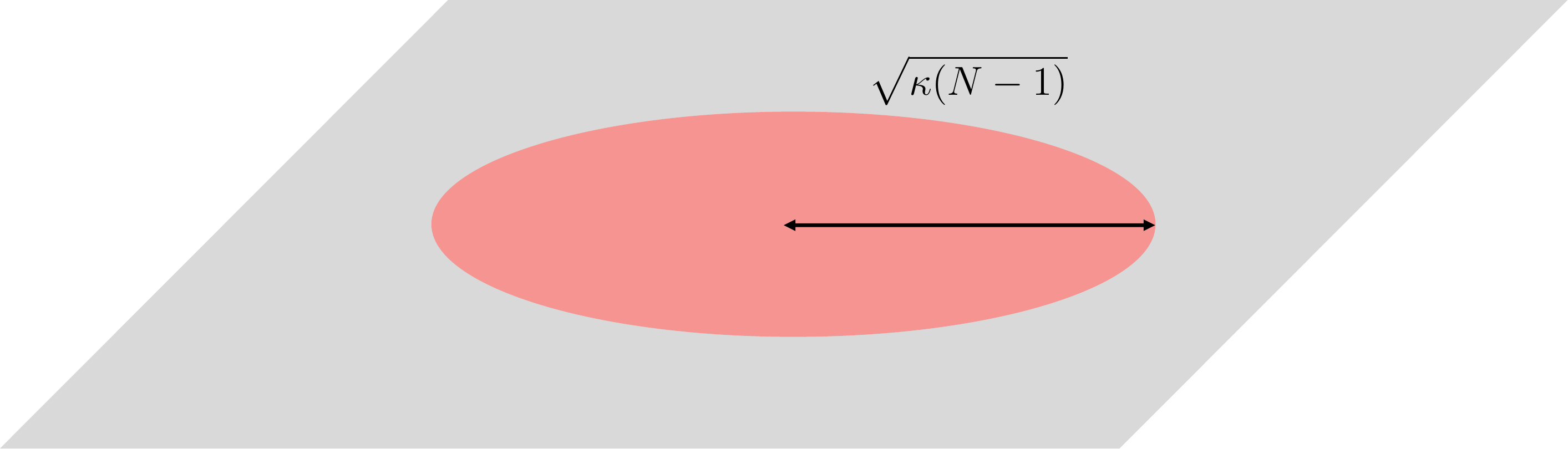}
\caption{A quantum droplet of radius $\sqrt{\kappa(N-1)}$ formed by $N$ electrons on the plane. }
\label{fig_csol1}
\end{center}
\end{figure}
In order to interpret our classical solutions (\ref{Rad_sola}) and (\ref{Rad_solb}), 
let us firstly consider a special case for $M=0$ 
where the supermatrix field $\widehat{Z}$ becomes the ordinary matrix $N\times N$ matrix $Z$ 
and $\widetilde{Z}=0$, $A=0$ and $B=0$. 
The non-trivial field configurations are given by 
\begin{align}
\label{B_sol1a}
Z&=\sqrt{\kappa}\left(
\begin{array}{cccccc}
0&\sqrt{1}&& && \\
&0&\sqrt{2}&&&\\
&&\ddots&\ddots&&\\
&&&&\sqrt{N-2}&\\
&&&&0&\sqrt{N-1}\\
&&&&&0\\
\end{array}
\right), & 
\phi&=\left(
\begin{array}{c}
0\\
0\\
\vdots\\
\vdots\\
0\\
x\\
\end{array}
\right),& 
\psi&=\left(
\begin{array}{c}
0\\
0\\
\vdots\\
\vdots\\
0\\
y\\
\end{array}
\right)
\end{align}
and we have the constraint 
\begin{align}
\label{B_sol1b}
|x|^{2}-y^{\dag}y&=\kappa N. 
\end{align}
In this case, the gauge symmetry is an ordinary $U(N)$ symmetry, 
but we still have a supergroup $SU(p|q)$ flavour symmetry. 
We have already presented the generalized Calogero model for this case in
section~\ref{subsec2Calogero}, but here we find the classical ground state. 
In the particular case when $q=0$, the supervector fields become 
$\phi=\sqrt{\kappa N}(0,0,\cdots, 1)^{T}$,  $\psi=0$ 
so that (\ref{B_sol1b}) can be uniquely solved 
and our configuration is exactly same as the unique classical ground state in \cite{Polychronakos:2001mi}.

One can obtain a physical interpretation of the resulting configurations 
in the description of the fractional quantum Hall effect \cite{Polychronakos:2001mi}. 
The solution (\ref{B_sol1a}) corresponds to the round quantum Hall droplet (see Figure \ref{fig_csol1}). 
The radius squared of the disk formed by $N$ electrons or $N$ vortices is given by 
the maximum eigenvalue of $Z^{\dag}Z$ 
\begin{align}
\label{R2_1}
R^2&=\kappa (N-1). 
\end{align}
The total energy is given by
\begin{align}
\label{energy_1}
E&=
\omega \Tr(Z^{\dag}Z)
=\kappa\omega \frac{N(N-1)}{2}
\end{align}
and depends on the size of the system consisting of $N$ vortices.

As discussed in \cite{Polychronakos:2001mi}, 
the non-zero values $x$ and $y$ which absorb the anomaly 
of the commutators are required to realize the finite droplet 
and they are interpreted as boundary terms. 
In general, when $p+q>1$ such terms may be associated to certain additional internal degrees of freedom in the system. 
For example, when we consider multi-layered two-dimensional systems, 
e.g.\ multi-layered graphene (with $q=0$), 
they correspond to the so-called valley degeneracies, which label $p$ multi-layers of vortices in our case. 
Also note that in the solutions above with $p=q=1$ the $SU(p|q)$ symmetry acts
on $x$ and $y$ (preserving $x^2 - y^{\dagger} y$) while all other fields are
invariant (noting that $z=0$ for these solutions).

Now for a general case with $M\neq 0$ the radius squared of the disk formed by electrons is given by 
the maximum eigenvalue of $Z^{\dag}Z$ 
\begin{align}
\label{R2_2}
R^2&=\kappa (N-M-1). 
\end{align}
This indicates that 
$(N-M)$ electrons form a round disk of area $\pi \kappa(N-M-1)$. 
The total energy is 
\begin{align}
\label{energy_2}
E&=
\omega\Tr (Z^{\dag}Z+B^{\dag}B)
=\kappa\omega\left[ \frac{(N-M)(N-M-1)}{2}-M \right]. 
\end{align}
The first term is the energy of $(N-M)$ electrons or vortices as in (\ref{energy_1}). 
An interesting result is the second term with negative contribution to the energy. 
This is the interaction energy of vortices and antivortices. 
As we have argued in section \ref{subsec_VAmlayers}, 
this indicates the negative binding energy of parallel polarized
vortex-antivortex pairs (see Figure \ref{fig_csol3}). 
Accordingly this classical configuration is expected to be associated 
with $M$ vortex-antivortex pairs with parallel polarization.

\begin{figure}
\begin{center}
\includegraphics[width=12.5cm]{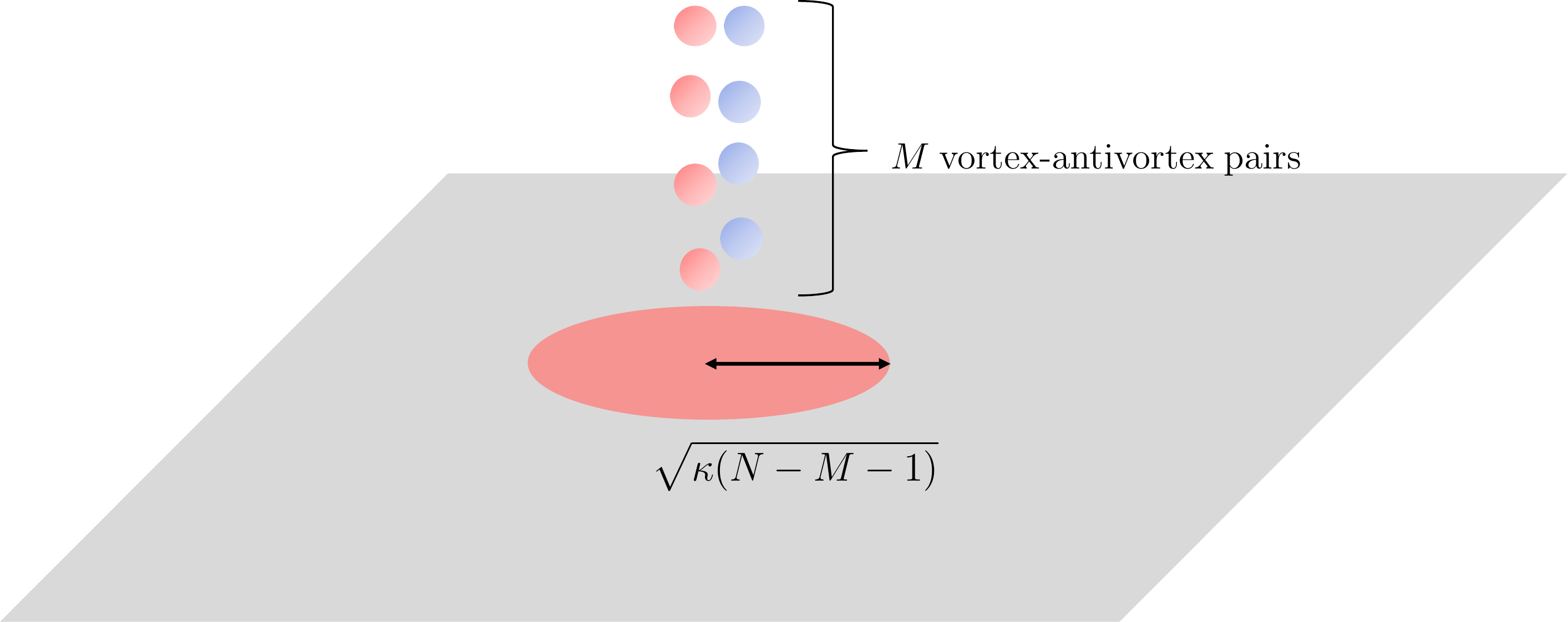}
\caption{Parallel polarized vortex-antivortex pair ground state. 
$(N-M)$ vortices form a droplet of radius $\sqrt{\kappa (N-M)}$. 
$M$ vortex-antivortex pairs are created 
and the cores of vortices and antivortices approach each other on spiraling orbits 
and meet in the center. }
\label{fig_csol3}
\end{center}
\end{figure}
%
%
%
%
%


\item \textbf{Antiparallel vortex-antivortex ground state}

Next consider the case of $r=M$. 
The configuration (\ref{Beta_conf1}) is given by 
\begin{align}
\label{Beta_conf1b}
\beta_a&=
\begin{cases}
(N+M+1-2a)\omega& a=1,\cdots, M\cr
(N-a)\omega&  a=M+1, \cdots, N \cr
\end{cases} \nonumber\\
\widetilde{\beta}_{\alpha}
&=(N+M-2\alpha)\omega. 
\end{align}
In these configurations all the values of 
$\beta_{a}$ and $\widetilde{\beta}_{\alpha}$ are different 
as required for no enhanced symmetry. 
As in the previous case, 
the configurations (\ref{Beta_conf1b}) admit $(N-M-1)+0+M+M=N+M-1$ non-trivial components of 
supermatrix field $\widehat{Z}$, in which there are $2M$ fermionic components. 
They correspond to the set of roots
\begin{align}
\label{root_Rot}
\Delta_{\overline{0}}&=
\left\{
\epsilon_{M+1}-\epsilon_{M+2}, 
\epsilon_{M+2}-\epsilon_{M+3}
\cdots, 
\epsilon_{N-1}-\epsilon_{N}
\right\},\nonumber\\
\Delta_{\overline{1}}&=
\left\{
\epsilon_{1}-\delta_{1}, 
\epsilon_{2}-\delta_{2}, 
\cdots, 
\epsilon_{M}-\delta_{M}
\right\}
\cup 
\left\{
\delta_{1}-\epsilon_{2}, 
\delta_{2}-\epsilon_{3}, 
\cdots, 
\delta_{M}-\epsilon_{M+1}
\right\}. 
\end{align}
In this case, the component fields of supermatrix $\widehat{Z}$ may take the form 
\begin{align}
\label{Rot_Gr_form}
Z&=Z_{a (a+1)},& a&=M+1,\cdots, N-1, \nonumber\\
\widetilde{Z}&=0, \nonumber\\
A_{a\alpha}&=A_{\alpha \alpha}\delta_{a \alpha},& 
B_{\alpha a}&=B_{\alpha (\alpha+1)}\delta_{a (\alpha+1)}
\end{align}
for $\alpha=1,\cdots, M$. 
Plugging the expressions (\ref{Rot_Gr_form}) into the first set (\ref{c_gauss_bb})
of the Gauss law constraints, 
we get
\begin{align}
\label{Rot_gauss_bb1}
A_{aa}A_{aa}^{\dag}&=B_{a (a-1)}^{\dag}B_{(a-1) a}+\kappa,& a&=1,\cdots, M, \\
\label{Rot_gauss_bb2}
|Z_{(M+1)(M+2)}|^{2}&=B_{(M+1)M}^{\dag}B_{M(M+1)}+\kappa, \\
\label{Rot_gauss_bb3}
|Z_{a (a+1)}|^{2}&=|Z_{(a-1) a}|^{2}+\kappa,& a&=M+2,\cdots, N-1,\\
\label{Rot_gauss_bb4}
|Z_{(N-1) N}|^{2}&=\kappa(N-M-1). 
\end{align}
Thus we find 
\begin{align}
\label{F_sol0a}
|Z_{a (a+1)}|^{2}&=\kappa(a-M),\\
\label{F_sol0b}
B_{M (M+1)}&=0
\end{align}
for $a=M+1,\cdots, N-1$. 
From the Gauss law conditions (\ref{c_gauss_ff}) one finds
\begin{align}
\label{F_gauss_ff1}
A_{\alpha\alpha}A_{\alpha\alpha}^{\dag}&=B_{(\alpha+1) \alpha}^{\dag}B_{\alpha (\alpha+1)}+\kappa
\end{align}
where $\alpha=1,\cdots, M$. 
It then follows that 
\begin{align}
\label{F_gauss_ff2}
B_{\alpha (\alpha+1)}&=0,& 
A_{\alpha\alpha}A_{\alpha\alpha}^{\dag}&=\kappa. 
\end{align}
Putting all together, the classical solution is given by 
\begin{align}
\label{Rot_sola}
Z&=\sqrt{\kappa}
\left(
\begin{array}{ccccc|cccccc}
&&&&&&&&& \\
&&\textrm{\Large 0}&&&&&&\textrm{\Large 0}& \\
&&&&&&&& \\ \hline 
&&&&&0&\sqrt{1}&&& \\
&&&&&&0&\sqrt{2}&& \\
&&\textrm{\Large 0}&&&&&\ddots&\ddots& \\
&&&&&&&&&\sqrt{N-M-1} \\
&&&&&&&&&0 \\
\end{array}
\right),& \widetilde{Z}&=0, \nonumber\\
A&=
\left(
\begin{array}{cccc}
A_{11}&&&\\
&A_{22}&&\\
&&\ddots&\\
&&&A_{MM} \\ \hline 
&&& \\
&&\textrm{\Large 0}& \\ 
&&& \\
\end{array}
\right)
,& 
B&=0, \nonumber\\
\phi&=\left(
\begin{array}{c}
0\\
0\\
\vdots\\
\vdots\\
0\\
x\\
\end{array}
\right),& 
\psi&=\left(
\begin{array}{c}
0\\
0\\
\vdots\\
\vdots\\
0\\
y\\
\end{array}
\right)
\end{align}
so that 
\begin{align}
\label{Rot_solb}
A_{\alpha\alpha}A_{\alpha\alpha}^{\dag}&=\kappa,& 
|x|^{2}-y^{\dag}y&=\kappa (N-M). 
\end{align}

The radius squared of the disk formed by electrons is given by 
the maximum eigenvalue of $Z^{\dag}Z$ 
\begin{align}
\label{R2_3}
R^2&=\kappa (N-M-1). 
\end{align}
Again this implies that 
$(N-M)$ vortices form a round disk of area $\pi \kappa(N-M-1)$. 
The total energy is 
\begin{align}
\label{energy_3}
E&=
\omega\Tr (Z^{\dag}Z-A^{\dag}A)
=\kappa\omega\left[ \frac{(N-M)(N-M-1)}{2}+M \right]. 
\end{align}
The first term is the energy of $(N-M)$ vortices as in (\ref{energy_2})
However, unlike the previous result (\ref{energy_2}), 
the second term has positive contributions to the energy. 
This would correspond to the positive energy of antiparallel polarized vortex-antivortex pairs, 
where the $M$ vortex-antivortex pairs lose the energy due to the emission of sound waves (see Figure \ref{fig_csol2}). 
Thus this classical solution would be the vortex-antivortex pairs of antiparallel polarization. 

%
%
%
%
%
\begin{figure}
\begin{center}
\includegraphics[width=12.5cm]{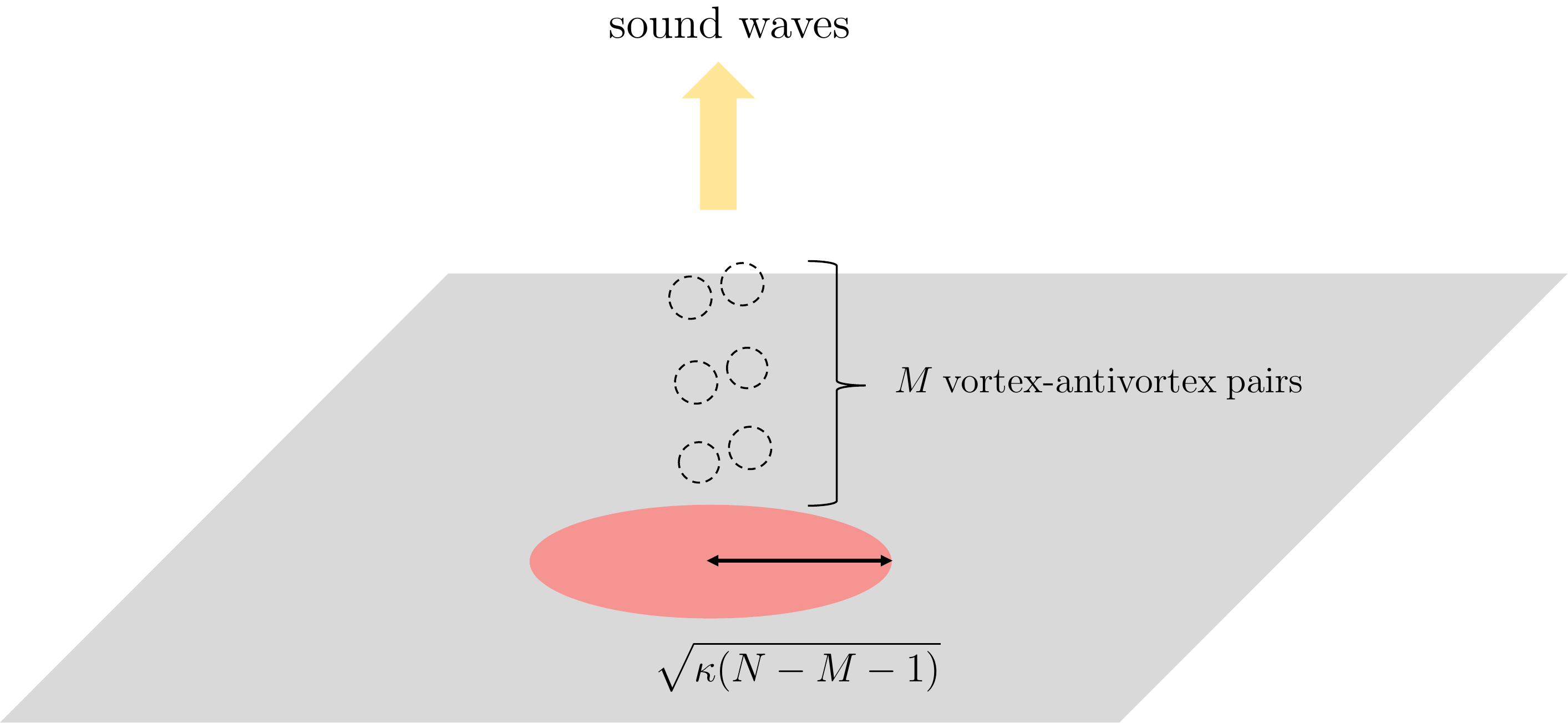}
\caption{
Antiparallel polarized vortex-antivortex pair ground state. 
$(N-M)$ electrons form a droplet of radius $\sqrt{\kappa (N-M-1)}$ on the plane. 
The annihilation of $M$ vortex-antivortex pairs releases the energy via the emission of sound wave. }
\label{fig_csol2}
\end{center}
\end{figure}

\end{enumerate}

\section{Quantum states}
\label{sec2quantize}

\subsection{Quantization}
\label{subsecquantize}
The elements of supermatrix $\widehat{Z}$ and 
the components of supervectors $\widehat{\Phi}$ become operators upon quantization. 
The action (\ref{sgqm1c}) specifies the
canonical commutation relations
\begin{align}
\label{sgqmcom1}
[Z_{ab}, Z_{cd}^{\dag}]&=\delta_{ad}\delta_{bc},&
[\widetilde{Z}_{\alpha \beta}, \widetilde{Z}_{\gamma \delta}^{\dag}]&= -\delta_{\alpha \delta}\delta_{\beta \gamma}, \nonumber\\
\{ A_{a\beta}, A_{\gamma d}^{\dag} \} &= -\delta_{ad}\delta_{\beta\gamma},&
\{ B_{\alpha b}, B_{c\delta}^{\dag} \} &=\delta_{\alpha\delta}\delta_{bc}, \nonumber \\
[\phi_{ai},\phi_{jb}^{\dag}]&=\delta_{ij}\delta_{ab}, &
[\widetilde{\phi}_{\alpha \lambda},\widetilde{\phi}_{\rho\beta}^{\dag}]&= -\delta_{\lambda\rho}\delta_{\alpha\beta}, \nonumber\\
\{ \psi_{a \lambda},\psi_{\rho b}^{\dag} \} &= -\delta_{\lambda\rho}\delta_{ab}, &
\{ \widetilde{\psi}_{\alpha i},\widetilde{\psi}_{j\beta}^{\dag} \} &=\delta_{ij}\delta_{\alpha\beta}, 
\end{align}
which can also be expressed in terms of superbrackets as
\begin{align}
\label{sgqmcom1S}
\left[ \widehat{Z}_{AB}, \widehat{Z}_{CD}^{\dag} \right]_S & \equiv \widehat{Z}_{AB} \widehat{Z}_{CD}^{\dag} - (-1)^{(A+B)(C+D)} \widehat{Z}_{CD}^{\dag} \widehat{Z}_{AB} = (-1)^B \delta_{AD}\delta_{BC} , \\
\left[ \widehat{\Phi}_{AI}, \widehat{\Phi}_{JB}^{\dag} \right]_S & \equiv \widehat{\Phi}_{AI} \widehat{\Phi}_{JB}^{\dag} - (-1)^{(I+A)(J+B)} \widehat{\Phi}_{JB}^{\dag} \widehat{\Phi}_{AI} = (-1)^I\delta_{IJ}\delta_{AB}. 
\end{align}

When $M=0$ and $q=0$, (\ref{sgqmcom1}) simplifies to
\begin{align}
\label{qmxcom1}
[Z_{ab}, Z_{cd}^{\dag}]&=\delta_{ad}\delta_{bc},& 
[\phi_{ai},\phi_{jb}^{\dag}]&=\delta_{ab}\delta_{ij}. 
\end{align}
Given the canonical commutation relations (\ref{qmxcom1}), 
the quantization of the matrix Chern-Simons model (\ref{qmx1}) is performed in 
\cite{Polychronakos:2001mi, Dorey:2016mxm} 
by introducing a reference state $|0\rangle$ that obeys 
\begin{align}
\label{qmxcom2}
Z_{ab}|0\rangle&=\phi_{i}|0\rangle=0, 
\end{align}
and by acting on $|0\rangle$ with $Z^{\dag}$ and $\varphi_{i}^{\dag}$. 
When requiring that all physical states satisfy the Gauss law constraints
there are operator ordering ambiguities. 
They are fixed as 
\begin{align}
\label{qmxgauss1}
:[Z,Z^{\dag}]:+\sum_{i=1}^{p}\phi_{i}\phi_{i}^{\dag}
&=\kappa\mathbb{I}_{N}. 
\end{align}
Combining the commutation relations (\ref{qmxcom1}) and 
the trace part of the constraint (\ref{qmxgauss1}), 
we find that 
\begin{align}
\label{qmxgauss2}
\sum_{a=1}^{N}\sum_{i=1}^{p}\phi_{ia}^{\dag}\phi_{ai}&=(\kappa -p)N. 
\end{align}
This means that 
all physical states have charge $(\kappa -p)$ under the $U(1)\subset U(N)$. 
Alternatively it demands that 
all physical states involve $(\kappa -p )N$ copies of $\phi^{\dag}$. 
Meanwhile the traceless part of the constraint (\ref{qmxgauss1}) 
demands that they are $SU(N)$ singlets. 

However, in order to perform the quantization 
for the canonical commutation relations (\ref{sgqmcom1}), 
one needs to determine which operators are realized by multiplication 
and which operators by differentiation on the quantum states. 
This prescription is called the polarization, 
which leads to the division of the phase space into coordinates and momenta. 
Here we encounter the issue on quantization due to the non-trivial polarization. 
There exist two polarizations which include 
the proposed quantization (\ref{qmxcom2}) in the ordinary matrix Chern-Simons model \cite{Polychronakos:2001mi, Dorey:2016mxm}.

\begin{enumerate}

\item Holomorphic polarization

Let us choose a polarization by introducing a reference state $|0\rangle$ that obeys
\begin{align}
\label{sgqmcom2}
\widehat{Z}_{AB}|0\rangle&=\widehat{\Phi}_{AI}|0\rangle=0
\end{align}
and construct the Hilbert space
by acting on $|0\rangle$ with $\widehat{Z}^{\dag}_{AB}$ and $\widehat{\Phi}_{IA}^{\dag}$.

Due to the minus signs for the $\widetilde{\phi}$, $\psi$, $\widetilde{Z}$ and $A$ quantization conditions,
we define the following number operators to count the number of each type of
creation operator acting on the reference state: 
\begin{align}
\label{sgqmNumOps}
N_Z & = \Tr(Z^{\dagger}Z) , & N_{\widetilde{Z}} = & - \Tr(\widetilde{Z}^{\dagger}\widetilde{Z}) , \\
N_A & = - \Tr(A^{\dagger}A) , & N_B = & \Tr(B^{\dagger}B) , \\
N_{\phi} & = \Tr(\phi^{\dagger}\phi) , & N_{\widetilde{\phi}} = & -\Tr(\widetilde{\phi}^{\dagger}\widetilde{\phi}) , \\
N_{\psi} & = -\Tr(\psi^{\dagger}\psi) , & N_{\widetilde{\psi}} = & \Tr(\widetilde{\psi}^{\dagger}\widetilde{\psi}) .
\end{align}
We can also define total number operators
\begin{align}
\label{sgqmNumOpsS}
N_{\widehat{Z}} & = N_Z + N_B + N_{\widetilde{Z}} + N_A = \Str \left(\widehat{Z}^{\dagger} \widehat{Z} \right), \\
N_{\widehat{\Phi}} & = N_{\phi} + N_{\widetilde{\phi}} + N_{\psi} + N_{\widetilde{\psi}} = (-1)^I \widehat{\Phi}^{\dagger}_{IA} \widehat{\Phi}_{AI}. 
\end{align}

Then the Hamiltonian is
\begin{align}
H & = \omega
\Tr
\left(
Z^{\dag}Z+B^{\dag}B-\widetilde{Z}^{\dag}\widetilde{Z} - A^{\dag}A
\right) \\
 & \equiv \omega \left( N_Z + N_B + N_{\tilde{Z}} + N_A \right) \equiv \omega N_{\widehat{Z}}
\label{Ham}
\end{align}
which is manifestly non-negative.

The system is a free set of bosonic and fermionic oscillators, subject to the
Gauss law constraints on physical states. From the form of the Hamiltonian,
we see that the ground state is the physical state with the minimal total
$\widehat{Z}$ number, irrespective of contributions from $\widehat{\Phi}$.

To analyze the Gauss law constraints, we can just replace the expressions
(\ref{gaussbb})-(\ref{gaussbf}) with the corresponding operators expressions.
There is the question of normal ordering. This is only relevant for the diagonal
constraints, i.e.\ the diagonal parts of (\ref{gaussbb}) and (\ref{gaussff}).
Following \cite{Dorey:2016hoj} we can normal order the terms coming from
$\widehat{Z}$. Choosing to do this or not to do this is equivalent to shifting
the value of $\kappa$ by $(N-M)$. E.g. taking the trace of (\ref{gaussbb})
without normal ordering gives the constraint
\begin{align}
N_{\phi} + N_{\psi} + N_A - N_B = N(\kappa - N + M - p + q)
\end{align}
whereas if we had normal ordered $\widehat{Z}$ terms we would get
\begin{align}
N_{\phi} + N_{\psi} + N_A - N_B = N(\kappa - p + q). 
\end{align}
The latter expression is more convenient when taking a large $N$, $M$ limit.
Similarly, we could normal order the terms from $\widehat{\Phi}$ but this would
just result in a shift of $\kappa$ by $(p-q)$ and we are considering those to
be fixed in the large $N$, $M$ limit. These possibilities can all be encoded in
a relation between $\kappa$ appearing the the action and $k$ defined so that
in the quantum Gauss law constraints (\ref{gaussbb}) and (\ref{gaussff}) we
just take (\ref{gaussbb}) and (\ref{gaussff}) to be completely
normal ordered and replace $\kappa$ with $k$.

Now, taking the trace of (\ref{gaussbb}) gives
\begin{align}
\label{sgauss1a}
N_{\phi} + N_{\psi} + N_A - N_B = Nk
\end{align}
while taking the trace of (\ref{gaussff}) gives
\begin{align}
\label{sgauss1b}
 - N_{\widetilde{\phi}} - N_{\widetilde{\psi}} + N_A - N_B = Mk .
\end{align}
Taking the difference of these equations gives
\begin{align}
\label{sgqmNumPhi}
N_{\widehat{\Phi}} \equiv N_{\phi} + N_{\psi} + N_{\widetilde{\phi}} + N_{\widetilde{\psi}} = (N-M)k
\end{align}
which is the supergroup analogue of (\ref{qmxgauss2}).

\item Super polarization

Taking account into the superbracket 
\begin{align}
\widehat{Z}_{AB}\widehat{Z}_{CD}^{\dag}-(-1)^{(A+B)(C+D)}
\widehat{Z}_{CD}^{\dag}\widehat{Z}_{AB}&=(-1)^{B}\delta_{AD}\delta_{BC}\nonumber\\
&=(-1)^{C}\delta_{AD}\delta_{BC}
\end{align}
and identifying the annihilation operator $\widehat{Z}_{AB}$ or 
$\widehat{Z}^{\dag}_{CD}$ as we have $[a,a^{\dag}]=1$ or $\left\{a,a^{\dag}\right\}=1$, 
we can consider the reference state defined by
\begin{align}
(-1)^{B+1}\widehat{Z}_{AB}|0\rangle&=\widehat{Z}_{AB}|0\rangle, \nonumber \\
(-1)^{C}\widehat{Z}_{CD}^{\dag}|0\rangle&=\widehat{Z}_{CD}^{\dag}|0\rangle. 
\end{align}
Similarly we could impose
\begin{align}
(-1)^{I+1}\widehat{\Phi}_{AI}|0\rangle&=\widehat{\Phi}_{AI}|0\rangle, \nonumber \\
(-1)^{I}\widehat{\Phi}_{IA}^{\dag}|0\rangle&=\widehat{\Phi}_{IA}^{\dag}|0\rangle.  
\end{align}
We call this procedure of quantization the super polarization. 
We will leave more detailed investigation of this super polarization to future work.

\end{enumerate}

\subsection{Quantum ground states}
\label{subsecGS}

We will first review the quantum ground states for the models without supergroup symmetries. 
These states can be constructed as (a power of) a determinant of a matrix of operators acting on the reference state. 
This motivates similar constructions in the more general supergroup case, 
generally involving superdeterminants. 
However, there are different candidate states. 
The states constructed all solve the Gauss Law constraints, but we do not have a proof 
that there are no physical states with lower energy. In addition, one 
complication is that in some cases the construction may give the zero state due to the 
possibility of constructing nilpotent operators in the supergroup case. 
This results in the possibility that a construction may produce the ground state for low enough values of $k$, 
but for larger $k$ will simply produce the zero state.

\subsubsection{Determinant states}
\label{subsec2MqzeroQGS}
In the analysis of the ordinary matrix Chern-Simons theory (\ref{qmx1}), 
which is regarded as our model for $M=0$ and $q=0$, 
the quantum physical states are constructed \cite{Polychronakos:2001mi, Dorey:2016mxm}. 
Let us firstly review the construction. 
For $p=1$ the ground state can be constructed by acting with $kN$ copies of $\phi^{\dag}$ 
while keeping the number of $Z^{\dag}$ to a minimum. 
In addition, the ground state should be the $SU(N)$ singlet. 
Defining a baryon operator by
\begin{align}
\label{baryon1}
\mathcal{B}&\equiv
\epsilon^{a_{1}\cdots a_{N}}
\left(Z^{l_{1}}\phi\right)_{a_{1}}^{\dag}
\cdots \left(Z^{l_{N}}\phi\right)_{a_{N}}^{\dag}
\end{align}
where all the exponents 
$l_{a}$ are distinct because of the antisymmetrization factor 
$\epsilon^{a_{1}\cdots a_{N}}$, 
the baryon generator with the lowest energy 
\begin{align}
\label{baryon2}
\mathcal{B}_{\min}&\equiv 
\epsilon^{a_{1}\cdots a_{N}}
\left(Z^{0}\phi\right)_{a_{1}}^{\dag}
\left(Z^{1}\phi\right)_{a_{2}}^{\dag}
\cdots \left(Z^{N-1}\phi\right)_{a_{N}}^{\dag}
\end{align}
gives the ground state as $k$ multiple $\mathcal{B}_{\mathrm{min}}$'s \cite{Polychronakos:2001mi, Hellerman:2001rj}
\begin{align}
\label{ground1}
|\mathrm{ground}\rangle_{k}&=
\left[
\epsilon^{a_{1}\cdots a_{N}}
\left(Z^{0}\phi\right)_{a_{1}}^{\dag}
\left(Z^{1}\phi\right)_{a_{2}}^{\dag}
\cdots \left(Z^{N-1}\phi\right)_{a_{N}}^{\dag}
\right]^{k}|0\rangle,
\end{align}
which carries $kN$ copies of $\phi^{\dag}$ and $k$ charges of the $U(1)\subset U(N)$. 
Note that the baryon generator (\ref{baryon2}) with the lowest energy 
is in one to one correspondence with the Vandermonde determinant
\begin{align}
\label{vander_det}
\Delta&=\epsilon^{a_{1}\cdots a_{N}}
z_{a_{1}}^{0}\cdots z_{a_{N}}^{N-1}
=\prod_{a<b}(z_{a}-z_{b}). 
\end{align}

When $N$ is divisible by $p\ge 2$, the ground state is also uniquely determined. 
One can build up the $SU(p)$ singlet by collecting $p$ creation operators $\phi^{\dag}_{i}$ 
into the baryon operator 
\begin{align}
\label{baryonp2a}
\mathcal{B}(r)_{a_{1}\cdots a_{p}}^{\dag}
\equiv
\epsilon^{i_{1}\cdots i_{p}}
\left(Z^{r}\phi\right)_{i_{1}a_{1}}^{\dag}
\cdots \left(Z^{r}\phi\right)_{i_{p}a_{p}}^{\dag}.
\end{align}
This is a singlet under the $SU(p)$ transforming 
as the $p$-th antisymmetric representation 
of the $U(N)$ gauge symmetry group. 
To construct the $SU(N)$ singlet with $kN$ charge of 
the $U(1)\subset U(N)$, 
we furthermore collect $N/p$ baryon operators 
$\mathcal{B}(0)_{a_{1}\cdots a_{p}}^{\dag}$, 
$\mathcal{B}(1)_{a_{p+1}\cdots a_{2p}}^{\dag}$, 
$\cdots$, 
$\mathcal{B}(\frac{N}{p}-1)^{\dag}_{a_{N-p+1}\cdots a_{N}}$ 
by introducing a baryon of baryons
\begin{align}
\label{baryon2p1}
\mathcal{B}_{\mathrm{min}}\equiv 
\epsilon^{a_{1}\cdots a_{N}}
\mathcal{B}(0)^{\dag}_{a_{1}\cdots a_{p}}
\mathcal{B}(1)^{\dag}_{a_{p+1}\cdots a_{2p}}
\cdots 
\mathcal{B}\left(\frac{N}{p}-1\right)^{\dag}_{a_{N-p+1}\cdots a_{N}}
\end{align}
and find the ground state \cite{Dorey:2016mxm}
\begin{align}
\label{ground3}
|\mathrm{ground}\rangle_{k}&=
\Biggl[
\epsilon^{a_{1}\cdots a_{N}}
\mathcal{B}(0)^{\dag}_{a_{1}\cdots a_{p}}
\mathcal{B}(1)^{\dag}_{a_{p+1}\cdots a_{2p}}
\cdots 
\mathcal{B}\left(\frac{N}{p}-1\right)^{\dag}_{a_{N-p+1}\cdots a_{N}}
\Biggr]^{k}
|0\rangle 
\end{align}
whose energy is 
\begin{align}
\label{o_g_energy1}
E&=\omega k p\sum_{r=0}^{\frac{N}{p}-1}r 
=\omega k\frac{N(N-p)}{2p}. 
\end{align}

Note that the ground state can also be written in the form
\begin{align}
\label{ground3_det1}
|\mathrm{ground}\rangle_{k}&=
\left( \Det\left(S_{Ia}\right) \right)^k |0\rangle
\end{align}
where the elements of the $N \times N$ matrix are
\begin{align}
\label{ground3_det2}
S_{Ia} & = \left[ \phi^{\dagger}(Z^{\dagger})^r \right]_{ia}
\end{align}
with the notation
\begin{align}
\label{ground3_det3}
I = rp + i ,& \;   i \in \left\{1, 2, \ldots , p\right\} , \; r \in \left\{0, 1, \ldots , \frac{N}{p} - 1 \right\} .
\end{align}

When $N$ is not divisible by $p\ge 2$, 
there is no $SU(p)$ singlet ground state.
Let us express $N=mp+n$, $m, n\in \mathbb{Z}_{>0}$. 
Then the ground state is constructed as 
\begin{align}
\label{ground4}
|\mathrm{ground}\rangle_{k}&=
\prod_{l=1}^{k}
\Biggl[
\epsilon^{a_{1}\cdots a_{N}}
\mathcal{B}(0)_{a_{1}\cdots a_{p}}^{\dag}
\mathcal{B}(1)_{a_{p+1}\cdots a_{2p}}^{\dag}\cdots 
\mathcal{B}(m-1)_{a_{N-p-n+1}\cdots a_{N-n}}^{\dag}\nonumber\\
&\left(
Z^{m}\phi_{i_{(l,1)}}
\right)^{\dag}_{a_{N-n+1}}\cdots \left(
Z^{m}\phi_{i_{(l,n)}}
\right)^{\dag}_{a_{N}}
\Biggr]
|0\rangle,
\end{align}
where the indices $i_{(l,\alpha)}$ with $l=1,\cdots, k$ $\alpha=1,\cdots, n$ 
label the degenerate ground states. 
The ground state energy is 
\begin{align}
\label{G_energy_Nmpq}
E&=\omega k \left(
p\sum_{r=0}^{m-1}r+nm
\right)
=\omega k \left[
\frac{pm(m-1)}{2}+mn
\right]. 
\end{align}

\subsubsection{Superdeterminant states -- Case 1}
\label{subsec_sdet1}
Now we look for solutions to the Gauss law constraints, allowing non-zero $M$ and $q$. As reviewed in
section~\ref{subsec2MqzeroQGS} in the case of ordinary Lie groups, the ground state
was given by a determinant acting on the reference state. As may be expected
the generalization to supergroups requires the use of a superdeterminant.
Necessarily this is a rather formal expression since superdeterminants are
not polynomial functions of the matrix elements. However, for now we simply
show that at a formal level this gives a solution of the Gauss law constraints.
Explicitly, we conjecture a potential class of ground states given by
\begin{align}
\label{SG1ground}
\ket{SG gs1} & = S^k \ket{0} .
\end{align}
Here we have defined
\begin{align}
\label{SDetGS}
S & \equiv \Sdet (S_{\mathcal{I}A})
\end{align}
where the elements of the $(N+M) \times (N+M)$ matrix are defined as follows:
\begin{align}
\label{S_IA}
S_{IA} & = \left[ \widehat{\Phi}^{\dag}
\left(\widehat{Z}^{\dagger}\right)^r \right]_{iA} , \\
S_{\Lambda A} & = \left[ \widehat{\Phi}^{\dag}
\left(\widehat{Z}^{\dagger}\right)^r \right]_{\lambda A} .
\end{align}
In these expressions $I$ label the $N$ even components while $\Lambda$
label the $M$ odd components indexed by $\mathcal{I}$. Although not
explicitly labelled as such, the exponents $r$, and indices $i$ and $\lambda$
in each expression are determined by $I$ or $\Lambda$. Specifically,
(taking for now the simplest case where $N$ is a multiple of $p$ and $M$ is a
multiple of $q$) we have
\begin{align}
\label{indexSdet}
I = rp + i , \;  & i \in \{1, 2, \ldots , p\} , \; r \in \left\{0, 1, \ldots , \frac{N}{p} - 1 \right\}, \\
\Lambda = rq + \lambda , \; & \lambda \in \{1, 2, \ldots , q\} , \; r \in \left\{0, 1, \ldots , \frac{M}{q} - 1 \right\}. 
\end{align}

The normal ordered Gauss law constraints are given by
\begin{align}
\label{SGGauss}
\left( \widehat{G}_{AB} - k \delta_{AB} \right) \ket{\textrm{phys}} & = 0
\end{align}
where 
\begin{align}
\widehat{G}_{AB} & = (-1)^{(A+C)(C+B)}\widehat{Z}^{\dagger}_{CB} \widehat{Z}_{AC} - \widehat{Z}^{\dagger}_{AC} \widehat{Z}_{CB} + (-1)^{(I+A)(I+B)} \widehat{\Phi}^{\dagger}_{IB} \widehat{\Phi}_{AI}
\end{align}
is the quantum Gauss law constraint operator.

It is then straightforward to check that the state $\ket{SGgs1}$ defined in
equation~(\ref{SG1ground}) is indeed a physical state. One method is to note
that superdeterminants can be written as ratios of ordinary determinants, and
then the commutation relations can be used to find the commutators of
$\widehat{Z}_{AB}$ and $\widehat{\Phi}_{AI}$ with $S$. E.g.\ using the standard
even/odd split form of a supermatrix we have
\begin{align}
\label{sdet_1}
\Sdet\left( \begin{array}{cc} A & B \\ C & D \end{array} \right) = \det(A - BD^{-1}C) (\det(D))^{-1} .
\end{align}
Note that while this is invariant under $U(N|M)$ transformations, 
it is not invariant under $SU(p|q)$ transformations except in the case where 
$N/p = M/q$, since only then are all $p+q$ components 
$\left[ \widehat{\Phi}^{\dag} \left(\widehat{Z}^{\dagger}\right)^r \right]_{iA}$
and 
$\left[ \widehat{\Phi}^{\dag} \left(\widehat{Z}^{\dagger}\right)^r \right]_{\lambda A}$
(for each $A$) included in the superdeterminant expression for all values of $r$. 

A short calculation shows that (relative to the reference state $\ket{0}$)
the potential ground state has energy
\begin{align}
\label{SGgs1Energy}
E_{SGgs1} & = \omega k \left( \frac{N(N-p)}{2p} - \frac{M(M-q)}{2q} \right). 
\end{align}
This reproduces the ground state energy (\ref{o_g_energy1}) for
$M=q=0$. We expect that this gives the ground state energy in some more general
cases with non-zero $M$ and $q$. In the case where $N$ is not a multiple of $p$
or $M$ is not a multiple of $q$, a superdeterminant generalisation of
(\ref{ground4}) will lead to an expression for the potential ground state energy generalising (\ref{G_energy_Nmpq}). However, there will be cases where different
constructions give physical states with lower energy, particularly for small
values of $k$. We consider explicitly the case of $M=0$ but $q \ne 0$,
in which case, at least for $k=1$, the construction in this section does not give the ground state energy. We now consider this possibility and later
generalize that construction to $M \ne 0$.

\subsubsection{Analysis for $U(N)$}
\label{subsec2MzeroQGS}

In the case where $M=0$ but $q \ne 0$, we note that the previous construction could be used.
However, the state would not involve any of the fermionic creation operators
$\psi^{\dagger}_{\lambda a}$. This motivates us to consider
another possibility to construct the candidate ground state from the reference state.
To do this, we consider the candidate ground state 

\begin{align}
\label{Mzground}
\ket{GS_{M=0}} & \equiv S^k \ket{0} .
\end{align}
We have defined
\begin{align}
\label{MzGS}
S & \equiv \sqrt{\Det ((S^T S)_{ab})}
\end{align}
where
\begin{align}
\label{MzStS}
(S^T S)_{ab} & = S^T_{a\mathcal{I}} S_{\mathcal{I}b} \equiv S_{\mathcal{I}a} S_{\mathcal{I}b}
\end{align}
and the elements of the $N \times N$ matrix $S_{\mathcal{I}a}$ are:
\begin{align}
\label{S_Ia}
S_{Ia} & = \left[ \widehat{\Phi}^{\dag}
\left(Z^{\dagger}\right)^r \right]_{ia} , \\
S_{\Lambda a} & = \left[ \widehat{\Phi}^{\dag}
\left(Z^{\dagger}\right)^r \right]_{\lambda a} .
\end{align}

Clearly $S^k = (S^2)^{k/2}$ is polynomial for even $k$, while for odd $k$ we
do need to consider the square root in (\ref{MzGS}).
In these expressions $I$ label even components while $\Lambda$
label odd components indexed by $\mathcal{I}$. Although not
explicitly labelled as such, the exponents $r$, and indices $i$ and $\lambda$
in each expression are determined by $I$ or $\Lambda$. Specifically,
taking for now the simplest case where $N$ is a multiple of $p+q$, we have
\begin{align}
\label{indexMzDet}
I = rp + i , \;  & i \in \left\{1, 2, \ldots , p\right\} , \; r \in \left\{0, 1, \ldots , \frac{N}{p+q} - 1 \right\} , \\
\Lambda = rq + \lambda , \; & \lambda \in \left\{1, 2, \ldots , q\right\} , \; r \in \left\{0, 1, \ldots , \frac{N}{p+q} - 1 \right\} .
\end{align}
Note that for $q=0$ this again reproduces the previous determinant construction of the ground state. 

The normal ordered Gauss law constraints are given by
\begin{align}
\label{MzGauss}
\left( G_{ab} - k \delta_{ab} \right) \ket{\textrm{phys}} & = 0
\end{align}
where
\begin{align}
G_{ab} & = Z^{\dagger}_{cb} Z_{ac} - Z^{\dagger}_{ac} Z_{cb} + (-1)^I \widehat{\Phi}^{\dagger}_{Ib} \widehat{\Phi}_{aI}
\end{align}
is the quantum Gauss law constraint operator.

It is then straightforward to check that the state $\ket{GS_{M=0}}$ defined in
equation~(\ref{Mzground}) is indeed a physical state.
Another short calculation shows that (relative to the reference state $\ket{0}$)
the potential ground state has energy
\begin{align}
\label{MzGSEnergy}
E_{GS, M=0} & = \omega k \frac{N(N-p-q)}{2(p+q)} .
\end{align}

Note that the above claim for the potential ground state energy is made on the assumption that $\ket{GS_{M=0}}$ is a normalizable state. In fact this state will vanish for sufficiently large $k$. To see this simply note that if we set all the fermions to zero the matrix $S_{\mathcal{I}a}$ would have $qN/(p+q)$ rows of zeros, so
consequently $S = 0$. Reintroducing the fermions we see that all terms in $S^2$
must involve a product of at least $qN/(p+q)$ fermions. Since we only have
$qN$ independent fermionic components of $\widehat{\Phi}_{a\lambda}$ we see that certainly $S^k = 0$ for all $k > p+q$. In fact, we may have $S^k = 0$ for
lower values of $k$. It is not clear what the interpretation of this bound is
but it does indicate a richer structure appears for the supergroup models. 
The simplest case to explore the issue of the constructions producing the zero
state is $N=2$ and $p=q=1$.

We find a ground state for $k=1$ described by
\begin{align}
\label{211gs1}
S & \equiv \Det\left( \begin{array}{cc} \phi^{\dagger}_{1} & \phi^{\dagger}_{2} \\   \psi^{\dagger}_{1} & \psi^{\dagger}_{2} \end{array} \right) =
 \phi^{\dagger}_{1}\psi^{\dagger}_{2} - \phi^{\dagger}_{2}\psi^{\dagger}_{1}
\end{align}
which is clearly generically non-vanishing. We have dropped the redundant labels $i=1$ and $\lambda =1$ in this example.

However, it is easy to check that
$S^2 = 0$ so we have only found the ground state for $k=1$ and it indeed has
energy zero as expected. Now, in this example we can try to explicitly construct
the ground state for $k=2$. However, the result shown below is that no such state exists,
i.e.\ that for $k=2$ there is no extremal energy eigenstate arising as a
polynomial of creation operators acting on the reference state.

First note that the Gauss law constraints can be expressed as conditions on $S$
\begin{align}
k\delta_{ab}S = [G_{ab}, S] \equiv \left( Z^{\dagger}_{cb}\frac{\partial}{\partial Z^{\dagger}_{ca}} - Z^{\dagger}_{ac}\frac{\partial}{\partial Z^{\dagger}_{bc}} + \phi^{\dagger}_{b}\frac{\partial}{\partial \phi^{\dagger}_{a}} + \psi^{\dagger}_{b}\frac{\partial}{\partial \psi^{\dagger}_{a}} \right)S .
\end{align}

It is a simple task to check that there is no zero energy state for $k \ge 2$,
at least assuming it is constructed as a polynomial of the creation operators acting on the reference state. In this case the requirement of zero energy is
simply that no $Z^{\dagger}$ operators are used to create the state. Then the
two Gauss law constraints 
\begin{align}
[G_{11}, S] = [G_{22}, S] & = kS
\end{align}
impose the constraint that $S$ must have the form
\begin{align}
S & = (\phi^{\dagger}_1)^{k-1} (\phi^{\dagger}_2)^{k-1} \left( c_{00} \phi^{\dagger}_1\phi^{\dagger}_2 + c_{01} \psi^{\dagger}_1\phi^{\dagger}_2 + c_{10} \phi^{\dagger}_1\psi^{\dagger}_2 + c_{11} \psi^{\dagger}_1\psi^{\dagger}_2 \right) .
\end{align}
Now we still have to impose the constraints
\begin{align}
[G_{12}, S] = [G_{21}, S] & = 0
\end{align}
but clearly this means that $G_{12}$ and $G_{21}$ must annihilate the term with
coefficient $c_{00}$, and separately the term with coefficient $c_{11}$. This
is only possible if $c_{00}=0$ and (except for $k=1$) $c_{11}=0$. Then we are left with
\begin{align}
[G_{12}, S] & = c_{01} (k-1) (\phi^{\dagger}_1)^{k-2} (\phi^{\dagger}_2)^{k+1} \psi^{\dagger}_1 + (c_{01} + k c_{10})(\phi^{\dagger}_1)^{k-1} (\phi^{\dagger}_2)^{k} \psi^{\dagger}_2 , \\
[G_{21}, S] & = c_{10} (k-1) (\phi^{\dagger}_1)^{k+1} (\phi^{\dagger}_2)^{k-2} \psi^{\dagger}_2 + (c_{10} + k c_{01})(\phi^{\dagger}_1)^{k} (\phi^{\dagger}_2)^{k-1} \psi^{\dagger}_1 .
\end{align}
These equations have no solution unless $S=0$ or $k=1$.

We can relax the condition that the energy vanishes, but explicit calculation
shows that there is still no solution for the case $k=2$.
Of course, the superdeterminant construction of the previous section provides a formal solution, but not polynomial 
in the creation operators acting on the reference state.

\subsubsection{Superdeterminant states -- Case 2}
\label{subsec_sdet2}
The previous considerations for $M=0$ lead to an alternative proposal for the
ground states.
We can define an $(N+M) \times (N+M)$ matrix with elements
\begin{align}
\label{S_IA_SG2}
S_{IA} & = \left[ \widehat{\Phi}^{\dag}
\left(Z^{\dagger}\right)^r \right]_{iA} , \\
S_{\Lambda A} & = \left[ \widehat{\Phi}^{\dag}
\left(Z^{\dagger}\right)^r \right]_{\lambda A} .
\end{align}
Here the simplest construction is when $M+N$ is a multiple of
$p+q$, in which case
\begin{align}
\label{indexSG2SDet}
I = rp + i , \;  & i \in \left\{1, 2, \ldots , p\right\} , \; r \in \left\{0, 1, \ldots , \frac{N+M}{p+q} - 1 \right\} , \\
\Lambda = rq + \lambda , \; & \lambda \in \left\{1, 2, \ldots , q\right\} , \; r \in \left\{0, 1, \ldots , \frac{N+M}{p+q} - 1 \right\} .
\end{align}

Then we propose the potential ground states
\begin{align}
\label{SG2ground}
\ket{SGgs2} & \equiv S^k \ket{0}
\end{align}
where
\begin{align}
\label{SG2GS}
S & \equiv \sqrt{\Sdet ((S^T S)_{ab})}
\end{align}
and
\begin{align}
\label{SG2StS}
(S^T S)_{AB} & = S^T_{A\mathcal{I}} S_{\mathcal{I}B} \equiv S_{\mathcal{I}A} S_{\mathcal{I}B}
\end{align}

It is then straightforward to check that the state $\ket{SGgs2}$ defined in
equation~(\ref{SG2ground}) is indeed a physical state and
that (relative to the reference state $\ket{0}$) it
has energy
\begin{align}
\label{SG2GSEnergy}
E_{SGgs2} & = \omega k \frac{(N-M)(N+M-p-q)}{2(p+q)} .
\end{align}

Note that this state is by construction the same as the state $\ket{GS_{M=0}}$
defined in equation~(\ref{Mzground}) in the case $M=0$, and it is also exactly the same as the state $\ket{SGgs1}$ defined in
equation~(\ref{SG1ground}) in the case where $N/p = M/q$. However, in
general the states $\ket{SGgs1}$ and $\ket{SGgs2}$ are different, and it seems
likely that the states $\ket{SGgs2}$ are the better candidate ground states.
One particular feature the states $\ket{SGgs2}$ have is that (when $N+M$ is
a multiple of $p+q$) they respect the $SU(p|q)$ symmetry.

Finally, we note that when both superdeterminant constructions are compared
(assuming $N/p$, $M/q$ and $(N+M)/(p+q)$ are all integer) the difference in
energies is
\begin{equation}
E_{SGgs1} - E_{SGgs2} = \frac{\omega k}{2} \left( q^2N^2 - p^2M^2 \right)
\end{equation}
which can be positive, negative or zero.

In terms of a potential relation to WZW models, as demonstrated for
$M=0$ and $q=0$ \cite{Dorey:2016hoj} we must consider a generalization of the
large-$N$ limit with fixed $p$ and now also fixed $q$. Two natural choices
are to take $M=0$, or to scale $N$ and $M$ in the same ratio as $p:q$. In
the $M=0$ case we believe $\ket{SGgs2}$ is the ground state, while in the
other case $\ket{SGgs1} = \ket{SGgs2}$.


\section{Kac-Moody algebra}
\label{seckc}
When $M=0$, $q=0$, it was demonstrated \cite{Dorey:2016hoj} that
the matrix degrees of freedom lead to the affine Lie algebra $\widehat{\mathfrak{su}}(p)$ in the large $N$ limit.
We will firstly review the argument of \cite{Dorey:2016hoj}.
Then we conjecture that the result generalizes to $q \ne 0$, leading to a
$\widehat{\mathfrak{su}}(p|q)$ current algebra. We show this in the case $M=0$
but expect it also holds in a large $N$ and large $M$ limit.

\subsection{Affine Lie algebra}
In the case of $M=0$, $q=0$, i.e.\ $U(N)$ and $SU(p)$ symmetry, in
\cite{Dorey:2016hoj} the current operators were defined as
\begin{align}
\label{j1a}
\widetilde{J}^{m}&=\Tr Z^{m},\\
\label{j1b}
\widetilde{J}_{ij}^{m}
&=i\left(
\phi_{i}^{\dag}Z^{m}\phi_{j}
-\frac{1}{p}\delta_{ij}\phi_{k}^{\dag}Z^{m}\phi_{k}
\right)
\end{align}
with $i,j,k=1,\cdots,p$ and $m\ge0$. 

It is then straightforward to show that for $m$ and $n$ either both non-negative or both non-positive
\begin{align}
\label{j1c}
\left[
\widetilde{J}^{m}_{ij},\widetilde{J}_{kl}^{n}
\right]&=i\left(
\delta_{il}\widetilde{J}_{kj}^{m+n}-\delta_{kj}\widetilde{J}_{il}^{m+n}
\right) .
\end{align}

For $m<0$ the negative graded currents are defined by
\begin{align}
\label{j1d}
\widetilde{J}^{m}_{ij}=\widetilde{J}_{ji}^{|m|\dag}.
\end{align}
Now the commutator of a current at positive level with one at negative level is
more involved as it requires evaluating commutators of powers of $Z$ with
powers of $Z^{\dagger}$. In particular, for non-negative $m$ and $n$
\begin{align}
\label{km1a}
\left[
\widetilde{J}_{ij}^{m},
\widetilde{J}_{kl}^{-n}
\right]
&=
\left[
\phi_{i}^{\dag}Z^{m}\phi_{j}, 
\phi_{k}^{\dag}Z^{\dag n}\phi_{l}
\right]\nonumber\\
&=\delta_{jk}\phi_{i}^{\dag}
Z^{m}Z^{\dag n}\phi_{l}
-\delta_{il}\phi_{k}^{\dag n}Z^{m}\phi_{j}+\phi_{ia}^{\dag}\phi_{kb}
\left[
Z_{ac}^{m},Z_{bd}^{\dag n}
\right]\phi_{jc}\phi_{ld}. 
\end{align}

Additionally the current algebra is expected only when
acting on physical states, so the Gauss law constraint must be imposed. In
fact, the $\widehat{\mathfrak{su}}(p)$ current algebra is only correctly
reproduced in a large $N$ limit.

The combination of imposing the Gauss law constraints and the large $N$ limit
is carried out using knowledge of the classical and quantum ground states.
Specifically, results such as $Z^{\dagger} \phi_i = 0$ for classical ground states
are taken to imply that this relation holds for all physical states (at least
those with sufficiently low energy) to leading order in a $1/N$ expansion.
Using such considerations it is possible to identify the leading large $N$ behavior of various terms and retaining only the leading non-vanishing order in
expressions greatly simplifies the results. The discussion of classical
solutions is applicable to calculations of Poisson brackets, but this is expected
to carry over to quantum commutation relations. We refer the reader to
\cite{Dorey:2016hoj} for more details, although we also make some more detailed
comparisons when generalizing the results to $q \ne 0$.

%
%
%
%

The result is that after a rescaling\footnote{Note that this rescaling is
trivial in (\ref{j1c}) as that equation is homogeneous provided $m$ and $n$
have the same sign.} of the currents
\begin{align}
\label{redefcurrent1a}
J^{m}_{ij}&=\left(
\frac{(\kappa)N}{p}
\right)^{-\frac{|m|}{2}}\widetilde{J}_{ij}^{m}
\end{align}
the Kac-Moody algebra is produced at leading large $N$ order
\begin{align}
\label{kacmoody1}
\left[
J_{ij}^{m},J^{n}_{kl}
\right]
\sim 
i\left(
\delta_{il}J_{kj}^{m+n}-\delta_{kj}J_{il}^{m+n}
\right)+ \kappa m\left(
\delta_{jk}\delta_{il}-\frac{1}{p}\delta_{ij}\delta_{kl}. 
\right) .
\end{align}
Actually, this is the classical Poisson bracket result, but it was argued
\cite{Dorey:2016hoj} to
hold also for quantum commutators up to the replacement $\kappa \rightarrow k$
in the central term.

\subsection{Affine Lie superalgebra}

We now consider the case with $q \ne 0$ where we expect to get an affine Lie
superalgebra.
The case where $M=0$ is simpler than $M \ne 0$ although the calculations are
similar, so we present this first and briefly comment on the large $M$ case in the next section.

We define the supertraceless (in the $IJ$ indices) currents
\begin{align}
\label{M0Currents}
J^m_{IJ} & \equiv \left( \widehat{\Phi}^{\dagger}_I Z^m \widehat{\Phi}_J - \frac{(-1)^M}{p-q} \delta_{IJ} \widehat{\Phi}^{\dagger}_M Z^m \widehat{\Phi}_M\right) , \\
J^{-m}_{IJ} & = \left( J^m_{JI} \right)^{\dagger}
\end{align}
for $m \ge 0$.
In defining the current, note that
it is the supertrace,
not the trace, which is invariant under $SU(p|q)$ transformations
$\widehat{\Phi}_I \rightarrow \widehat{\Phi}_K M_{KI}$.
Also, $\delta_{IJ}$ and not $\eta_{IJ} \equiv (-1)^I \delta_{IJ}$ is invariant
under the same transformation as
$\widehat{\Phi}^{\dagger}_I Z^m \widehat{\Phi}_J$.

We now calculate explicitly for $m \ge 0$ and $n \ge 0$, using the quantization
conditions. The superbracket of currents is easily calculated to give
\begin{align}
\left[ J^m_{IJ}, J^n_{KL} \right]_S & \equiv J^m_{IJ}J^n_{KL} - (-1)^{(I+J)(K+L)} J^n_{KL}J^m_{IJ} \\
 & = \eta_{JK}J^{m+n}_{IL} - (-1)^{(I+J)(K+L)} \eta_{IL}J^{m+n}_{KJ}
\end{align}

Now we calculate (up to terms ensuring supertracelessness in $IJ$ and in $KL$)
\begin{align}
\left[ J^m_{IJ}, J^{-n}_{KL} \right]_S = &
 (-1)^{JK} \widehat{\Phi}_{Ia}^{\dagger} \widehat{\Phi}_{Kb}^{\dagger} [(Z^m)_{ac}, (Z^{\dagger n})_{bd} ] \widehat{\Phi}_{Jc} \widehat{\Phi}_{Ld} \nonumber \\
 & + \eta_{JK} \widehat{\Phi}_I^{\dagger} \left( Z^{\dagger n}Z^m + [Z^m, Z^{\dagger n}] \right) \widehat{\Phi}_L
 - (-1)^{(I+J)(K+L)} \eta_{IL} \widehat{\Phi}_K^{\dagger}Z^{\dagger n}Z^m\widehat{\Phi}_J. 
\end{align}
which generalizes (2.7) in \cite{Dorey:2016hoj}.

We can analyze this similarly to \cite{Dorey:2016hoj}. E.g.\ \cite{Dorey:2016hoj} (2.10) onwards becomes
\begin{align}
\widehat{\Phi}_I^{\dagger} [Z^m, Z^{\dagger n}] \widehat{\Phi}_L \sim & - \sum_{r=0}^{m-1} \sum_{s=0}^{n-1} \left( \widehat{\Phi}_I^{\dagger} Z^r Z^{\dagger s} \widehat{\Phi}_M \right) \left( \widehat{\Phi}_M^{\dagger} Z^{\dagger n-1-s} Z^{m-1-r} \widehat{\Phi}_L \right) \nonumber \\
 & + \kappa n \sum_{r=0}^{m-1} \widehat{\Phi}_I^{\dagger} Z^r Z^{\dagger n-1} Z^{m-1-r} \widehat{\Phi}_L , \\
\widehat{\Phi}_{Ia}^{\dagger} \widehat{\Phi}_{Kb}^{\dagger} [(Z^m)_{ac}, (Z^{\dagger n})_{bd} ] \widehat{\Phi}_{Jc} \widehat{\Phi}_{Ld} \sim &
 \sum_{r=0}^{m-1} \sum_{s=0}^{n-1} (-1)^{(J+K)L} \left( \widehat{\Phi}_I^{\dagger} Z^r Z^{\dagger s} \widehat{\Phi}_L \right) \left( \widehat{\Phi}_K^{\dagger} Z^{\dagger n-1-s} Z^{m-1-r} \widehat{\Phi}_J \right) .
\end{align}

Now we introduce some notation to split $\widehat{\Phi}_I^{\dagger} Z^r Z^{\dagger s} \widehat{\Phi}_J$ and $\widehat{\Phi}_I^{\dagger} Z^{\dagger r} Z^s \widehat{\Phi}_J$ into their supertrace and supertraceless parts
\begin{align}
X^{r,s}_{IJ} \equiv & \widehat{\Phi}_I^{\dagger} Z^r Z^{\dagger s} \widehat{\Phi}_J - \delta_{IJ} X^{r,s} , \\
X^{r,s} \equiv & \frac{(-1)^M}{p-q} \widehat{\Phi}_M^{\dagger} Z^r Z^{\dagger s} \widehat{\Phi}_M , \\
Y^{r,s}_{IJ} \equiv & \widehat{\Phi}_I^{\dagger} Z^{\dagger r} Z^s \widehat{\Phi}_J - \delta_{IJ} Y^{r,s} , \\
Y^{r,s} \equiv & \frac{(-1)^M}{p-q} \widehat{\Phi}_M^{\dagger} Z^{\dagger r} Z^s \widehat{\Phi}_M
\end{align}
so that the terms above with four $\Phi$s can be written as
\begin{align}
\Theta \sim & \sum_{r=0}^{m-1} \sum_{s=0}^{n-1} \left(
 (-1)^{(J+K)L+JK} (X^{r,s}_{IL} + \delta_{IL} X^{r,s}) (Y^{n-1-s,m-1-r}_{KJ} + \delta_{KJ} Y^{n-1-s,m-1-r}) \right. \nonumber \\
 & \left. - \eta_{JK} (X^{r,s}_{IM} + \delta_{IM} X^{r,s}) (Y^{n-1-s,m-1-r}_{ML} + \delta_{ML} Y^{n-1-s,m-1-r})
 \right) \nonumber \\
 = & \sum_{r=0}^{m-1} \sum_{s=0}^{n-1} \left(
 (-1)^{(J+K)L+JK} \delta_{IL} X^{r,s} Y^{n-1-s,m-1-r}_{KJ}
 - \eta_{JK} X^{r,s} Y^{n-1-s,m-1-r}_{IL} \right. \nonumber \\
 & + \left. (-1)^{(J+K)L+JK}X^{r,s}_{IL} Y^{n-1-s,m-1-r}_{KJ} - \eta_{JK} X^{r,s}_{IM} Y^{n-1-s,m-1-r}_{ML}
 \right)
\end{align}
noting that all four terms containing $Y^{n-1-s,m-1-r}$ cancel. Now
we need to apply the large $N$ limit with generalizations of Identities
$1$ and $2$ in \cite{Dorey:2016hoj}.

We apply the large $N$ limit as in \cite{Dorey:2016hoj}. The classical solutions for $M=0$
are similar in nature to those of \cite{Dorey:2016hoj} where also $q=0$. In particular the
form of $Z$ is the same and $Z^{\dagger}\widehat{\Phi}_I = 0$ in the ground
state. We also assume this implies that $Z^{\dagger}\widehat{\Phi}_I$ is suppressed
compared to the naive expectations based on the order of $Z^{\dagger}$ and
$\widehat{\Phi}$ at large $N$. This means that the same results as in
\cite{Dorey:2016hoj} hold, up to possible signs. In
particular, to leading order in $N$
\begin{align}
\sum_{r=0}^{m-1} \widehat{\Phi}_I^{\dagger} Z^r Z^{\dagger n-1} Z^{m-1-r} \widehat{\Phi}_L \sim & \frac{1}{p-q} \delta_{IL} \sum_{r=0}^{m-1} (-1)^M \widehat{\Phi}_M^{\dagger} Z^r Z^{\dagger n-1} Z^{m-1-r} \widehat{\Phi}_M \nonumber \\
 \sim & \frac{(-1)^M}{p-q} \delta_{IL} \widehat{\Phi}_M^{\dagger} Z^{\dagger n-1} Z^{m-1} \widehat{\Phi}_M .
\end{align}
Collecting terms we find
\begin{align}
\left[ J^m_{IJ}, J^{-n}_{KL} \right]_S \sim & \sum_{r=0}^{m-1} \sum_{s=0}^{n-1} \left(
 (-1)^{(J+K)L + JK} \delta_{IL} X^{r,s} Y^{n-1-s,m-1-r}_{KJ}
 - \eta_{JK} X^{r,s} Y^{n-1-s,m-1-r}_{IL} \right. \nonumber \\
 & + (-1)^{(J+K)L+JK}X^{r,s}_{IL} Y^{n-1-s,m-1-r}_{KJ} - \eta_{JK} X^{r,s}_{IM} Y^{n-1-s,m-1-r}_{ML} \nonumber \\
 & \left. + \kappa n\eta_{JK}\delta_{IL} Y^{n-1, m-1}
+ \eta_{JK}Y^{n,m}_{IL} - (-1)^{(I+J)(K+L)}\eta_{IL}Y^{n,m}_{KJ} \right). 
\label{JJSM0}
\end{align}

The second line of (\ref{JJSM0}) is subleading at large $N$ compared to the first line
so we drop it from now on.
Also, in the above sums over $r$ and $s$, the dominant terms come from $r=s=0$
since $Z^{\dagger}\widehat{\Phi}_I$ and $\widehat{\Phi}^{\dagger}_I Z$ are
suppressed.

Noting that the Gauss law constraints are
\begin{align}
[Z, Z^{\dagger}] + \widehat{\Phi}_M \widehat{\Phi}_M^{\dagger} & = \kappa \mathcal{I}
\end{align}
a slight generalization of the derivation of Identity $1$ in Appendix~A of \cite{Dorey:2016hoj} (particularly (A.4) and the equation directly above it) gives
\begin{align}
Y^{n, n+m}_{IL} & \sim \left(\frac{\kappa N}{p-q}\right)^n Y^{0, m}_{IL} +
 \widehat{\Phi}_I^{\dagger}\widehat{\Phi}_J\widehat{\Phi}_J^{\dagger}Z^{\dagger n-1}Z^{m-1+n}\widehat{\Phi}_L \nonumber \\
 & \sim \left(\frac{\kappa N}{p-q}\right)^n Y^{0, m}_{IL} +
 \left(\frac{\kappa N}{p-q}\right) Y^{n-1, n+m-1}_{IL}
\end{align}
where the terms kept are those at leading order in $N$ and consistent with the
fact that the expression is traceless. This can be rewritten as
\begin{align}
Y^{n, n+m}_{IL} - X^{0,0} Y^{n-1, n+m-1}_{IL} & \sim \left(\frac{\kappa N}{p-q}\right)^n Y^{0, m}_{IL} .
\label{Id1M0}
\end{align}

Now the dominant terms in the first line of (\ref{JJSM0}) are from
the case $r=s=0$, and these combine with the last two terms in the final line
in exactly the combination in (\ref{Id1M0}) so we have in the case $m \ge n$
\begin{align}
\left[ J^m_{IJ}, J^{-n}_{KL} \right]_S \sim &
 \eta_{JK} \left(\frac{\kappa N}{p-q}\right)^n Y^{0, m-n}_{IL}
 - (-1)^{(I+J)(K+L)}\eta_{IL} \left(\frac{\kappa N}{p-q}\right)^n Y^{0, m-n}_{KJ} \nonumber \\
 & + \kappa n\eta_{JK}\delta_{IL} Y^{n-1, m-1} .
\end{align}

Now, rescaling the currents $J^m$ by a factor $(\kappa N/(p-q))^{-|m|/2}$ we
get
\begin{align}
\left[ J^m_{IJ}, J^{-n}_{KL} \right]_S \sim &
 \eta_{JK} J^{m-n}_{IL} - (-1)^{(I+J)(K+L)}\eta_{IL} J^{m-n}_{KJ}
 + \kappa n\eta_{JK}\delta_{IL} \delta_{mn}
\end{align}
using a generalization of Identity $2$ in \cite{Dorey:2016hoj} and noting that
terms from $Y^{n-1, n-1}$ with $n \ne m$ are suppressed and so do not contribute to this result for large $N$.

Finally, we correct this expression to ensure it is supertraceless in both $IJ$ and
$KL$ so we have
\begin{align}
\left[ J^m_{IJ}, J^{-n}_{KL} \right]_S \sim &
 \eta_{JK} J^{m-n}_{IL} - (-1)^{(I+J)(K+L)}\eta_{IL} J^{m-n}_{KJ}
 + \kappa n \left( \eta_{JK}\delta_{IL} - \frac{1}{p-q}\delta_{IJ}\delta_{KL} \right) \delta_{mn} .
\end{align}

Presumably we would also have a quantum version of the generalization of
Identity 2 in \cite{Dorey:2016hoj} which would have the effect of introducing
a factor of $k/\kappa$ in the central term, in which case the final result
would be
\begin{align}
\left[ J^m_{IJ}, J^{-n}_{KL} \right]_S \sim &
 \eta_{JK} J^{m-n}_{IL} - (-1)^{(I+J)(K+L)}\eta_{IL} J^{m-n}_{KJ}
 + k n \left( \eta_{JK}\delta_{IL} - \frac{1}{p-q}\delta_{IJ}\delta_{KL} \right) \delta_{mn} .
\end{align}

\subsection{Generalization to $U(N|M)$}
We expect the results of the previous section for $M=0$ generalize to $M \ne 0$
provided we take a combined large $N$ and large $M$ limit. Specifically,
we would expect that the natural limit is to take large $N/p = M/q$. As the
arguments are essentially the same as in the previous section, we just present
the definitions and result. We also note the decomposition under
$U(N) \times U(M) \in U(N|M)$ but calculations are most naturally carried out
using superbrackets without such a decomposition.

For $p=q$ we need to take more care but otherwise we can propose the
following supertraceless $SU(p|q)$ supercurrents
\begin{align}
\label{sj0a}
\widehat{J}_{IJ}^{m}
=i\left(
\widehat{\Phi}_{I}^{\dag}\widehat{Z}^{m}\widehat{\Phi}_{J}
-\frac{(-1)^K}{p-q}\delta_{IJ} \widehat{\Phi}_{K}^{\dag}\widehat{Z}^{m}\widehat{\Phi}_{K}
\right)
\equiv
\left(
\begin{array}{cc}
\widehat{J}_{ij}^{m}&\widehat{J}_{i \kappa}^{m}\\
\widehat{J}_{\lambda j}^{m}&\widehat{J}_{\lambda \kappa}^{m}\\
\end{array}
\right) .
\end{align}

Denoting elements of $\widehat{Z}^{m}$ by
\begin{align}
\label{zm0}
\widehat{Z}^{m}&=
\left(
\begin{array}{cc}
\widehat{Z}_{B}^{m}& \widehat{Z}_{F}^{m}\\
\widehat{Z}_{\widetilde{F}}^{m}&\widehat{Z}_{\widetilde{B}}^{m}\\
\end{array}
\right),
\end{align}
one can express the currents (\ref{sj0a}) by
\begin{align}
\label{zm0a}
\widehat{J}_{ij}^{m}
= & i\left[
\phi_{i}^{\dag}\widehat{Z}^{m}_{B}\phi_{j}
+\widetilde{\psi}_{i}^{\dag}\widehat{Z}_{\widetilde{F}}^{m}\phi_{j}
+\phi^{\dag}_{i}\widehat{Z}_{F}^{m}\widetilde{\psi}_{j}
+\widetilde{\psi}_{i}^{\dag}\widehat{Z}_{\widetilde{B}}^{m}\widetilde{\psi}_{j}
\right]
-\frac{i}{p-q}\delta_{ij} \widehat{K}^{m} , \\
\label{zm0b}
\widehat{J}_{\lambda \kappa}^{m}
= & i\left[
\widetilde{\phi}_{\lambda}^{\dag}\widehat{Z}_{\widetilde{B}}^{m}\widetilde{\phi}_{\kappa}
+\psi_{\lambda}^{\dag}\widehat{Z}_{F}^{m}\widetilde{\phi}_{\kappa}
+\widetilde{\phi}_{\lambda}^{\dag}\widehat{Z}^{m}_{\widetilde{F}}\psi_{\kappa}
+\psi_{\lambda}^{\dag}\widehat{Z}_{B}^{m}\psi_{\kappa}
\right]
 - \frac{i}{p-q}\delta_{\lambda \kappa} \widehat{K}^{m} , \\
\label{zm0c}
\widehat{J}_{i\kappa}^{m}
= & i\left[
\phi_{i}^{\dag}\widehat{Z}_{B}^{m}\psi_{\kappa}
+\widetilde{\psi}_{i}^{\dag}\widehat{Z}^{m}_{\widetilde{F}}\psi_{\kappa}
+\phi^{\dag}_{i}\widehat{Z}^{m}_{F}\widetilde{\phi}_{\kappa}
+\widetilde{\psi}_{i}^{\dag}\widehat{Z}_{\widetilde{B}}^{m}\widetilde{\phi}_{\kappa}
\right],\\
\label{zm0d}
\widehat{J}_{\lambda j}^{m}
= & i\left[
\psi_{\lambda}^{\dag}\widehat{Z}_{B}^{m}\phi_{j}
+\widetilde{\phi}_{\lambda}^{\dag}\widehat{Z}_{\widetilde{F}}^{m}\phi_{j}
+\psi_{\lambda}^{\dag}\widehat{Z}_{F}^{m}\widetilde{\psi}_{j}
+\widetilde{\phi}_{\lambda}^{\dag}\widehat{Z}_{B}^{m}\widetilde{\psi}_{j}
\right], \\
\widehat{K}^{m}
= & \left[
\phi_{k}^{\dag}\widehat{Z}^{m}_{B}\phi_{k}
+\widetilde{\psi}_{k}^{\dag}\widehat{Z}_{\widetilde{F}}^{m}\phi_{k}
+\phi^{\dag}_{k}\widehat{Z}_{F}^{m}\widetilde{\psi}_{k}
+\widetilde{\psi}_{k}^{\dag}\widehat{Z}_{\widetilde{B}}^{m}\widetilde{\psi}_{k}
\right] \nonumber \\
 & - \left[
\widetilde{\phi}_{\rho}^{\dag}\widehat{Z}_{\widetilde{B}}^{m}\widetilde{\phi}_{\rho}
+\psi_{\rho}^{\dag}\widehat{Z}_{F}^{m}\widetilde{\phi}_{\rho}
+\widetilde{\phi}_{\rho}^{\dag}\widehat{Z}^{m}_{\widetilde{F}}\psi_{\rho}
+\psi_{\rho}^{\dag}\widehat{Z}_{B}^{m}\psi_{\rho}
\right].
\label{zm0e}
\end{align}

Assuming the large $N$ and now also large $M$ properties hold, along with
generalizations of the classical and quantum identities described in the
previous section, we will find the same affine Lie superalgebra result
\begin{align}
\label{affinesuper}
[\widehat{J}_{IJ}^{m}, \widehat{J}_{KL}^{-n}]_S
&\sim
\left(
\eta_{JK}\widehat{J}_{IL}^{m-n}-
(-1)^{(I+J)(K+L)}\eta_{IL}\widehat{J}_{KJ}^{m-n}
\right)+ k m
\left(
\eta_{JK}\delta_{IL}
-\frac{1}{p-q}\delta_{IJ}\delta_{KL}
\right) \delta_{mn}
\end{align}
for large $N/p = M/q$.

\section{Partition function}
\label{secpfn}

\subsection{Definition}
\label{subsecpfndef}
In this section we study the spectrum of the system 
by computing the partition function of the supermatrix Chern-Simons model (\ref{sgqm1a}). 
Let us consider the modified Hamiltonian 
\begin{align}
\label{spfn1a1}
H'&=\omega \Str \left(\widehat{Z}^{\dag}\widehat{Z} \right)
-\sum_{I,A}\widehat{\mu}_{I} (-1)^I \widehat{\Phi}_{IA}^{\dag}\Phi_{AI}\nonumber\\
&=\omega \Tr \left(Z^{\dag}Z+B^{\dag}B -\widetilde{Z}^{\dag}\widetilde{Z}-A^{\dag}A\right)\nonumber\\
&-\sum_{i=1}^{p}\sum_{a=1}^{N}\mu_{i}\phi_{ia}^{\dag}\phi_{ai}
+\sum_{\lambda=1}^{q}\sum_{\alpha=1}^{M}\widetilde{\mu}_{\lambda}\widetilde{\phi}_{\lambda \alpha}^{\dag}\widetilde{\phi}_{\alpha \lambda}
-\sum_{i=1}^{p}\sum_{\alpha=1}^{M}\mu_{i}\widetilde{\psi}_{i \alpha}^{\dag}\widetilde{\psi}_{\alpha i}
+\sum_{\lambda=1}^{q}\sum_{a=1}^{N}\widetilde{\mu}_{\lambda}\psi_{\lambda a}^{\dag}\psi_{a \lambda}. 
\end{align}
Here we have introduced the chemical potential $-\sum \widehat{\mu}_{I}(-1)^{I}\widehat{\Phi}^{\dag}_{IA}\widehat{\Phi}_{AI}$ 
where $\widehat{\mu}_{I}=\left\{\mu_{i}, \widetilde{\mu}_{\lambda}\right\}$ is a set of coupling constants. 
It counts the number of $\hat{Z}^{\dag}$ and $\widehat{\Phi}_{I}^{\dag}$ excitations with 
weights $\omega$ and $\mu_{I}$. 
When evaluated on the physical state $|\mathrm{phys}\rangle$, 
the modified Hamiltonian gives 
\begin{align}
\label{spfn1a2}
H'|\mathrm{phys}\rangle
&=\left(
\omega N_{\widehat{Z}}
-\sum_{i=1}^{p}\mu_{i}J_{i}
-\sum_{\lambda=1}^{q}\widetilde{\mu}_{\lambda}\widetilde{J}_{\lambda}
\right)|\mathrm{phys}\rangle 
\end{align}
where 
\begin{align}
\label{qnumber_1}
N_{\widehat{Z}}=N_{Z}+N_{B}+N_{\widetilde{Z}}+N_{A}
\end{align}
is the total number of excitations of $\widehat{Z}^{\dag}$ 
and 
\begin{align}
\label{qnumber_2}
J_{i}&=N_{\phi_{i}}+N_{\widetilde{\psi}_{i}},& 
\widetilde{J}_{\lambda}=N_{\widetilde{\phi}_{\lambda}}+N_{\psi_{\lambda}}
\end{align}
is the total number of excitations of fundamental fields $\phi_{i}$ and $\widetilde{\psi}_i$ 
and that of fundamental fields $\widetilde{\phi}_{\lambda}$ and $\psi_{\lambda}$. 
The partition function of the modified Hamiltonian is given by
\begin{align}
\label{spfn1a3}
\mathcal{Z}(q,x,y)&=\Tr e^{-\beta H'}
=\Tr q^{N_{\widehat{Z}}} \prod_{i=1}^{p}x_{i}^{J_{i}}\prod_{\lambda=1}^{q}y_{\lambda}^{\widetilde{J}_{\lambda}}
\end{align}
where the trace is taken over the physical states $|\mathrm{phys}\rangle$ 
and $\beta$ is the inverse temperature. 
We have defined parameters $q:=e^{-\beta \omega}$, $x_{i}:=e^{\beta \mu_{i}}$ and $y_{\lambda}:=e^{\beta\widetilde{\mu}_{\lambda}}$. 

To compute this partition function, 
we firstly collect all states and then project out the non-physical states by requiring that 
the physical states are gauge invariant so that they obey the Gauss law constraints. 
The Lie superalgebra $\mathfrak{u}(N|M)$ is a $\mathbb{Z}_{2}$-graded space $V$ 
decomposed into a direct sum of $\mathbb{Z}_{2}$-graded subspaces $V_{\overline{0}}$ and $V_{\overline{1}}$. 
As we have the supertrace form on $\mathfrak{u}(N|M)$, 
a supersymmetric bilinear form on $V$ is defined so that 
$V_{\overline{0}}$ and $V_{\overline{1}}$ are orthogonal and 
the restriction of the bilinear form to $V_{\overline{0}}$ is symmetric 
and to $V_{\overline{1}}$ is skew-symmetric. 
Identifying the Cartan subalgebra $\mathfrak{h}$ with the root space $\mathfrak{h}^{*}$ via this bilinear form, 
we have \cite{Kac:1977em} 
\begin{align}
\label{gauge_ch1}
\epsilon_{a}&=E_{aa},& 
\delta_{\alpha}&=-E_{\alpha\alpha}
\end{align}
where $\epsilon_{a}, a=1,\cdots, N$ and 
$\delta_{\alpha}, \alpha=1,\cdots, M$ are a basis 
of the root space $\mathfrak{h}^{*}$ 
while $E_{ab}$ and $E_{\alpha\beta}$ are the basis 
of the Cartan subalgebra $\mathfrak{h}$. 
Since (\ref{gauge_ch1}) defines the gauge charges, 
the relative minus sign for the $\mathfrak{u}(M)$ subalgebra would require an additional sign to 
read off the correct $U(1)$ charges for the $U(M)$ symmetry from the related excitation modes. 

We will focus on the holomorphic polarized quantization 
where the Hilbert space is constructed by acting with $\widehat{Z}_{AB}^{\dag}$ and $\widehat{\Phi}_{IA}^{\dag}$ on 
the reference states $|0\rangle$. 
All the physical states are characterized by the number operators 
$N_{Z}$, $N_{\widetilde{Z}}$, $N_{A}$, $N_{B}$, $N_{\phi_{i}}$, $N_{\widetilde{\phi}}$, $N_{\psi}$ and $N_{\widetilde{\psi}}$. 
Their quantum numbers $N_{\widehat{Z}}$, $J_{i}$ and $\widetilde{J}_{\lambda}$ appearing in the partition function (\ref{spfn1a3})
are determined from 
(\ref{qnumber_1}) and (\ref{qnumber_2}). 
Their gauge charges are determined from the trace parts (\ref{sgauss1a}) and (\ref{sgauss1b}) of the quantum Gauss law conditions 
by noting the relation (\ref{gauge_ch1}).  
Let $\mathfrak{q}$ 
and $\widetilde{\mathfrak{q}}$ be the 
%
diagonal $U(1)$ charges 
for $U(1)^{N}\subset U(N)$ and $U(1)^{M} \subset U(M)$ of the associated excitation modes respectively. 
Then they read
\begin{align}
\label{gauge_ch2N1}
\mathfrak{q}[Z]&=0,& \mathfrak{q}[\widetilde{Z}]&=0,& \mathfrak{q}[A]&=1,& \mathfrak{q}[B]&=-1\\
\label{gauge_ch2N2}
\mathfrak{q}[\phi]&=1,& \mathfrak{q}[\widetilde{\phi}]&=0,& \mathfrak{q}[\psi]&=1,& \mathfrak{q}[\widetilde{\psi}]&=0\\
\label{gauge_ch2M1}
\widetilde{\mathfrak{q}}[Z]&=0, &\widetilde{\mathfrak{q}}[\widetilde{Z}]&=0, &\widetilde{\mathfrak{q}}[A]&=-1, &\widetilde{\mathfrak{q}}[B]&=1\\
\label{gauge_ch2M2}
\widetilde{\mathfrak{q}}[\phi]&=0, &\widetilde{\mathfrak{q}}[\widetilde{\phi}]&=1, &\widetilde{\mathfrak{q}}[\psi]&=0, &\widetilde{\mathfrak{q}}[\widetilde{\psi}]&=1
\end{align}
In the following we will introduce $\omega_{a}$ and $\widetilde{\omega}_{\alpha}$ as the fugacity parameters 
for each Cartan element of the gauge symmetries $U(1)^{N}\subset U(N)$ and $U(1)^{M}\subset U(M)$ respectively. 
Taking account into these charges and fugacity parameters, we can collect all the contributions to the partition function as follows:

\begin{enumerate}
\item $\mathcal{Z}_{\widehat{Z}}$

Supermatrix field $\widehat{Z}_{AB}$ consists of 
bosonic fields $Z_{ab}$, $\widetilde{Z}_{\alpha\beta}$ 
and fermionic fields $A_{a\beta}$ and $B_{\alpha b}$. 
According to (\ref{qnumber_1}), each of the associated excitations carries quantum number $N_{\widehat{Z}}=1$. 
In addition, these component fields have two units of gauge charges as they involve two gauge indices. 
According to (\ref{gauge_ch2N1}) and (\ref{gauge_ch2M1}), 
$A_{a\beta}$ and $B_{\alpha b}$ have quantum numbers of $\frac{\omega_{a}}{\widetilde{\omega}_{\beta}}$ 
and $\frac{\widetilde{\omega}_{\alpha}}{\omega_{b}}$ respectively. 
Although the total gauge charges of $Z_{ab}$ and $\widetilde{Z}_{\alpha\beta}$ are zero, 
as we are now turning on the gauge fugacity parameter for each of the Cartan elements, 
$Z_{ab}$ and $\widetilde{Z}_{\alpha\beta}$ carry quantum numbers of 
$\frac{\omega_{a}}{\omega_{b}}$ and $\frac{\widetilde{\omega}_{\alpha}}{\widetilde{\omega}_{\beta}}$.  
The contribution to the partition function from the operators$\widehat{Z}_{AB}$ is given by 
\begin{align}
\label{spfn1a4}
\mathcal{Z}_{\widehat{Z}}&=
\prod_{a,b=1}^{N}
\frac{1}{1-q\frac{\omega_{a}}{\omega_{b}}}
\prod_{\alpha,\beta=1}^{M}
\frac{1}{1-q\frac{\widetilde{\omega}_{\alpha}}{\widetilde{\omega}_{\beta}}}
\prod_{c=1}^{N}\prod_{\gamma=1}^{M}
\left(1+q\frac{\widetilde{\omega}_{\gamma}}{\omega_{c}} \right)
\left(1+q\frac{\omega_{c}}{\widetilde{\omega}_{\gamma}}\right)
\end{align}
where the first two factors come from 
the bosonic fields $Z_{ab}$, $\widetilde{Z}_{\alpha\beta}$ 
and the latter two from the fermionic fields $A_{a \beta}$ and $B_{\alpha b}$. 

\item $\mathcal{Z}_{\widehat{\Phi}}$

The operators $\widehat{\Phi}_{AI}$ involve 
$\phi_{ai}$, $\widetilde{\phi}_{\alpha\lambda}$, 
$\psi_{a \lambda}$ and $\widetilde{\psi}_{\alpha i}$. 
While $\phi_{ai}$ and $\widetilde{\psi}_{\alpha i}$ carry quantum numbers $J_{i}=1$, 
$\widetilde{\phi}_{\alpha\lambda}$ and $\psi_{a\lambda}$ have quantum numbers $\widetilde{J}_{\lambda}=1$. 
Unlike the supermatrix field, these fields are labelled by a single gauge index. 
As seen from the gauge charges (\ref{gauge_ch2N2}) and (\ref{gauge_ch2M2}), 
$\phi_{ai}$ and $\psi_{a\lambda}$ have quantum numbers of $\omega_{a}$,
while $\widetilde{\phi}_{\alpha\lambda}$ and $\widetilde{\psi}_{\alpha i}$ have those of $\widetilde{\omega}_{\alpha}$. 
The contribution to the partition function from the operators $\widehat{\Phi}_{AI}$ is given by 
\begin{align}
\label{spfn1a5}
\mathcal{Z}_{\widehat{\Phi}}&=
\prod_{a=1}^{N}\prod_{i=1}^{p}
\frac{1}{1-x_{i}\omega_{a}}
\prod_{\alpha=1}^{M}\prod_{\lambda=1}^{q}
\frac{1}{1-y_{\lambda}\widetilde{\omega}_{\alpha}}
\prod_{b=1}^{N}\prod_{\rho=1}^{q}
\left(1+y_{\rho}\omega_{b}\right)
\prod_{\beta=1}^{M}\prod_{j=1}^{p}
\left(1+x_{j}\widetilde{\omega}_{\beta}\right)
\end{align}
where the first two terms correspond to bosonic excitations of $\phi_{ai}$ and $\widetilde{\phi}_{\alpha\lambda}$ 
while the others are fermionic contributions of $\psi_{a\lambda}$ and $\widetilde{\psi}_{\alpha i}$.

\end{enumerate}

To project onto the physical states we will carry out a contour integration over the gauge fugacity parameters 
$\omega_{a}$ and $\widetilde{\omega}_{\alpha}$ in such a way that 
only gauge invariant states are picked up as a contour integration allows us to compute infinite sums 
by reducing them to finite sums of residues at poles. 

According to the trace parts (\ref{sgauss1a}) and (\ref{sgauss1b}) of the Gauss law constraints 
and the sign factor (\ref{gauge_ch1}), 
the physical states should carry charge $k$ for each of the Cartan of the $U(N)$ 
and charge $k$ for each of the Cartan of the $U(M)$. 
Therefore we introduce poles of order $k+1$ and $k+1$ 
by adding the factors $\prod_{a}\frac{1}{\omega_{a}^{k}}$ and $\prod_{\alpha}\frac{1}{\widetilde{\omega}_{\alpha}^{k}}$ respectively. 

As we deal with integration with respect to the elements $\omega_{a}$ and $\widetilde{\omega}_{\alpha}$ of the $U(N|M)$ supermatrix, 
we will introduce the $U(N|M)$ Berezinian measure \cite{MR914369}. 
Taking these additional factors into the product of the two contributions 
(\ref{spfn1a4}) and (\ref{spfn1a5}), 
one can express the partition function as
\begin{align}
\label{spfn1a6}
\mathcal{Z}&=\frac{1}{N!}\frac{1}{M!}
\left(\prod_{a=1}^{N}\frac{1}{2\pi i}\oint_{\Gamma} \frac{d\omega_{a}}{\omega_{a}^{k+1}}\right)
\left(\prod_{\alpha=1}^{M}\frac{1}{2\pi i}\oint_{\widetilde{\Gamma}} \frac{d\widetilde{\omega}_{\alpha}}{\widetilde{\omega}_{\alpha}^{k+1}}\right)
%
\left(
\frac{
\prod_{b\neq c}^{N}\left(1-\frac{\omega_{b}}{\omega_{c}}\right)
\prod_{\beta\neq \gamma}^{M}\left(1-\frac{\widetilde{\omega}_{\beta}}{\widetilde{\omega}_{\gamma}}\right)
}
{\prod_{d=1}^{N}\prod_{\delta=1}^{M}
\left(1+\frac{\widetilde{\omega}_{\delta}}{\omega_{d}}\right)
\left(1+\frac{\omega_{d}}{\widetilde{\omega}_{\delta}}\right)
}
\right)
\nonumber\\
&
\times 
\left(
\frac{\prod_{a=1}^{N}\prod_{\alpha=1}^{M} \left(1+q\frac{\widetilde{\omega}_{\alpha}}{\omega_{a}}\right) \left(1+q\frac{\omega_{a}}{\widetilde{\omega}_{\alpha}}\right)}
{\prod_{a,b=1}^{N}\left(1-q\frac{\omega_{a}}{\omega_{b}}\right) \prod_{\alpha,\beta=1}^{M}\left(1-q\frac{\widetilde{\omega}_{\alpha}}{\widetilde{\omega}_{\beta}}\right)}
\right)
%
%
\left(
\frac{\prod_{a=1}^{N}\prod_{\lambda=1}^{q}\left(1+y_{\lambda}\omega_{a}\right) \prod_{\alpha=1}^{M}\prod_{i=1}^{p}\left(1+x_{i}\widetilde{\omega}_{\alpha}\right)}
{\prod_{a=1}^{N}\prod_{i=1}^{p}\left(1-x_{i}\omega_{a}\right) \prod_{\alpha=1}^{M}\prod_{\lambda=1}^{q}\left(1-y_{\lambda}\widetilde{\omega}_{\alpha}\right)}
\right)
\end{align}
where $\Gamma$ and $\widetilde{\Gamma}$ are the $N$-dimensional cycle and $M$-dimensional cycle respectively.

Using the completeness relation \cite{moens2007supersymmetric}
\begin{align}
\label{sschur_relation1}
\frac{\prod_{a=1}^{N}\prod_{\lambda=1}^{q}\left(1+y_{\lambda}\omega_{a}\right) \prod_{\alpha=1}^{M}\prod_{i=1}^{p}\left(1+x_{i}\widetilde{\omega}_{\alpha}\right)}
{\prod_{a=1}^{N}\prod_{i=1}^{p}\left(1-x_{i}\omega_{a}\right) \prod_{\alpha=1}^{M}\prod_{\lambda=1}^{q}\left(1-y_{\lambda}\widetilde{\omega}_{\alpha}\right)}
&=\sum_{\lambda}s_{\lambda}(x/y)s_{\lambda}(\omega/\widetilde{\omega})
\end{align}
of the supersymmetric Schur polynomial $s_{\lambda}(x/y)$ 
and the definition 
\begin{align}
\label{hl1a}
P_{\lambda}(x;q)&:=\frac{1}{v_{\lambda}}
\sum_{w\in S_{N}}
w \left\{
x_{1}^{\lambda_{1}}x_{2}^{\lambda_{2}}\cdots x_{N}^{\lambda_{N}}
\prod_{i>j}\frac{\left(1-q\frac{x_{i}}{x_{j}}\right)}{\left(1-\frac{x_{i}}{x_{j}}\right)}
\right\}\\
\label{hl1b}
&=\sum_{w\in S_{N}\setminus S_{N}^{\lambda}}w
\left\{
x_{1}^{\lambda_{1}}\cdots x_{N}^{\lambda_{N}}
\prod_{\lambda_{i}<\lambda_{j}}\frac{\left(1-q\frac{x_{i}}{x_{j}}\right)}{\left(1-\frac{x_{i}}{x_{j}}\right)}
\right\}
\end{align}
of the Hall-Littlewood polynomial,
where 
\begin{align}
\label{hl1c}
v_{\lambda}&:=\frac{\varphi_{N-l(\lambda)}\prod_{j\ge 1}\varphi_{m_{j}(\lambda)}}{(1-q)^{N}}, 
& 
\varphi_{m}&:=\prod_{i=1}^{m}(1-q^{i}) ,
\end{align}
$S_{N}^{\lambda}$ is the set of permutations that fix $\lambda$, 
$l(\lambda)$ is the length of the partition $\lambda$, 
and $m_{j}(\lambda)$ is the multiplicity of the partition $\lambda$, 
we can write 
\begin{align}
\label{spfn1a7}
\mathcal{Z}_{\widehat{\Phi}}
&=\sum_{\lambda}s_{\lambda}(x/y)s_{\lambda}(\omega/\widetilde{\omega}), \\
\label{spfn1a7a}
\prod_{a=1}^{N}\frac{1}{\omega_{a}^{k}}&=P_{(k^{N})}(\omega^{-1};q), \ \ \ \ \ \ \ \ \ \ 
\prod_{\alpha=1}^{M}\frac{1}{\widetilde{\omega}_{\alpha}^{k}}=P_{(k^{M})}(\widetilde{\omega}^{-1}; q). 
\end{align}
Making use of the relations (\ref{sschur_relation1}), (\ref{spfn1a7}) and (\ref{spfn1a7a}), 
we can express the partition function (\ref{spfn1a6}) as
\begin{align}
\label{spfn1a8}
\mathcal{Z}&=
\sum_{\lambda}s_{\lambda}(x/y)
\frac{1}{N!}\frac{1}{M!}
\left(
\prod_{a=1}^{N}\frac{1}{2\pi i}\oint_{\Gamma}\frac{d\omega_{a}}{\omega_{a}}
\right)
\left(
\prod_{\alpha=1}^{M}\frac{1}{2\pi i}\oint_{\widetilde{\Gamma}} \frac{d\widetilde{\omega}_{\alpha}}{\widetilde{\omega}_{\alpha}}
\right)
s_{\lambda}(\omega/\widetilde{\omega}) 
P_{(k^{N})}(\omega^{-1};q)P_{(k^{M})}(\widetilde{\omega}^{-1};q)
\nonumber\\
&\times 
%
\left(
\frac{
\prod_{b\neq c}^{N}\left(1-\frac{\omega_{b}}{\omega_{c}}\right)
\prod_{\beta\neq \gamma}^{M}\left(1-\frac{\widetilde{\omega}_{\beta}}{\widetilde{\omega}_{\gamma}}\right)
}
{\prod_{d=1}^{N}\prod_{\delta=1}^{M}
\left(1+\frac{\widetilde{\omega}_{\delta}}{\omega_{d}}\right)
\left(1+\frac{\omega_{d}}{\widetilde{\omega}_{\delta}}\right)
}
\right)
\left(
\frac{\prod_{a=1}^{N}\prod_{\alpha=1}^{M} \left(1+q\frac{\widetilde{\omega}_{\alpha}}{\omega_{a}}\right) \left(1+q\frac{\omega_{a}}{\widetilde{\omega}_{\alpha}}\right)}
{\prod_{a,b=1}^{N}\left(1-q\frac{\omega_{a}}{\omega_{b}}\right) \prod_{\alpha,\beta=1}^{M}\left(1-q\frac{\widetilde{\omega}_{\alpha}}{\widetilde{\omega}_{\beta}}\right)}
\right)
\end{align}
Although we have not precisely yet understood the issue of choice of integration contour, 
it would be very important as we are now considering the symmetries of Lie superalgebra whose representation and (super)characters have rather rich structures. 
While the integration contour of simple unit circles would give us partition function contributed from singlet sectors, 
other non-trivial contour picking up specific poles may realize non-singlet sectors. 
In the next subsection, we will give an explicit computation for $M=0$ by taking simply unit circles and comment on general cases in subsection \ref{subsec_SHL}.

\subsection{Computation for $U(N)$}
\label{subsecpfn_M0}
Let us consider the case where the gauge symmetry is ordinary $U(N)$ 
and the coupling $\omega$ is very large. 
The contributions (\ref{spfn1a4}) from the supermatrix field $\widehat{Z}^{\dag}$ simplify as 
\begin{align}
\label{spfn_m0a}
\mathcal{Z}_{Z}&=\prod_{a,b=1}^{N}\frac{1}{1-q\frac{\omega_a}{\omega_b}}
\end{align}
and the contributions (\ref{spfn1a5}) from the supervector field $\widehat{\Phi}$ only contain two parts 
\begin{align}
\label{spfn_m0b1}
\mathcal{Z}_{\phi}&=\prod_{a=1}^{N}\prod_{i=1}^{p}\frac{1}{1-x_i \omega_a}
=\sum_{\lambda}s_{\lambda}(x)s_{\lambda}(\omega),\\
\label{spfn_m0b2}
\mathcal{Z}_{\psi}&=\prod_{a=1}^{N}\prod_{\lambda=1}^{q}(1+y_{\lambda}\omega_{a})
=\sum_{\mu}s_{\mu}(y)s_{\mu'}(\omega)
\end{align}
where $\mu'$ is the conjugate of a partition $\mu$ 
whose Young diagram is the transpose of that of $\mu$. 
Thus the integral expression (\ref{spfn1a8}) reduces to 
\begin{align}
\label{spfn1a8_M0}
\mathcal{Z}&=
\sum_{\lambda,\mu}s_{\lambda}(x)s_{\mu}(y)
\frac{1}{N!}
\prod_{a=1}^{N}\frac{1}{2\pi i}\oint_{C}\frac{d\omega_{a}}{\omega_{a}} 
\frac{\prod_{b\neq c}^{N}\left(1-\frac{\omega_{b}}{\omega_{c}} \right)}
{\prod_{a,b=1}^{N}\left(1-q\frac{\omega_{a}}{\omega_{b}}\right)}
s_{\lambda}(\omega)s_{\mu'}(\omega)
P_{(k^{N})}(\omega^{-1};q)\nonumber\\
&=
\sum_{\lambda,\mu,\eta,\rho}s_{\lambda}(x)s_{\mu}(y)
\frac{1}{N!}
\prod_{a=1}^{N}\frac{1}{2\pi i}\oint_{C}\frac{d\omega_{a}}{\omega_{a}} 
\frac{\prod_{b\neq c}^{N}\left(1-\frac{\omega_{b}}{\omega_{c}} \right)}
{\prod_{a,b=1}^{N}\left(1-q\frac{\omega_{a}}{\omega_{b}}\right)}
\sum_{\eta,\rho} c^{\eta}_{\lambda\mu'}K_{\eta,\rho}(q)P_{\rho}(\omega;q)
P_{(k^{N})}(\omega^{-1};q). 
\end{align}
On the second line we have used the relation 
\begin{align}
\label{spfn_m0c}
s_{\lambda}(\omega)s_{\mu'}(\omega)
&=\sum_{\eta} c^{\eta}_{\lambda\mu'}s_{\eta}(\omega)
=\sum_{\eta,\rho} c^{\eta}_{\lambda\mu'}K_{\eta,\rho}(q)P_{\rho}(\omega;q) 
\end{align}
where $c_{\mu \nu}^{\lambda}$ are the Littlewood-Richardson coefficients \cite{MR1354144} 
and $K_{\lambda,\mu}(q)$ are the Kostka polynomials which are defined by 
\begin{align}
\label{kostka_DEF}
s_{\lambda}(x)=\sum_{\mu}K_{\lambda,\mu}(q) P_{\mu}(x;q). 
\end{align}

Furthermore, using the orthogonal property
\begin{align}
\label{hl1e}
\langle P_{\lambda}(x;q), P_{\mu}(x^{-1};q)\rangle_{P}&=\frac{1}{v_{\lambda}}\delta_{\lambda,\mu} 
\end{align}
of the Hall-Littlewood polynomials with respect to the following inner product
\begin{align}
\label{hl1d}
\langle f_{\lambda}(\omega;q), g_{\mu}(\omega^{-1};q)\rangle_{P}
&:=\frac{1}{N!}
\prod_{a=1}^{N}\frac{1}{2\pi i}\oint_{C}\frac{d\omega_{a}}{\omega_{a}}
\frac{\prod_{a\neq b}\left(1-\frac{\omega_{a}}{\omega_{b}}\right)}
{\prod_{a\neq b}\left(1-q\frac{\omega_{a}}{\omega_{b}}\right)}
f_{\lambda}(\omega)g_{\mu}(\omega^{-1}), 
\end{align}
we can rewrite the partition function (\ref{spfn1a8_M0}) as
\begin{align}
\label{spfn_m0d}
\mathcal{Z}&=
\prod_{i=1}^{N}\frac{1}{(1-q^{i})}
\sum_{\lambda, \mu, \eta}c_{\lambda \mu'}^{\eta}K_{\eta, k^{N}}(q)s_{\lambda}(x)s_{\mu}(y).
\end{align}
According to the relation \cite{moens2007supersymmetric}
\begin{align}
\label{spfn_m0e}
s_{\lambda}(x/y)&=\sum_{\mu,\nu}c^{\lambda}_{\mu\nu}
s_{\mu}(x)s_{\nu'}(y), 
\end{align}
the expression (\ref{spfn_m0d}) can be expressed as
\begin{align}
\label{spfn_m0f}
\mathcal{Z}&=
\prod_{i=1}^{N}\frac{1}{(1-q^{i})}
\sum_{\mu}K_{\mu,k^{N}}(q)s_{\mu}(x/y)
\end{align}
where $s_{\mu}(x/y)$ is the supersymmetric Schur polynomial. 
Here the summation is taken over the partitions $\mu$ which obey 
\begin{align}
\label{p_sum1}
|\mu|&=kN,\\
\label{p_sum2}
q\le l(\mu)&\le p+q,\\
\label{p_sum3}
\mu_{p+1}&\le q. 
\end{align}
The first condition (\ref{p_sum2}) is required for non-trivial $K_{\mu, k^{N}}(q)$ 
whereas the other conditions (\ref{p_sum2}) and (\ref{p_sum3}) are for non-zero valued $s_{\mu}(x/y)$ 
as the supersymmetric Schur polynomial $s_{\mu}(x/y)$ vanishes when $\mu_{p+1}>q$.  


As the modified Hall-Littlewood polynomials $Q_{\mu}'(x;q)$ are defined by 
\begin{align}
\label{mod_HL1}
Q_{\mu}'(x;q)&:=
\sum_{\lambda}K_{\lambda,\mu}(q)s_{\lambda}(x), 
\end{align}
we will define the supersymmetric modified Hall-Littlewood polynomial $Q_{\mu}'(x/y;q)$ by 
\begin{align}
\label{mod_sHL1}
Q_{\mu}'(x/y;q)&:=
\sum_{\lambda}K_{\lambda,\mu}(q)s_{\lambda}(x/y).  
\end{align}
Then the partition function is expressed as 
\begin{align}
\label{spfn_m0g}
\mathcal{Z}&=\prod_{i=1}^{N}\frac{1}{(1-q^{i})}Q_{k^{N}}'(x/y; q). 
\end{align}
Further study of properties of the supersymmetric modified Hall-Littlewood polynomials (\ref{mod_HL1}) is intriguing. 
In particular, it would be desirable to understand 
the large $N$ behavior of the Kostka polynomials as the branching coefficient 
of $\widehat{\mathfrak{su}}(p|q)/\mathfrak{su}(p|q)$ as in the ordinary case \cite{Kirillov:1994en, Nakayashiki:1995bi}.

\subsection{
Comments on general case}
\label{subsec_SHL}
Although it would be important to study the residues for different choices of contours of the integral (\ref{spfn1a8}), 
we will not get into any details of these issues in this paper. 
Instead, we will comment on some implications of the resulting expression (\ref{spfn1a8}). 
To have a well-defined partition function from the integration (\ref{spfn1a8}), 
it is expected that the integration can be performed 
by using the orthogonal property of certain functions 
with respect to $\omega$ and $\widetilde{\omega}$. 
Provided that the supersymmetric Schur polynomial $s_{\lambda}(\omega/\widetilde{\omega})$ 
in (\ref{spfn1a8}) is expanded in terms of the supersymmetric Hall-Littlewood polynomial $P_{\mu}(x/y;q)$ 
\footnote{This is the supersymmetric generalization of the definition (\ref{kostka_DEF}) of the Kostka polynomial. }
\begin{align}
\label{SHL_def}
s_{\lambda}(\omega/\widetilde{\omega})&=\sum_{\mu}K_{\lambda\mu}(q) 
P_{\mu}(\omega/\widetilde{\omega};q),
\end{align}
the second line in (\ref{spfn1a8}),
equipped with the expressions (\ref{hl1a}) in terms of permutation of variables,
would be regarded as the dual of $P_{\mu}(\omega/\widetilde{\omega};q)$. 
In fact, it takes the form of a generalization of the Berele-Regev formula \cite{berele1987hook}
\begin{align}
\label{berele_ragev}
s_{\lambda}(\omega/\widetilde{\omega})
&= 
\prod_{a=1}^{N}\prod_{\alpha=1}^{M} (\omega_a + \widetilde{\omega}_{\alpha}) s_{\lambda}(\omega) s_{\lambda'}(\widetilde{\omega}). 
\end{align}
It would be also interesting to observe that 
the supersymmetric Schur polynomial $s_{\lambda}(x/y)$ is alternatively expanded in terms of two Hall-Littlewood polynomials 
\cite{MR869577, MR1048507}
\begin{align}
\label{Super_Kostka}
s_{\lambda}(x/y)&=\sum_{\mu,\eta}K_{\lambda,\mu|\eta}(q)
P_{\mu}(x;q)P_{\eta}(y; q), 
\end{align}
which defines the Kostka polynomial $K_{\lambda,\mu|\eta}(q)$ 
and that 
it is expanded in terms of supersymmetric monomial functions $m_{\mu}(x/y)$ \cite{moens2007supersymmetric} 
\begin{align}
\label{Super_mon2}
s_{\lambda}(x/y)&=\sum_{\mu}K_{\lambda\mu}m_{\mu}(x/y)
\end{align}
where $K_{\lambda\mu}$ is the Kostka number. 
Since the supersymmetric Hall-Littlewood polynomials $P_{\mu}(x/y;q)$ may interpolate between 
the supersymmetric Schur polynomials when $q=0$ and 
the supersymmetric monomial functions when $q=1$, 
these relations may help us proceed to further survey of the supersymmetric Hall-Littlewood polynomials $P_{\mu}(x/y;q)$.

In the partition function (\ref{spfn1a4}) all states 
constructed from the supermatrix field $\widehat{Z}_{AB}$ have been picked up.
However, there are distinguished operators with different structures of the contracted gauge indices:
among themselves, with antisymmetric invariant tensor, with the supervector fields. 
There will exist operators 
$\left(Z^{\dag n}\right)_{ab}$, $\left(\widetilde{Z}^{\dag n}\right)_{\alpha\beta}$ 
as a product of $Z^{\dag}$'s with gauge indices contracted among them 
so that the antifundamental index of each operator is contracted with 
the fundamental index of the following operator.
If we start with a set of states with minimal basis constructed by the operator $\left\{Z^{\dag}_{ab}\right\}$ 
and next count a set of states with $\left\{Z^{\dag}_{ab}\right\}$ being replaced with $\left\{\left(Z^{\dag n}\right)_{ab}\right\}$ as they have 
the same gauge charges but $n$ units of the energy of $\left\{Z^{\dag}_{ab}\right\}$, 
then the corresponding partition function may take the form of 
\begin{align}
\label{spfn1a4_MORE}
\mathcal{Z}_{\widehat{Z}}&=
\prod
\frac{1}{1-q^{n}\frac{\omega_{a}}{\omega_{b}}}
\frac{1}{1-q^{n}\frac{\widetilde{\omega}_{\alpha}}{\widetilde{\omega}_{\beta}}}
\left(1+q^{n}\frac{\widetilde{\omega}_{\alpha}}{\omega_{b}} \right)
\left(1+q^{n}\frac{\omega_{a}}{\widetilde{\omega}_{\beta}}\right)
\end{align}
by taking some appropriate constrained product to avoid over counting. 
This has the same form as the affine Weyl denominator $\widehat{R}$ (divided by Weyl denominator $R$) \cite{Kac:1994kn}
\begin{align}
\label{aff_denom}
\widehat{R}
&:=R\prod_{n=1}^{\infty}
\left[
(1-q^{n})^{l}\frac{\prod_{\alpha\in \Delta_{0}} (1-q^{n}e^{\alpha})}
{\prod_{\alpha\in \Delta_{1}}(1+q^{n}e^{\alpha})}
\right]
\end{align}
where $R$ is the Weyl denominator defined by \cite{Kac:1994kn} 
\begin{align}
\label{weyl_den}
R
&:=\frac{\prod_{\alpha\in \Delta_{0}} (1-e^{\alpha})}
{\prod_{\alpha\in \Delta_{1}}(1+e^{\alpha})}
\end{align}
and $l$ is the quantum number of the Virasoro generator $L_{0}$, which is equal to the rank for $N\neq M$, 
under the identifications $\omega_{a}:=e^{-\epsilon_{a}}$ and $\widetilde{\omega}_{\alpha}:=e^{-\delta_{\alpha}}$
where $\epsilon_{a}, a=1,\cdots, N$ and $\delta_{\alpha}, \alpha=1,\cdots, M$ is a basis of the root space
(see (\ref{even_root1}) and (\ref{odd_root1})).
We also note that the factor of the Berezinian measure 
\begin{align}
\label{weyl_den0}
\frac{
\prod_{b\neq c}^{N}\left(1-\frac{\omega_{b}}{\omega_{c}}\right)
\prod_{\beta\neq \gamma}^{M}\left(1-\frac{\widetilde{\omega}_{\beta}}{\widetilde{\omega}_{\gamma}}\right)
}
{\prod_{d=1}^{N}\prod_{\delta=1}^{M}
\left(1+\frac{\widetilde{\omega}_{\delta}}{\omega_{d}}\right)
\left(1+\frac{\omega_{d}}{\widetilde{\omega}_{\delta}}\right)
}
\end{align}
in (\ref{spfn1a6}) has a close similarity with the Weyl denominator.
Since the affine Weyl denominator is associated to Ramanujan's mock theta function,
as pointed out by Kac and Wakimoto \cite{Kac:1994kn,MR1810948} (also see \cite{MR969247}),
the partition function (\ref{spfn1a4_MORE}) would indicate the property of mock modularity.
%
%
%
%


\section{Discussion}
\label{sec_dis}
We have studied a $(0+1)$-dimensional $U(N|M)$ matrix Chern-Simons quantum mechanics with an $SU(p|q)$ global symmetry. 
We have proposed it as a description of a system consisting of $N$ vortices and $M$ antivortices with $SU(p|q)$ spin degrees of freedom. 
At the classical level, we have seen that 
the model can be viewed as a generalized Calogero model with $SU(p|q)$ spin degrees of freedom. 
We have also found two types of classical ground states which admit non-trivial configuration of fermionic matrix fields. 
They are similar to the two types of vortex-antivortex pairs; 
parallel polarized vortex-antivortex pairs with negative energy 
and antiparallel polarized vortex-antivortex pairs with positive energy. 
Meanwhile we have provided a general expression of the partition function in an integral form 
and we have found that the expression can be explicitly written in terms of Kostka polynomials and super-Schur polynomials
as a generalization of \cite{Dorey:2016hoj}.


It is physically important to obtain further understanding of vortex-antivortex systems from the $U(N|M)$ matrix Chern-Simons models.  
In particular, it is intriguing to find new explanations and predictions in quantum Hall physics beyond the well-known features of the Laughlin theory. 
For instance, as in the ordinary matrix Chern-Simons models \cite{Polychronakos:2001mi, Dorey:2016mxm}, 
we would like to understand the level quantization and its relation to the filling fractions of the quantum Hall states. 
Besides, it would be interesting to construct and understand generalized wavefunctions valid for the superdeterminant states  
which we found in this work.

Further understanding of the mathematical structure would be intriguing. 
Although we have found that the current operators constructed from 
matrix degrees of freedom give rise to the affine Lie superalgebra in the large $N$ limit, 
we would like to support our results with a rigorous treatment of the partition function. 
In addition, for general supergroup we have not found an explicit expression in terms of polynomials. 
This is due to the lack of knowledge of the orthogonal properties and 
we expect that it could be achieved by defining supersymmetric Hall-Littlewood polynomials.
But we leave this problem for future study. 

In addition, it is an open question even for the ordinary Lie algebra to understand 
the underlying larger algebra without taking a large $N$ limit. 
Interestingly it has been argued \cite{hikami1995yangian, hikami2000supersymmetric} that in the related Polychronakos spin chain model \cite{Polychronakos:1993wc}
the Yangian symmetry can be embedded in the WZW model. Specifically,
the partition function becomes the character for the WZW model at level one in the large $N$ limit,
as the first Yangian invariant operator is identified with the Virasoro generator. 


Another attractive future direction is to explore 
the gravitational dual of the $(0+1)$-dimensional matrix Chern-Simons quantum mechanics 
as it may be useful to understand the holographic dual description of generally conjectured infinite dimensional symmetry in two-dimensional gravity. 
Geroch showed \cite{Geroch:1972yt} that 
a hidden symmetry in two-dimensional gravity is infinite dimensional, known as the Geroch group, 
which indicates that Einstein gravity is integrable after reducing to two-dimensions. 
Julia demonstrated \cite{Hawking:1981bu,Julia:1981wc} that the Geroch group is the affine Lie algebra $A_{1}^{(1)}$.
Since Dorey, Tong and Turner's recent work \cite{Dorey:2016hoj} and our result show that 
the quantum mechanical systems with $N$ degrees of freedom realize the affine Kac-Moody symmetry 
in the large $N$ limit, it may help us to understand the underlying infinite-dimensional symmetry structure in two-dimensional gravity 
and further lifted symmetry in higher dimensional gravity. 
There has also been recent work on matrix $U(N)$ Chern-Simons quantum mechanics systems with $N_f$ fundamental and anti-fundamental fields \cite{Betzios:2017yms}. These models, also related to Calogero systems, describe FZZT branes in Liouville theory and also two-dimensional blackholes. It was also shown \cite{Betzios:2017yms} that these models exhibit a phase transition at large $N$ and $N_f$, and an intriguing relation of the grand-canonical partition function to the Toda intergrable hierarchy was found. It will be interesting to explore these issues for our supergroup models.

Further possible applications of the matrix Chern-Simons model could be found in string and M-theory. 
In the type IIB string theory the D1-branes which end on the intersecting
D3-branes are vortices in the effective 3d gauge theory, and the relation
between the vortex D1-branes and the matrix Chern-Simons model has been examined
in \cite{Tong:2003vy}. In \cite{Mikhaylov:2014aoa} intersecting D3-branes and
NS5-branes in curved spacetime are shown to correspond to supergroup
Chern-Simons theory. It would be interesting to explore the relation between
further attached vortex-like D1-branes involving the supergroup symmetry and
our supergroup Chern-Simons matrix model. 
In M-theory intersecting M2-branes can be viewed as vortices in the Chern-Simons matter theory. 
In this brane setup the large $N$ limit of the Chern-Simons matrix model 
corresponds to an infinite number of intersecting M2-branes, 
which would lead to an M5-brane as a condensate of M2-branes. 
In \cite{Okazaki:2015fiq, Okazaki:2016pne} we found that 
a certain configuration of intersecting M2-M5 branes on a two-dimensional plane can be effectively described 
by the supergroup WZW model associated to the affine Lie superalgebra. 
Since we have found a connection to the affine Lie superalgebra in this work, 
we believe that further physical explanation and application can be available in string and M-theory.

\subsection*{Acknowledgements}
We would like to thank Heng-Yu Chen, Chong-Sun Chu, Nick Dorey, Amihay Hanany,
Taro Kimura, Anatoli Kirillov, Alexios Polychronakos, David Tong and
Carl Turner for useful discussions and comments. 
This is the version published in JHEP, including
several corrections and clarifications in response to helpful comments from the referee. 
In particular, Section 4.2 is revised and extended, and Sections 6.2 and 6.3 are revised. 
TO thanks the organizers of the 
\textit{Topological Field Theories, String theory and Matrix Models} in Moscow 
for hospitality during the course of the work. 
TO is supported by MOST under the Grant No.106-2811-M-002-053. 
DJS thanks the Taiwan NCTS Physics Division for support and hospitality to enable a
productive visit to NTHU and also NTU where some of this work was carried out.
DJS is supported in part by the STFC Consolidated Grant ST/P000371/1.

\appendix

\section{More classical ground states}
\label{app_classicsol}
%
%
%
%
%
%
%
Consider the case with $\alpha_{NN}=\widetilde{\alpha}_{MM}=0$ as the solution to 
(\ref{csol1b2}) and take 
\begin{align}
\label{csol1d}
\beta_{a}&=(N-a)\omega,& 
\widetilde{\beta}_{\alpha}&=(M-\alpha)\omega.
\end{align}
As $\alpha_{NN}=\widetilde{\alpha}_{MM}$, this would imply the occurrence of enhanced symmetry. 
The configurations (\ref{csol1d}) tell us that the fields $Z_{ab}$, $\widetilde{Z}_{\alpha\beta}$, 
$A_{a\alpha}$ and $B_{\alpha a}$ have general forms
\begin{align}
\label{csol1e}
Z&=\left(
\begin{array}{ccccc}
0&Z_{12}&&& \\
&0&Z_{23}&& \\
&&\ddots&\ddots& \\
&&&0&Z_{N-1 N}\\
&&&&0\\
\end{array}
\right),& 
\widetilde{Z}&=\left(
\begin{array}{cccc}
0&\widetilde{Z}_{12}&& \\
&\ddots&\ddots& \\
&&0&\widetilde{Z}_{M-1 M} \\
&&&0\\
\end{array}
\right), \nonumber\\
A&=\left(
\begin{array}{ccc}
0&0&0\\
\vdots&\vdots&\vdots\\
A_{(N-M) 1}&0&0\\
0&\ddots&0\\
&&A_{(N-1) M}\\
&&0\\
\end{array}
\right),& 
B&=\left(
\begin{array}{ccccc}
0&\cdots&B_{1(N-M+2)}&&\\
&&&\ddots&\\
&&&&B_{(M-1) N}\\
&&&&0\\
\end{array}
\right)
\end{align}
where the elements 
$Z_{a (a+1)}$, $\widetilde{Z}_{\alpha (\alpha+1)}$, 
$A_{a (M-N+1+a)}$, $B_{\alpha (N-M+1+\alpha)}$ are the only components allowed to have non-zero values. 
Then the Gauss law constraints (\ref{gaussbb})-(\ref{gaussbf}) become 
\begin{align}
&\label{csol1f}
\left(
|Z_{a (a+1)}|^{2}-|Z_{a (a-1)}|^{2}
+A_{a(a-N+M+1)}A^{\dag}_{(a-N+M+1) b}
-B^{\dag}_{a (a-N+M-1)}B_{(a-N+M-1) b}
\right)\delta_{ab}\nonumber\\
&+(\kappa(N-M) + |z|^{2})\delta_{aN}\delta_{bN}=\kappa \delta_{ab},\\
\label{csol1g}
&
\left(
|\widetilde{Z}_{\alpha (\alpha+1)}|^{2}-|\widetilde{Z}_{\alpha (\alpha-1)}|^{2}
+B_{\alpha(\alpha+N-M+1)}B^{\dag}_{(a+N-M+1) b}
-A^{\dag}_{\alpha (\alpha+N-M-1)}A_{(\alpha+N-M-1) \beta}
\right)\delta_{\alpha\beta}\nonumber\\
& + |z|^{2} \delta_{\alpha M}\delta_{\beta M}=\kappa \delta_{\alpha\beta},\\
\label{csol1h}
&B_{\alpha (\alpha+N-M+1)}Z_{(\alpha+N-M+1) \alpha+N-M}^{\dag}
-\widetilde{Z}_{\alpha (\alpha-1)}^{\dag}B_{(\alpha-1)(\alpha+N-M)}
\nonumber\\
&-A_{\alpha (\alpha+N-M-1)}^{\dag}Z_{(\alpha+N-M-1)(\alpha+N-M)} 
+\widetilde{Z}_{\alpha (\alpha+1)}A^{\dag}_{(\alpha+1)(\alpha+N-M)}
+zy^{\dag}\delta_{\alpha M}=0.
\end{align}
We define the following quantities:
\begin{align}
z_{a}&= \frac{1}{\kappa} |Z_{a(a+1)}|^2 \; , \;\; a \in \{ 1, 2, \ldots , N-1 \},\nonumber\\
\widetilde{z}_{\alpha}&= \frac{1}{\kappa} |\widetilde{Z}_{\alpha(\alpha+1)}|^2 \; , \;\; \alpha \in \{ 1, 2, \ldots , M-1 \},\nonumber\\
\mathcal{A}_{a}&= \frac{1}{\kappa} A_{a(a-N+M+1)} A_{(a-N+M+1)a}^{\dagger} \; , \;\; a \in \{ N-M, N-M+1, \ldots , N-1 \},\nonumber\\
\mathcal{B}_{a}&= \frac{1}{\kappa} B_{a(a-N+M-1)}^{\dagger} B_{(a-N+M-1)a} \; , \;\; a \in \{ N-M+2, N-M+3, \ldots , N \}. 
\end{align}


Above we have assumed $N > M$. In the case $N=M$ note that there is one fewer
$\mathcal{A}_a$ as clearly the value $a=0$ is not allowed as in (\ref{csol1e})
there is no component $A_{(N-M)1} = A_{01}$.

Then the Gauss law constraints become, assuming $N \ge M + 2$
\begin{align}
\widetilde{z}_{\alpha}&= \alpha - \sum_{\gamma = N-M}^{\alpha+N-M-1} \mathcal{A}_{\gamma} + \sum_{\gamma = N-M+2}^{\alpha+N-M+1} \mathcal{B}_{\gamma} \; , \;\; \alpha \in \{ 1, 2, \ldots , M-1 \}, \\
\frac{1}{\kappa}|z|^2 & = M - \sum_{\gamma = N-M}^{N-1} \mathcal{A}_{\gamma} + \sum_{\gamma = N-M+2}^{N} \mathcal{B}_{\gamma}, \\
\label{cgaussza1}
z_a & = a \; , \;\; a \in \{ 1, 2, \ldots , N-M-1 \}, \\
\label{cgausszaNM}
z_{N-M} & = N - M - \mathcal{A}_{N-M}, \\
z_{N-M+1} & = N - M +1 - \mathcal{A}_{N-M} - \mathcal{A}_{N-M+1}, \\
z_a & = a - \sum_{\gamma = N-M}^{a} \mathcal{A}_{\gamma} + \sum_{\gamma = N-M+2}^{a} \mathcal{B}_{\gamma} \; , \;\; a \in \{ N-M+2, N-M+3, \ldots , N-1 \}, \\
|x|^2 - y^{\dagger}y - |z|^2 & = (N-M)\kappa
\end{align}
along with (\ref{csol1h}).

In the case $N=M+1$ we don't have equation~(\ref{cgaussza1}) while in the case
$N=M$ the Gauss law constraints are instead
\begin{align}
\widetilde{z}_1 & = 1 + \mathcal{B}_2, \\
\widetilde{z}_{\alpha}&= \alpha - \sum_{\gamma = 1}^{\alpha-1} \mathcal{A}_{\gamma} + \sum_{\gamma = 2}^{\alpha+1} \mathcal{B}_{\gamma} \; , \;\; \alpha \in \{ 2, \ldots , N-1 \}, \\
\frac{1}{\kappa}|z|^2 & = N - \sum_{\gamma = 1}^{N-1} \mathcal{A}_{\gamma} + \sum_{\gamma = 2}^{N} \mathcal{B}_{\gamma}, \\
z_{1} & = 1 - \mathcal{A}_{1}, \\
z_a & = a - \sum_{\gamma = 1}^{a} \mathcal{A}_{\gamma} + \sum_{\gamma = 2}^{a} \mathcal{B}_{\gamma} \; , \;\; a \in \{ 2, 3, \ldots , N-1 \}, \\
\label{cgaussNNnorm}
|x|^2 - y^{\dagger}y - |z|^2 & = 0
\end{align}
along with (\ref{csol1h}).


\bibliographystyle{utphys}
\bibliography{ref}

\def\polhk#1{\setbox0=\hbox{#1}{\ooalign{\hidewidth
  \lower1.5ex\hbox{`}\hidewidth\crcr\unhbox0}}} \def\cprime{$'$}
\providecommand{\href}[2]{#2}\begingroup\raggedright\begin{thebibliography}{10}

\bibitem{Polychronakos:2001mi}
A.~P. Polychronakos, ``{Quantum Hall states as matrix Chern-Simons theory},''
  {\em JHEP} {\bf 04} (2001) 011,
  \href{http://xxx.lanl.gov/abs/hep-th/0103013}{{\tt hep-th/0103013}}.

\bibitem{Dorey:2016mxm}
N.~Dorey, D.~Tong, and C.~Turner, ``{Matrix model for non-Abelian quantum Hall
  states},'' {\em Phys. Rev.} {\bf B94} (2016), no.~8 085114,
  \href{http://xxx.lanl.gov/abs/1603.09688}{{\tt 1603.09688}}.

\bibitem{Susskind:2001fb}
L.~Susskind, ``{The Quantum Hall fluid and noncommutative Chern-Simons
  theory},'' \href{http://xxx.lanl.gov/abs/hep-th/0101029}{{\tt
  hep-th/0101029}}.

\bibitem{Laughlin:1983fy}
R.~B. Laughlin, ``{Anomalous quantum Hall effect: An Incompressible quantum
  fluid with fractionallycharged excitations},'' {\em Phys. Rev. Lett.} {\bf
  50} (1983) 1395.

\bibitem{Hellerman:2001rj}
S.~Hellerman and M.~Van~Raamsdonk, ``{Quantum Hall physics equals
  noncommutative field theory},'' {\em JHEP} {\bf 10} (2001) 039,
  \href{http://xxx.lanl.gov/abs/hep-th/0103179}{{\tt hep-th/0103179}}.

\bibitem{Karabali:2001xq}
D.~Karabali and B.~Sakita, ``{Chern-Simons matrix model: Coherent states and
  relation to Laughlin wavefunctions},'' {\em Phys. Rev.} {\bf B64} (2001)
  245316, \href{http://xxx.lanl.gov/abs/hep-th/0106016}{{\tt hep-th/0106016}}.

\bibitem{Karabali:2001pw}
D.~Karabali and B.~Sakita, ``{Orthogonal basis for the energy eigenfunctions of
  the Chern-Simons matrix model},'' {\em Phys. Rev.} {\bf B65} (2002) 075304,
  \href{http://xxx.lanl.gov/abs/hep-th/0107168}{{\tt hep-th/0107168}}.

\bibitem{Cappelli:2004xk}
A.~Cappelli and M.~Riccardi, ``{Matrix model description of Laughlin Hall
  states},'' {\em J. Stat. Mech.} {\bf 0505} (2005) P05001,
  \href{http://xxx.lanl.gov/abs/hep-th/0410151}{{\tt hep-th/0410151}}.

\bibitem{Rodriguez:2008nz}
I.~D. Rodriguez, ``{Edge excitations of the Chern Simons matrix theory for the
  FQHE},'' {\em JHEP} {\bf 07} (2009) 100,
  \href{http://xxx.lanl.gov/abs/0812.4531}{{\tt 0812.4531}}.

\bibitem{Dorey:2016hoj}
N.~Dorey, D.~Tong, and C.~Turner, ``{A Matrix Model for WZW},'' {\em JHEP} {\bf
  08} (2016) 007, \href{http://xxx.lanl.gov/abs/1604.05711}{{\tt 1604.05711}}.

\bibitem{Tong:2015xaa}
D.~Tong and C.~Turner, ``{Quantum Hall effect in supersymmetric Chern-Simons
  theories},'' {\em Phys. Rev.} {\bf B92} (2015), no.~23 235125,
  \href{http://xxx.lanl.gov/abs/1508.00580}{{\tt 1508.00580}}.

\bibitem{Tong:2003vy}
D.~Tong, ``{A Quantum Hall fluid of vortices},'' {\em JHEP} {\bf 02} (2004)
  046, \href{http://xxx.lanl.gov/abs/hep-th/0306266}{{\tt hep-th/0306266}}.

\bibitem{Hanany:2003hp}
A.~Hanany and D.~Tong, ``{Vortices, instantons and branes},'' {\em JHEP} {\bf
  07} (2003) 037, \href{http://xxx.lanl.gov/abs/hep-th/0306150}{{\tt
  hep-th/0306150}}.

\bibitem{Manton:1997tg}
N.~S. Manton, ``{First order vortex dynamics},'' {\em Annals Phys.} {\bf 256}
  (1997) 114--131, \href{http://xxx.lanl.gov/abs/hep-th/9701027}{{\tt
  hep-th/9701027}}.

\bibitem{Romao:2000ru}
N.~M. Romao, ``{Quantum Chern-Simons vortices on a sphere},'' {\em J. Math.
  Phys.} {\bf 42} (2001) 3445--3469,
  \href{http://xxx.lanl.gov/abs/hep-th/0010277}{{\tt hep-th/0010277}}.

\bibitem{Romao:2004df}
N.~M. Romao and J.~M. Speight, ``{Slow Schroedinger dynamics of gauged
  vortices},'' {\em Nonlinearity} {\bf 17} (2004) 1337--1355,
  \href{http://xxx.lanl.gov/abs/hep-th/0403215}{{\tt hep-th/0403215}}.

\bibitem{Manton:1981mp}
N.~S. Manton, ``{A Remark on the Scattering of BPS Monopoles},'' {\em Phys.
  Lett.} {\bf 110B} (1982) 54--56.

\bibitem{Kosterlitz:1973xp}
J.~M. Kosterlitz and D.~J. Thouless, ``{Ordering, metastability and phase
  transitions in two-dimensional systems},'' {\em J. Phys.} {\bf C6} (1973)
  1181--1203.

\bibitem{oshikawa2007topological}
M.~Oshikawa, Y.~B. Kim, K.~Shtengel, C.~Nayak, and S.~Tewari, ``Topological
  degeneracy of non-abelian states for dummies,'' {\em Annals of Physics} {\bf
  322} (2007), no.~6 1477--1498.

\bibitem{papanicolaou1999semitopological}
N.~Papanicolaou and P.~Spathis, ``Semitopological solitons in planar
  ferromagnets,'' {\em Nonlinearity} {\bf 12} (1999), no.~2 285.

\bibitem{barenghi2016primer}
C.~F. Barenghi and N.~G. Parker, {\em A Primer on Quantum Fluids}.
\newblock SpringerBriefs in Physics. Springer, Cham, first~ed., 2016.

\bibitem{komineas2007rotating}
S.~Komineas, ``Rotating vortex dipoles in ferromagnets,'' {\em Physical review
  letters} {\bf 99} (2007), no.~11 117202.

\bibitem{komineas2007dynamics}
S.~Komineas and N.~Papanicolaou, ``Dynamics of vortex-antivortex pairs in
  ferromagnets,'' \href{http://xxx.lanl.gov/abs/0712.3684}{{\tt 0712.3684}}.

\bibitem{hertel2006exchange}
R.~Hertel and C.~M. Schneider, ``Exchange explosions: Magnetization dynamics
  during vortex-antivortex annihilation,'' {\em Physical review letters} {\bf
  97} (2006), no.~17 177202.

\bibitem{lee2005radiation}
K.-S. Lee, S.~Choi, and S.-K. Kim, ``Radiation of spin waves from magnetic
  vortex cores by their dynamic motion and annihilation processes,'' {\em
  Applied Physics Letters} {\bf 87} (2005), no.~19 192502.

\bibitem{van2006magnetic}
B.~Van~Waeyenberge, A.~Puzic, H.~Stoll, K.~Chou, T.~Tyliszczak, R.~Hertel,
  M.~F{\"a}hnle, H.~Br{\"u}ckl, K.~Rott, G.~Reiss, {\em et.~al.}, ``Magnetic
  vortex core reversal by excitation with short bursts of an alternating
  field,'' {\em Nature} {\bf 444} (2006), no.~7118 461.

\bibitem{hertel2007ultrafast}
R.~Hertel, S.~Gliga, M.~F{\"a}hnle, and C.~Schneider, ``Ultrafast nanomagnetic
  toggle switching of vortex cores,'' {\em Physical review letters} {\bf 98}
  (2007), no.~11 117201.

\bibitem{tretiakov2007vortices}
O.~Tretiakov and O.~Tchernyshyov, ``Vortices in thin ferromagnetic films and
  the skyrmion number,'' {\em Physical Review B} {\bf 75} (2007), no.~1 012408.

\bibitem{eisenstein1992new}
J.~Eisenstein, G.~Boebinger, L.~Pfeiffer, K.~West, and S.~He, ``New fractional
  quantum hall state in double-layer two-dimensional electron systems,'' {\em
  Physical review letters} {\bf 68} (1992), no.~9 1383.

\bibitem{suen1994origin}
Y.~Suen, H.~Manoharan, X.~Ying, M.~Santos, and M.~Shayegan, ``Origin of the
  $\nu$= 1/2 fractional quantum hall state in wide single quantum wells,'' {\em
  Physical review letters} {\bf 72} (1994), no.~21 3405.

\bibitem{murphy1994many}
S.~Murphy, J.~Eisenstein, G.~Boebinger, L.~Pfeiffer, and K.~West, ``Many-body
  integer quantum hall effect: evidence for new phase transitions,'' {\em
  Physical review letters} {\bf 72} (1994), no.~5 728.

\bibitem{chae2012direct}
S.~Chae, N.~Lee, Y.~Horibe, M.~Tanimura, S.~Mori, B.~Gao, S.~Carr, and S.-W.
  Cheong, ``Direct observation of the proliferation of ferroelectric loop
  domains and vortex-antivortex pairs,'' {\em Physical review letters} {\bf
  108} (2012), no.~16 167603.

\bibitem{hierro2017deterministic}
A.~Hierro-Rodriguez, C.~Quir{\'o}s, A.~Sorrentino, R.~Valc{\'a}rcel,
  I.~Est{\'e}banez, L.~Alvarez-Prado, J.~Mart{\'\i}n, J.~Alameda, E.~Pereiro,
  M.~V{\'e}lez, {\em et.~al.}, ``Deterministic propagation of vortex-antivortex
  pairs in magnetic trilayers,'' {\em Applied Physics Letters} {\bf 110}
  (2017), no.~26 262402.

\bibitem{Brink:1997zi}
L.~Brink, A.~Turbiner, and N.~Wyllard, ``{Hidden algebras of the
  (super)Calogero and Sutherland models},'' {\em J. Math. Phys.} {\bf 39}
  (1998) 1285--1315, \href{http://xxx.lanl.gov/abs/hep-th/9705219}{{\tt
  hep-th/9705219}}.

\bibitem{BasuMallick:1999vm}
B.~Basu-Mallick, H.~Ujino, and M.~Wadati, ``{Exact spectrum and partition
  function of $SU(m|n)$ supersymmetric Polychronakos model},'' {\em J. Phys.
  Soc. Jap.} {\bf 68} (1999) 3219--3226,
  \href{http://xxx.lanl.gov/abs/hep-th/9904167}{{\tt hep-th/9904167}}.

\bibitem{Hikami:1999twa}
K.~Hikami and B.~Basu-Mallick, ``{Supersymmetric Polychronakos Spin Chain:
  Motif, Distribution Function, and Character},'' {\em Nucl. Phys.} {\bf B566}
  (2000) 511, \href{http://xxx.lanl.gov/abs/math-ph/9904033}{{\tt
  math-ph/9904033}}.

\bibitem{Basu-Mallick:2008bsa}
B.~Basu-Mallick and N.~Bondyopadhaya, ``{Spectral properties of supersymmetric
  Polychronakos spin chain associated with $A_{N-1}$ root system},'' {\em Phys.
  Lett.} {\bf A373} (2009) 2831, \href{http://xxx.lanl.gov/abs/0811.3110}{{\tt
  0811.3110}}.

\bibitem{Kac:1977em}
V.~G. Kac, ``{Lie Superalgebras},'' {\em Adv. Math.} {\bf 26} (1977) 8--96.

\bibitem{MR914369}
F.~A. Berezin, {\em Introduction to superanalysis}, vol.~9 of {\em Mathematical
  Physics and Applied Mathematics}.
\newblock D. Reidel Publishing Co., Dordrecht, 1987.
\newblock Edited and with a foreword by A. A. Kirillov, With an appendix by V.
  I. Ogievetsky, Translated from the Russian by J. Niederle and R. Koteck\'y,
  Translation edited by Dimitri Le\u\i tes.

\bibitem{moens2007supersymmetric}
E.~Moens, {\em Supersymmetric Schur functions and Lie superalgebra
  representations}.
\newblock PhD thesis, Ghent University, 2007.

\bibitem{MR1354144}
I.~G. Macdonald, {\em Symmetric functions and Hall polynomials}.
\newblock Oxford Mathematical Monographs. The Clarendon Press, Oxford
  University Press, New York, second~ed., 1995.
\newblock With contributions by A. Zelevinsky, Oxford Science Publications.

\bibitem{Kirillov:1994en}
A.~N. Kirillov, ``{Dilogarithm identities},'' {\em Prog. Theor. Phys. Suppl.}
  {\bf 118} (1995) 61--142, \href{http://xxx.lanl.gov/abs/hep-th/9408113}{{\tt
  hep-th/9408113}}.

\bibitem{Nakayashiki:1995bi}
A.~Nakayashiki and Y.~Yamada, ``{Kostka polynomials and energy functions in
  solvable lattice models},'' \href{http://xxx.lanl.gov/abs/q-alg/9512027}{{\tt
  q-alg/9512027}}.

\bibitem{berele1987hook}
A.~Berele and A.~Regev, ``Hook young diagrams with applications to
  combinatorics and to representations of lie superalgebras,'' {\em Advances in
  mathematics} {\bf 64} (1987), no.~2 118--175.

\bibitem{MR869577}
A.~N. Kirillov and N.~Y. Reshetikhin, ``The {B}ethe ansatz and the
  combinatorics of {Y}oung tableaux,'' {\em Zap. Nauchn. Sem. Leningrad. Otdel.
  Mat. Inst. Steklov. (LOMI)} {\bf 155} (1986), no.~Differentsial\cprime naya
  Geometriya, Gruppy Li i Mekh. VIII 65--115, 194.

\bibitem{MR1048507}
A.~N. Kirillov, ``On the {K}ostka-{G}reen-{F}oulkes polynomials and
  {C}lebsch-{G}ordan numbers,'' {\em J. Geom. Phys.} {\bf 5} (1988), no.~3
  365--389.

\bibitem{Kac:1994kn}
V.~G. Kac and M.~Wakimoto, ``{Integrable highest weight modules over affine
  superalgebras and number theory},''
  \href{http://xxx.lanl.gov/abs/hep-th/9407057}{{\tt hep-th/9407057}}.

\bibitem{MR1810948}
V.~G. Kac and M.~Wakimoto, ``Integrable highest weight modules over affine
  superalgebras and {A}ppell's function,'' {\em Comm. Math. Phys.} {\bf 215}
  (2001), no.~3 631--682.

\bibitem{MR969247}
D.~Hickerson, ``A proof of the mock theta conjectures,'' {\em Invent. Math.}
  {\bf 94} (1988), no.~3 639--660.

\bibitem{hikami1995yangian}
K.~Hikami, ``Yangian symmetry and virasoro character in a lattice spin system
  with long-range interactions,'' {\em Nuclear Physics B} {\bf 441} (1995),
  no.~3 530--548.

\bibitem{hikami2000supersymmetric}
K.~Hikami and B.~Basu-Mallick, ``Supersymmetric polychronakos spin chain:
  motif, distribution function, and character,'' {\em Nuclear Physics B} {\bf
  566} (2000), no.~3 511--528.

\bibitem{Polychronakos:1993wc}
A.~P. Polychronakos, ``{Exact spectrum of SU(n) spin chain with inverse square
  exchange},'' {\em Nucl. Phys.} {\bf B419} (1994) 553--566,
  \href{http://xxx.lanl.gov/abs/hep-th/9310095}{{\tt hep-th/9310095}}.

\bibitem{Geroch:1972yt}
R.~P. Geroch, ``{A Method for generating new solutions of Einstein's equation.
  2},'' {\em J. Math. Phys.} {\bf 13} (1972) 394--404.

\bibitem{Hawking:1981bu}
S.~W. Hawking and M.~Rocek, eds., {\em {Superspace and Supergravity.
  Proceedings, Nuffield Workshop, Cambridge, UK, June 16 - July 12, 1980}},
  1981.

\bibitem{Julia:1981wc}
B.~Julia, ``{Infinite Lie Algebras in Physics},'' in {\em {Unified field
  theories and beyond}}, pp.~23--41, 1981.

\bibitem{Betzios:2017yms}
P.~Betzios and O.~Papadoulaki, ``{FZZT branes and non-singlets of Matrix
  Quantum Mechanics},'' \href{http://xxx.lanl.gov/abs/1711.04369}{{\tt
  1711.04369}}.

\bibitem{Mikhaylov:2014aoa}
V.~Mikhaylov and E.~Witten, ``{Branes And Supergroups},'' {\em Commun. Math.
  Phys.} {\bf 340} (2015), no.~2 699--832,
  \href{http://xxx.lanl.gov/abs/1410.1175}{{\tt 1410.1175}}.

\bibitem{Okazaki:2015fiq}
T.~Okazaki and D.~J. Smith, ``{Topological M-Strings and Supergroup WZW
  Models},'' {\em Phys. Rev.} {\bf D94} (2016) 065016,
  \href{http://xxx.lanl.gov/abs/1512.06646}{{\tt 1512.06646}}.

\bibitem{Okazaki:2016pne}
T.~Okazaki and D.~J. Smith, ``{Mock Modular Index of M2-M5 Brane System},''
  {\em Phys. Rev.} {\bf D96} (2017), no.~2 026017,
  \href{http://xxx.lanl.gov/abs/1612.07565}{{\tt 1612.07565}}.

\end{thebibliography}\endgroup

\end{document}